\documentclass[10pt,a4paper]{article}
\pdfoutput=1
\usepackage{jcappub}

\usepackage{tabularx,varwidth,array,ragged2e,multirow}
\usepackage{graphicx}
\usepackage{float}
\usepackage{enumerate}
\usepackage{caption}
\usepackage[hang,flushmargin]{footmisc} 
\usepackage{amsfonts,amsmath,amssymb,amsthm}

\newcommand{\eq}[1]{Eq.~(\ref{#1})}
\newcommand{\eqs}[1]{Eqs.~(\ref{#1})}

\newcommand{\bea}{\begin{eqnarray}}
\newcommand{\eea}{\end{eqnarray}}
\newcommand{\bal}{\begin{alignedat}}
\newcommand{\eal}{\end{alignedat}}
\newcommand{\beq}{\begin{equation}}
\newcommand{\eeq}{\end{equation}}
\newcommand{\bit}{\begin{itemize}}
\newcommand{\eit}{\end{itemize}}
\newcommand{\benu}{\begin{enumerate}}
\newcommand{\eenu}{\end{enumerate}}
\newcommand{\bdes}{\begin{description}}
\newcommand{\edes}{\end{description}}
\newcommand{\nn}{\nonumber}

\newcommand{\hc}{\rm{h.c.}}

\newcommand{\pare}[1]{\left( #1 \right)}
\newcommand{\sqpare}[1]{\left[ #1 \right]}
\newcommand{\ang}[1]{\left\langle #1 \right\rangle}

\def\g5{\gamma_{5}}
\newcommand{\slashed}[1]{{#1}\hspace{-2mm}/}

\def \eV{\: {\rm eV}}
\def\keV{\: {\rm keV}}
\def\MeV{\: {\rm MeV}}
\def\GeV{\: {\rm GeV}}
\def\TeV{\: {\rm TeV}}

\def\gr {\: {\rm g}}

\def \cm{\: {\rm cm}}
\def \km{\: {\rm km}}

\def\kpc{\: {\rm kpc}}

\def \snd{\: {\rm s}}

\def \Gyr{\: {\rm Gyr}}

\def\a{\alpha}
\def\b{\beta}
\def\g{\gamma}
\def\d{\delta}
\def\e{\epsilon}
\def\z{\zeta}

\def\th{\theta}

\def\l{\lambda}
\def\m{\mu}
\def\n{\nu}
\def\ks{\xi}

\def\p{\pi}
\def\r{\rho}
\def\s{\sigma}

\def\f{\phi}

\def\w{\omega}

\def\fvar{\varphi}

\def\G{\Gamma}
\def\D{\Delta}

\def\S{\Sigma}

\def\W{\Omega}

\def\DM  {_{_{\rm DM}}}

\def\BD  {B_{_D}}
\def\LD  {L_{_D}}
\def\eD  {e_{_D}}
\def\eDL {e_{_{D,L}}}
\def\eDR {e_{_{D,R}}}
\def\pD  {p_{_D}}
\def\HD  {H_{_D}}
\def\gammaD  {\gamma_{_D}}
\def\fD  {\f_{_D}}
\def\fvarD {\fvar_{_D}}
\def\vD  {v_{_D}}
\def\aD  {\a_{_{D}}}
\def\muD {\mu_{_{D}}}
\def\MD  {M_{_{D}}}
\def\xD  {x_{_D}}
\def\mpD  {m_{\bf p}}
\def\meD  {m_{\bf e}}
\def\mH  {m_{_{\bf H}}}

\def\mpl {M_{\rm Pl}}

\def\vrel {v_{\rm rel}}

\def\TV {T_{_{\rm V}}}
\def\TD {T_{_{\rm D}}}
\def\gV {g_{_{\rm V}}}
\def\gD {g_{_{\rm D}}}

\title{Self-interacting asymmetric dark matter coupled to a light massive dark photon}

\author[a]{Kalliopi Petraki,}
\author[b]{Lauren Pearce} 
\author[b,c]{and Alexander Kusenko}

\affiliation[a]{Nikhef, Science Park 105, 1098 XG Amsterdam, The Netherlands}
\affiliation[b]{Department of Physics and Astronomy, University of California, Los Angeles, CA 90095-1547, USA}
\affiliation[c]{Kavli IPMU (WPI), University of Tokyo, Kashiwa, Chiba 277-8568, Japan}
\date{\today}

\abstract{
Dark matter (DM) with sizeable self-interactions mediated by a light species offers a compelling explanation of the observed galactic substructure; furthermore, the direct coupling between DM and a light particle contributes to the DM annihilation in the early universe.  If the DM abundance is due to a dark particle-antiparticle asymmetry, the DM annihilation cross-section can be arbitrarily large, and the coupling of DM to the light species can be significant.  We consider the case of asymmetric DM interacting via a light (but not necessarily massless) Abelian gauge vector boson, a dark photon. In the massless dark photon limit, gauge invariance mandates that DM be multicomponent, consisting of positive and negative dark ions of different species which partially bind in neutral dark atoms.  
We argue that a similar conclusion holds for light dark photons; in particular, we establish that the multi-component and atomic character of DM persists in much of the parameter space where the dark photon is sufficiently light to mediate sizeable DM self-interactions. We discuss the cosmological sequence of events in this scenario, including the dark asymmetry generation, the freeze-out of annihilations, the dark recombination and the phase transition which gives mass to the dark photon.   We estimate the effect of self-interactions in DM haloes, taking into account this cosmological history.  We place constraints based on the observed ellipticity of large haloes, and identify the regimes where DM self-scattering can affect the dynamics of smaller haloes, bringing theory in better agreement with observations. Moreover, we estimate the cosmological abundance of dark photons in various regimes, and derive pertinent bounds. 
}

\arxivnumber{1403.1077}

\begin{document}
\maketitle
 
\section{Introduction}
\label{sec:intro} 

The gravitational clustering of dark matter (DM) provides essential information for understanding its nature. It is now well established that the observed structure of the universe in galaxy-cluster and larger scales can be explained extremely well within the collisionless cold DM (CDM) paradigm. In contrast, it cannot be reproduced well under the assumption that the dominant component of DM is hot, with this possibility being therefore excluded. At smaller scales,  there are currently discrepancies between observations and the predictions of collisionless CDM. A number of problems have been identified. The subhaloes formed in collisionless CDM simulations of Milky-Way-size haloes exceed in number the observed dwarf galaxies of the Milky Way by a large number. More importantly, the most massive of these subhaloes are too dense to host the brightest dwarfs of the Milky Way~\cite{BoylanKolchin:2011de,BoylanKolchin:2011dk}. In addition, the subhalo density profiles are predicted to be cuspier than what observations of dwarf galaxies favour.  These discrepancies are now understood to emanate from the same problem: Numerical simulations of collissionless CDM predict too much mass in the central regions of haloes and subhaloes~\cite{Weinberg:2013aya}. Baryonic physics has been invoked to alleviate this problem. It is however unclear whether baryons can influence the DM halo dynamics in systems where they are greatly subdominant. The disagreement between observations and collisionless CDM predictions may in fact indicate the need for a shift from the collisionless CDM paradigm. Observations seem to favour a scenario which can reproduce the large-scale structure of the universe equally well, while suppressing the formation of structure at smaller scales.

Self-interacting DM has emerged as an alternative to collisionless CDM which can potentially successfully address the above issues~\cite{Spergel:1999mh,Wandelt:2000ad,Mohapatra:2001sx,Rocha:2012jg,Peter:2012jh,Vogelsberger:2012ku,Vogelsberger:2012sa,Zavala:2012us}. Various scenarios have been proposed~\cite{Faraggi:2000pv,Kusenko:2001vu,Foot:2014mia,Boddy:2014yra,Hochberg:2014dra,Alves:2010dd,CyrRacine:2012fz,Cyr-Racine:2013fsa,Kaplan:2009de,Kaplan:2011yj,Cline:2012is,Cline:2013pca,Cline:2013zca}.
The self-scattering of DM in haloes redistributes the energy and momentum among DM particles, thus heating the low-entropy material which would otherwise concentrate in the core of the galaxies. As a result, the inner density of DM haloes is reduced, their central density profiles become less cuspy, and the star formation rate is suppressed. While these features might be supported by observations of dwarf-galaxy-size haloes, larger haloes put constraints on how self-interacting DM can be. In particular, the self-scattering of DM, if too strong, can isotropise the haloes. This may potentially be in conflict with observations which show that larger galaxies are elliptical. Recent simulations of single-component self-interacting DM show that the dynamics of small haloes can be affected while the ellipticity of larger haloes is retained, if DM self-scatters with a velocity-independent cross-section per unit mass in a narrow range around $\s/m\DM \sim 0.5 \cm^2/\gr$~\cite{Rocha:2012jg,Peter:2012jh}. Baryonic effects may widen this range~\cite{Kaplinghat:2013xca}.
On the other hand, DM self-interactions whose cross-section decreases with increasing velocity can more easily reproduce observations~\cite{Vogelsberger:2012ku,Vogelsberger:2012sa,Zavala:2012us}. Indeed, such interactions can be efficient in the smaller haloes, which possess a small velocity dispersion, while they become ineffective in larger haloes, which have larger velocity dispersions~\cite{Feng:2009mn,Feng:2009hw,Loeb:2010gj,Vogelsberger:2012ku,Vogelsberger:2012sa,Zavala:2012us}.

Interactions whose strength decreases with increasing velocity are those mediated by light force carriers. In this case, the scattering amplitude is determined by the momentum transfer which dominates over the mass of the mediator. It is important of course that the long-range nature of these interactions be curtailed, such that there is no effect on the clustering of matter in very large distances. Depending on the specific nature of the interaction considered, long-range forces may be screened by a non-zero mediator mass, by the Debye length in neutral plasma, and/or by the formation of neutral bound states.

In this paper, we explore the cosmology of and the astrophysical implications of \emph{asymmetric} DM coupled to a massive albeit light vector boson, henceforth called the ``dark photon".  The limit of zero dark photon mass has been studied in Refs.~\cite{CyrRacine:2012fz,Cyr-Racine:2013fsa,Kaplan:2009de,Kaplan:2011yj,Cline:2012is,Cline:2013pca,Cline:2013zca}, and we use some of their results in our analysis. In this limit, gauge invariance dictates that dark matter is multi-component, consisting of positive and negative dark ions of different species which bind partially in neutral dark atoms.  We argue that if the dark photon is sufficiently light, gauge invariance still implies that significant abundances of both positively and negatively charged dark ions have survived until present.  We estimate the maximum dark photon mass for which this is inevitable.  Following this, we show that this regime encompasses much of the parameter space of interest, in which the following two conditions are satisfied: first, the dark photon is sufficiently light to mediate long-range DM self-scattering and, secondly the DM coupling to the dark photon is sufficiently strong for this scattering to affect the halo dynamics.  In this multi-component regime, DM self-interactions in haloes today are suppressed with respect to what would be expected if DM were single-component, due to the formation of neutral DM bound states (dark atoms) in the early universe. With respect to the massless dark photon case, there is additional screening of the ionised component self-interactions, due to the non-zero mass of the dark photon. Accounting properly for these effects, we estimate the impact of DM self-interaction in haloes. We circumscribe the parametric regimes disfavoured by the observed ellipticity of large haloes, and identify the regions where DM self-interactions can potentially affect the dynamics of smaller haloes. We demonstrate how the continuum of dark photon masses --from the regime where the dark photon is heavy and mediates an effectively short-range interaction, to the limit where the dark photon is exactly massless-- can produce viable scenarios.

We focus on asymmetric DM for two reasons. The direct coupling of DM to a light species, here the dark photon, contributes also to the annihilation of DM. Requiring that the coupling is sufficiently strong such that the DM self-interaction is sizeable yields a minimum contribution to the DM annihilation cross-section, which may in turn exceed the canonical value for symmetric thermal relic DM.   On the other hand, the asymmetric DM scenario can accommodate arbitrarily large annihilation cross-sections. Indeed, the relic abundance of asymmetric DM is determined by an excess of dark particles over antiparticles 
and by the DM mass, rather than by the DM annihilation cross-section. (For reviews on asymmetric DM, see Refs.~\cite{Davoudiasl:2012uw,Petraki:2013wwa,Zurek:2013wia,Boucenna:2013wba,Volkas:2013eia}.) In this sense, asymmetric DM  encompasses a much larger parameter space in which DM can exhibit sizeable self-interactions. Moreover, 
asymmetric DM can provide a dynamical explanation for the near coincidence of the dark and the ordinary matter abundances, which are observed to differ only by a factor of a few. While the similarity of the relic abundances is the most robust argument for considering a relation between the physics of DM and ordinary matter, the observed clustering of DM at small scales lends extra support to this idea. Indeed, as seen from the estimate given above, the DM self-scattering cross-section per unit mass required to affect the dynamics of small haloes is within one order of magnitude from the neutron self-scattering cross-section per unit mass. 

This paper is organised as follows.  In the next section we discuss further the motivation for considering asymmetric DM coupled to a gauge vector boson and introduce our model.  We describe briefly the case of a massless dark photon. Then we turn to the case of a dark photon acquiring a non-zero mass, and establish the multi-component and atomic character of DM in the case of small dark photon masses. We discuss this issue further in Sec.~\ref{sec:cosmo}, where we inspect in detail the cosmological sequence of events in the scenario under consideration. In Sec.~\ref{sec:self-inter}, we describe the DM self-interactions, and delineate the relevant parameter space. We conclude in Sec.~\ref{sec:conc}.

%%%%%%%%%%%%%%%%%%%%%%%%%%%%%%%%%%%%%%%
%%%%%%%%%%%%%%%%%%%%%%%%%%%%%%%%%%%%%%%
\section{Atomic dark matter}
\label{sec:atomic}

Thermal relic DM, whether symmetric or asymmetric, presupposes interactions which annihilate efficiently the DM population until it reduces to the observed DM density.  A particle-antiparticle asymmetry suppresses the overall annihilation rate, thus necessitating a larger annihilation cross-section than in the case of symmetric DM, albeit only by a factor of a few. In fact, the antiparticles reduce to less than $1\%$ of the total DM density if the annihilation cross-section is only 2.4 times larger than the canonical value for symmetric thermal relic DM~\cite{Graesser:2011wi}. This leaves the excess of DM particles as the dominant component of DM; obtaining the correct abundance fixes the product of the DM asymmetry and mass. While this means that asymmetric DM need be only weakly interacting, current bounds from colliders and direct detection experiments highly constrain the possibility of weak-scale annihilation of DM directly into Standard Model (SM) particles~\cite{Bai:2010hh,Buckley:2011kk,Fox:2012ee,MarchRussell:2012hi,Haisch:2012kf}. The constraints are already severe for symmetric DM, but become even more so for asymmetric DM which requires an at least somewhat larger annihilation cross-section. 

This motivates considering a dark interaction via which DM annihilates either into new stable light degrees of freedom (d.o.f.s), or into metastable species which subsequently decay into Standard Model (SM) particles.\footnote{Since the primary incentive for considering asymmetric DM is not any theoretical expectation of new physics related to the electroweak interactions of the SM, invoking dark interactions and dark light species is a completely natural possibility which does not remove any of the motivation for this class of theories.}  
A new Abelian gauge group under which DM is charged stands out as a minimal possibility, and  appears in many asymmetric DM models (see e.g.~\cite{Petraki:2011mv,vonHarling:2012yn,Kaplan:2011yj,Bell:2011tn}).  Not only can it provide for efficient annihilation of DM, but it also introduces structural complexity in the dark sector (in comparison to scalar or Yukawa couplings) which can result in the emergence of an accidental particle-number symmetry at low energies. The latter is of course an essential feature of asymmetric DM models, where the dark particle-antiparticle excess is maintained in the low-energy environment of today's universe due to a particle-number symmetry governing the low-energy interactions of DM. We shall refer to this symmetry as the {\sl dark baryon number} $\BD$, in analogy to the ordinary baryon-number symmetry of the SM which is responsible for the conservation of the baryon asymmetry of the universe and the relic abundance of ordinary matter.

\subsection{Massless dark photon}
\label{sec:massless}

Of course, any particle asymmetry under a global number has to be generated by gauge-invariant interactions which uphold the gauge-charge neutrality of the universe.  If DM is charged under an unbroken Abelian gauge group U(1)$_{_D}$, any particle-number asymmetry carried by a DM species must be compensated by an opposite gauge-charge asymmetry carried by (at least one) different species.\footnote{If asymmetric DM carries non-Abelian gauge charges, gauge-charge neutrality can often be ensured also by an appropriate combination of the various ``flavours'' or ``colours'' of the DM multiplet(s). Referring again to ordinary matter, the valence quarks of protons and neutrons form SU(3)$_c$-neutral combinations.}   
The stability of DM and the other species can be understood as a consequence of the fact that they are the lightest d.o.f.s charged under the global symmetry and the gauge symmetry respectively. This is analogous to the properties of ordinary matter: the baryonic asymmetry carried by the protons is inevitably associated with a net positive electric charge. This is in turn compensated by an asymmetric population of electrons. Protons, being the lightest baryons in the SM, are stable, and electrons are similarly stable as the lightest electrically charged particles.

Thus, in the simplest realisation of the scenario under consideration, involving the minimal assumptions of asymmetric DM coupled to a massless gauge vector boson, gauge invariance implies that DM consists of two stable and fundamental particle species, oppositely charged under U(1)$_{_D}$, which we shall assume here to be fermionic. Adopting partly the notation appearing in recent literature, we shall refer to them as the dark proton, $\pD$, and the dark electron, $\eD$, with masses $\mpD$ and $\meD$ and U(1)$_{_D}$ charges $q_{\bf p} = +1$ and $q_{\bf e} = -1$ respectively. We take $\mpD \geqslant \meD$. The low-energy effective Lagrangian is
\beq
{\cal L}_0 =  \bar{p}_{_D}(i \slashed{D} - \mpD) \pD + \bar{e}_{_D}(i \slashed{D} - \meD) \eD 
- \frac{1}{4} F_{_D \, \m\n} F_{_D}^{\m\n} \ ,
\label{eq:L0}
\eeq
where $F_{_D}^{\m\n} = \partial^\m A_{_D}^\n - \partial^\n A_{_D}^\m$, with $A_{_D}$ being the dark-photon field.  (As in QED, we will use $A_{_D}^\m$ for the field in the Lagrangian, and $\gammaD$ for the photon when discussing processes such as $\eD^+ + \eD^- \rightarrow \gammaD \gammaD$.) The covariant derivative for $\pD$ and $\eD$ is $D^\m  = \partial^\m + i q_{\bf i} g A_{_D}^\m$, where $q_{\bf i}$ is the respective charge and $g$ is the gauge coupling of the dark force.  In the following, we shall use instead the dark fine-structure constant $\aD \equiv g^2/4\p$.
Moreover, the generation of a $\BD$ asymmetry in the early universe implies the existence of high-energy interactions which may generate gauge-invariant and $\BD$-violating effective operators of the kind
\beq
{\cal L}_{\slashed{B}_{_D}} \supset  
(\overline{e^c_{_D}} \, \pD)^n  \, {\cal O}_{\rm GI} \ ,
\eeq
where ${\cal O}_{\rm GI}$ is an operator invariant under both U(1)$_{_D}$ and the SM gauge group. If ${\cal O}_{\rm GI} = 1$, then $n \geqslant 2$ is implied by the assumption of a low-energy global symmetry $\BD$.  For the dark and the ordinary baryonic asymmetries to have been related dynamically in the early universe, ${\cal O}_{\rm GI}$ must transform under the ordinary $(B-L)_{_V}$ symmetry of the SM.

The relic populations of $\pD$ and $\eD$ are asymmetric, with the amount of antiparticles having survived until today being entirely negligible. In this framework, the dark proton and the dark electron can form U(1)$_{_D}$-neutral bound states, dark Hydrogen atoms $\HD$, with mass $\mH \equiv \mpD + \meD -\D$, where $\D$ is the ground-state binding energy of the dark atoms, with $\D =  (1/2) \muD \aD^2$. Here $\muD \equiv \mpD \meD/(\mpD + \meD)$ is the reduced mass of the $\pD - \eD$ system, which satisfies  the consistency condition
\beq 4 \muD \leqslant \mH + \D \ , \label{eq:muD upper} \eeq
with the equality being realised for $\mpD = \meD$.  In this scenario, dark matter today consists in general of a mixture of dark ions, $\pD^+$ and $\eD^-$, and dark Hydrogen atoms, $\HD$. Considering its low energy phenomenology presupposes considering first the cosmology of atomic DM, most importantly the process of dark recombination.  This has been studied in detail in Ref.~\cite{CyrRacine:2012fz}. In the next section we review some important results.\footnote{Note that the atomic DM scenario we explore in this paper is different from the scenario in which DM consists of heavy particles carrying ordinary electromagnetic charge and forming bound states with ordinary atoms~\cite{Khlopov:2007ic,Khlopov:2008ty,Belotsky:2014haa}.}

The phenomenology of atomic DM can be quite rich. This fact in itself motivates the study of atomic DM, in addition to this being a minimal scenario arising in asymmetric DM models.  In particular, atomic DM is a scenario of multi-component DM with the various species -- dark ions and atoms -- having the same origin. This is in contrast to other scenarios invoking multi-component DM, in which the various components are typically unrelated and their existence in comparable amounts in the universe today has no obvious justification.  Moreover, as described in the introduction and we shall see in more detail in Sec.~\ref{sec:self-inter}, atomic DM can be self-interacting, with its various self-scattering cross-sections being strongly velocity-dependent. The coupling of the DM species to dark radiation can also result in late DM kinetic decoupling, which can suppress structure at small scales and can contribute, in a different way than DM self-scattering in haloes, to the resolution of the small-scale structure problems of the standard DM paradigm.  Finally, the cosmological abundance of massless dark photons can account for the excess of relativistic energy favoured by CMB data.

\subsection{Massive dark photon}
\label{sec:massive}

In the remainder of this paper, we generalise the above scenario to non-zero, albeit small dark photon masses $\MD$. As we shall show, a low dark photon mass ensures that the main features of the scenario --the multi-component and atomic nature of DM-- remain the same, while it may still result in different phenomenology.
A small $\MD$ (to be defined more precisely in the following) retains the long-range character of the DM self-scattering in haloes, while it screens the ion-ion interaction at longer distances. We shall explore this in detail. Moreover, a massive dark photon may result in distinct direct and indirect detection signals, which however do not explore in this work.

Dark photons may acquire mass either via the St\"{u}ckelberg mechanism, or via the Higgs mechanism after a cosmological phase transition which breaks U(1)$_{_D}$.  In the case of the St\"{u}ckelberg mechanism~\cite{Stueckelberg:1938zz}, the particles charged under U(1)$_{_D}$ couple to the dark photon via a conserved current~\cite{Kors:2005uz}. The discussion and the arguments of the previous section remain thus valid:  Gauge invariance still implies that dark matter is multi-component, consisting of asymmetric populations of two stable particle species, oppositely charged under U(1)$_{_D}$. This is in fact independent of how large the dark photon mass is.  In the following, we focus on the case of the dark photons acquiring mass due to the breaking of U(1)$_{_D}$ via the Higgs mechanism (as has been described for example in Ref.~\cite{Batell:2009yf}).

To break U(1)$_{_D}$, we introduce a complex scalar field $\fD$, with charge $q_\f$, interacting via the Lagrangian
\beq
{\cal L}_\f = D_\m \fD^\dagger \, D^\m \fD - \l \pare{ |\fD|^2 - \vD^2 }^2 \ .
\label{eq:Lf}
\eeq
where $D^\m\fD = (\partial^\m + i q_\f g A^\m)\fD$ and $q_\f$ is the charge of $\fD$ under U(1)$_{_D}$.
The field $\fD$ acquires a vacuum expectation value $\ang{\fD} = \vD$, as a result of which the dark photon acquires mass
\beq
\MD = (8\p q_\f^2 \aD)^{1/2} \: \vD \ .
\eeq 
Expanding around the vacuum, in the unitarity gauge,  $\fD = \vD + \fvarD/\sqrt{2}$, the self-couplings of the physical scalar field $\fvarD$ are
\beq
{\cal L}_\fvar = \frac{1}{2} (\partial_\m \fvarD)(\partial^\m \fvarD) - \sqpare{ 
\frac{1}{2} m_\fvar^2 \fvarD^2 + (\p q_\f^2 \aD)^{1/2} \, \frac{m_\fvar^2}{\MD} \, \fvarD^3 
+ \frac{\p q_\f^2 \aD}{2} \, \frac{m_\fvar^2}{\MD^2} \, \fvarD^4
} \ ,
\label{eq:L vf}
\eeq
where $m_\fvar^2 = 4\l \vD^2$ is the mass-squared of the physical scalar field $\fvar_{_D}$. The interaction of $\fvar$ with the dark photon is described by the terms
\beq
{\cal L}_{\rm int} = (2\p q_\f^2 \aD) \, \fvarD^2 \, A_\m A^\m + (4\p q_\f^2 \aD)^{1/2} \MD \, \fvarD \,  A_\m A^\m \ .
\label{eq:L int}
\eeq
We will use \eqs{eq:L vf} and \eqref{eq:L int} in estimating the cosmological abundances of $\fvarD$ and $\gammaD$ in Sec.~\ref{sec:dark photons}.

The above description applies to today's low-energy universe. However, thermal corrections to the scalar potential ensure that the U(1)$_{_D}$ symmetry was restored when the universe was at very high temperatures~\cite{Linde:1978px}. The transition of the universe to the U(1)$_{_D}$-broken vacuum occurred when the dark plasma was at temperature 
\beq 
T_{_{\rm D, \, PT}} \sim v_{_D} = \MD/(8\p q_\f^2 \aD)^{1/2} \ .
\label{eq:T_PT}
\eeq  
Clearly, the magnitude of the dark photon mass is correlated to the cosmological sequence of events, which includes the $\BD$ asymmetry generation, the freeze-out of the DM annihilations, the dark recombination and the U(1)$_{_D}$ phase transition. The cosmological sequence of these events is in turn critical in determining the composition and the phenomenology of DM today, as we discuss below.

For completeness, we mention that in general, $\fD$ may couple to the SM Higgs $H$ via the renormalisable operator~\cite{Foot:1991bp}
\beq \d {\cal L}_{_{\f H}} = \l_{\f H} |\fD|^2 |H|^2 \ . \label{eq:Higgs portal} \eeq 
The coupling $\l_{\f H}$ implies that after the electroweak and U(1)$_{_D}$ symmetry breaking, $\fD$ and $H$ mix.  Here we are mostly interested in light dark photons and therefore a light dark Higgs, with mass $m_\f \ll m_{_H} \simeq 126 \GeV$, where $m_{_H}$ is the SM Higgs mass. The hierarchy between $m_{_H}$ and $m_\f$ is stable if either $\l_{\f H}$ is sufficiently small~\cite{Volkas:1988cm,Foot:1991bp,Foot:2013hna}, or the theory is supersymmetric.  Since the focus of the present study is the DM self-interaction in haloes, which of course does not depend on the dark-ordinary sector couplings,  $\l_{\f H}$ may be taken to be arbitrarily small. 
An analysis of the Higgs mixing, and resulting bounds and observational signatures can be found e.g.~in Ref.~\cite{Ahlers:2008qc}. For a recent study of the signatures of a light scalar ($\sim 100 \MeV - 10 \GeV$) mixing with the SM Higgs, see Ref.~\cite{Clarke:2013aya}.

While not essential in our study, in the following we shall also allow for the dark gauge force to mix kinetically with the hypercharge via the renormalisable operator~\cite{Holdom:1985ag,Foot:1991kb}
\beq
\d {\cal L}_{\rm kin} =  \frac{\e }{ 2 } \: F_{Y  \m\n} \, F_D^{\m\n}  \ .
\label{eq:Lkin} 
\eeq
This coupling makes the massive dark photons unstable against decay into SM charged fermions, so long as $\MD > 1.022 \MeV$.\footnote{
For $\MD < 1.022 \MeV$ and $\e \neq 0$, dark photons are still unstable, albeit their decay rate is extremely suppressed in the mass range of interest.  Dark photons may decay via their mixing to the $Z$ boson into neutrinos, or via a charged-fermion loop into three photons. The corresponding rates are
\[
\G_{\gammaD \to \bar{\n}\n} = \frac{1}{3}\e^2 \, \a_{_{\rm EM}} \MD  \times \frac{ 3\MD^4}{4 \cos^2 \th_{_{\rm W}} m_Z^4} \ , \quad
\G_{\gammaD \to 3 \g} = \frac{17 \e^2 \a_{_{\rm EM}}^4 \MD^9}{2^7 3^6 5^3 \p^3 \tilde{m}_e^8} \ ,
\label{eq:gammaD to 3 gamma}
\]
where $\a_{_{\rm EM}} = 1/137$, $\tilde{m}_e$ is the ordinary electron mass, $m_Z$ is the $Z$ boson mass, and $\th_{_{\rm W}}$ is the Weinberg angle. The $\gammaD \to 3 \gamma$ rate has been calculated in Ref.~\cite{Pospelov:2008jk}.
}        
Their decay rate is (see e.g.~\cite{Batell:2009yf})
\beq
\G_{\gammaD \to f^+ f^-} =  f_{_{\rm EM}} \times \frac{1}{3}\,  \e^2 \a_{_{\rm EM}} \MD \ ,
\label{eq:gammaD to f+f-}
\eeq
where $f_{_{\rm EM}}$ accounts for the number of kinematically available channels.
% ($f_{_{\rm EM}} = 1$ for decay into $e^{\pm}$ only).  
If the cosmological abundance of dark photons is eliminated via decay, then the bounds on the dark-sector temperature are relaxed (see Sec.~\ref{sec:dark photons}). However, similarly to the scalar coupling $\l_{\f H}$, the kinetic mixing $\e$ may be vanishingly small.

\subsubsection{Asymmetry generation}

As described above, in the case of an unbroken U(1)$_D$, gauge invariance implies that equal asymmetries of dark protons and dark electrons are generated. 
This remains valid in the case of a mildly broken U(1)$_{_D}$.
The breaking of U(1)$_{_D}$ occurs at dark-sector temperature given by \eq{eq:T_PT}.  As long as the transition to the broken phase takes place after the $\BD$-asymmetry generation, gauge invariance still implies, according to the discussion in the previous section, that equal asymmetries of $\pD$ and $\eD$ must be generated.  $\BD$-asymmetry generation has to take place before annihilations diminish the abundance of DM below the observed DM density, i.e. while $Y_{\bf p} \equiv n_{\bf p} / s > \W\DM \r_c/(s_0 \mpD) \simeq 10^{-11} (100 \GeV /\mpD)$, where $n_{\bf p}$ is the number density of the dark protons, $s$ is the entropy density of the universe, $\r_c$ and $s_0$ are the critical energy density and the entropy density of the universe today, and $\W\DM \simeq 0.25$. For thermal DM, this implies $\mpD/ T_{\rm asym} < 25 + \ln \pare{\mpD /100 \GeV }$, where $T_{\rm asym}$ is the dark-sector temperature at the time of dark asymmetry generation. Realistically, $\BD$-genesis occurs at higher temperatures than this limit. Even in this unrealistic limit, the gauge symmetry is unbroken at the time of asymmetry generation, $T_{\rm asym} > T_{_{\rm D, \, PT}}$, if~\footnote{Here we assumed that the thermal bath of the dark and the ordinary sectors are at the same temperature at the time of asymmetry generation. This is indeed expected if a relation between the dark and ordinary asymmetries was established dynamically by high-energy processes in the early universe. If the two sectors were at different temperatures at the time of dark-asymmetry generation, and the ordinary sector dominated the energy density of the universe, then this would introduce only logarithmic corrections in the ratio of temperatures of the dark and the ordinary sectors, $\ks \equiv \TD/\TV$, to the above limit: $\mpD/T_{\rm asym} <  25 + \ln \pare{\mpD /100 \GeV } + \ln\ks^3$, with $\ks <1$. Thus, the condition of \eq{eq:MDasym} would in fact encompass additional parameter space.   \label{foot:Tasym} }    
\beq 
\MD \lesssim  (8\p q_\f^2 \aD)^{1/2} \, \mpD / x_{\rm asym} \ , \qquad \text{\rm where} \quad
x_{\rm asym} \simeq 25 + \ln \pare{\mpD /100 \GeV }
\ .
\label{eq:MDasym} 
\eeq 
The condition \eqref{eq:MDasym} implies the generation of an $\eD$ asymmetry along with the $\pD$ asymmetry; however, it does not alone ensure the survival of a significant $\eD$ density at late times. 
It is possible that after the U(1)$_{_D}$ breaking, the  $\eD$ asymmetry is washed out. To derive the conditions under which DM today contains a significant $\eD$ component, we first have to describe the various cosmological events that take place. We thus postpone a detailed discussion on the relic $\eD$ abundance until Sec.~\ref{sec:eD asym}.

\begin{figure}%[p]
\centering
\begin{tabular}{p{0.5\textwidth} p{0.5\textwidth}}
  \vspace{0pt} \includegraphics[width=0.45\textwidth]{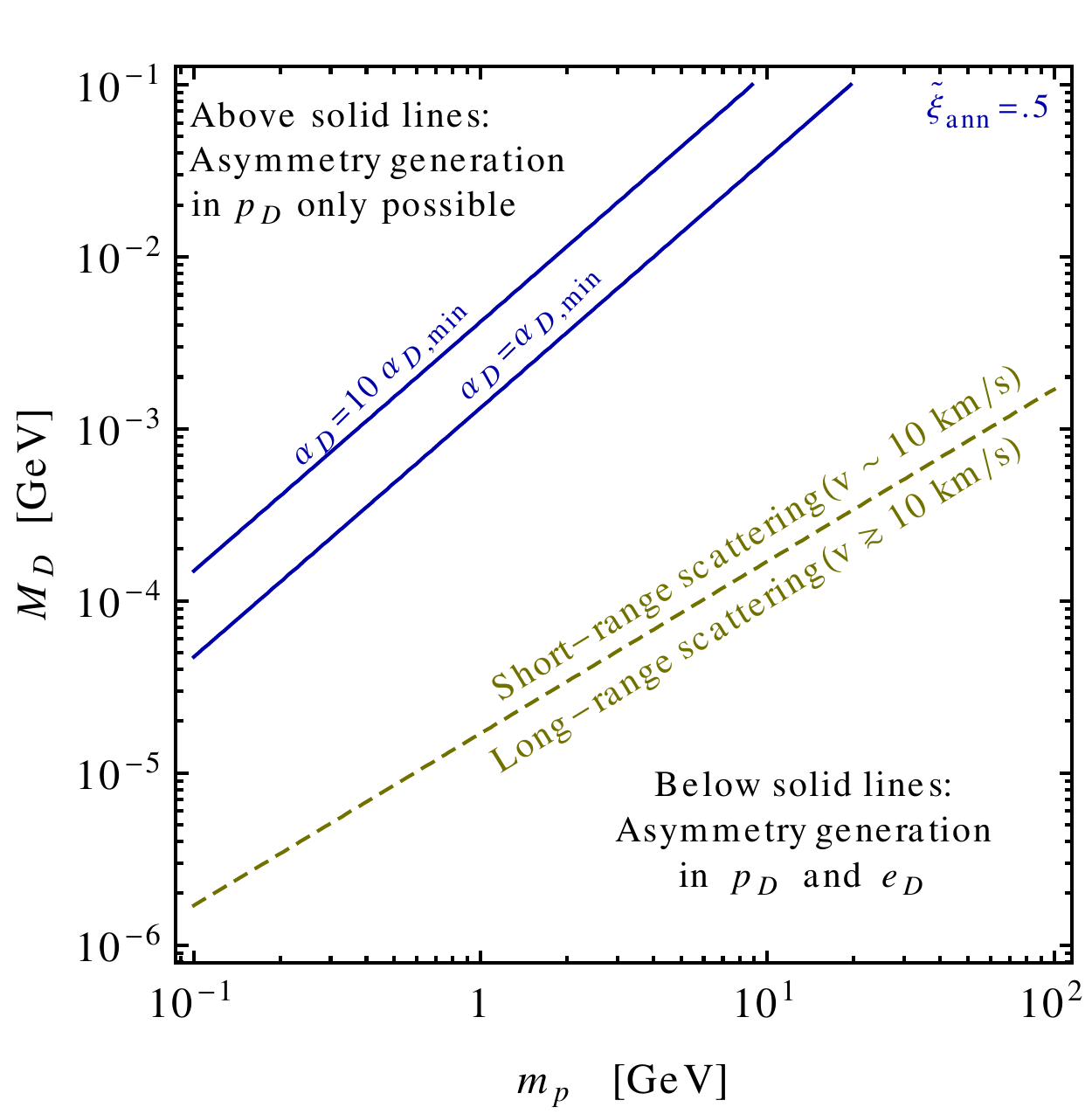} &
  \vspace{0pt} \includegraphics[width=0.45\textwidth]{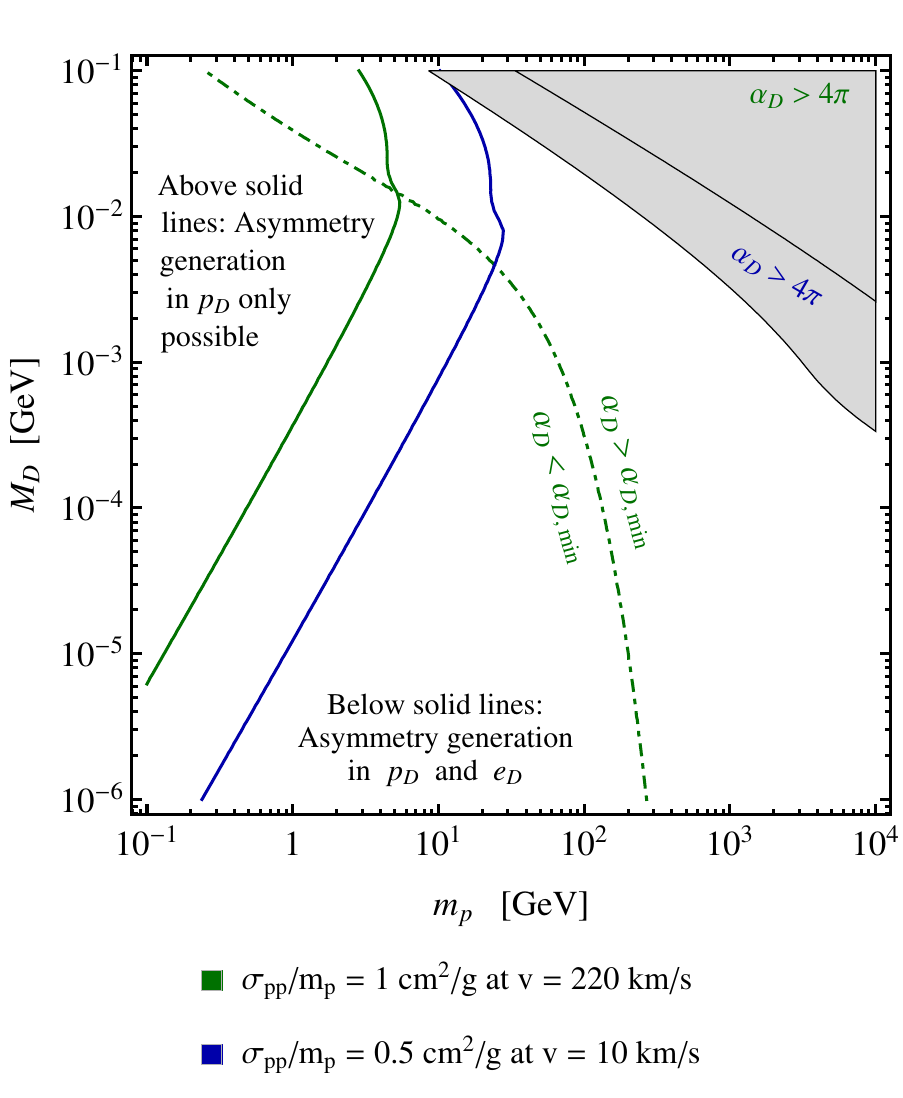}
\end{tabular}
\caption{\label{fig:conditions}  \footnotesize
To the right of the solid lines, the condition~\eqref{eq:MDasym} is satisfied: an asymmetry is generated both in $\pD$ and $\eD$. We have set $q_\phi = 1$.
\newline
\emph{Left:} We fix the fine structure constant to $\alpha_{_D} = \alpha_{_{D, \rm min}}(m_{\bf p})$ (lower solid blue line) and $\alpha_{_D} = 10 \, \alpha_{_{D, \rm min}}(m_{\bf p})$ (upper solid blue line), where $\alpha_{_{D, \rm min}}(m_{\bf p})$ is the minimum value which allows sufficient annihilation of the thermal population of dark protons in the early universe [c.f. Eq.~\eqref{eq:alpha ann}, where we set the dark-to-ordinary temperature ratio at the time of freeze-out to $\ks_{\rm ann} = 0.5$].   Obviously, larger values of $\alpha_{_D}$ imply that more parameter space is encompassed in the multi-component DM realisation. To the right of the yellow dashed line, the $p_{_D} - p_{_D}$ collisions manifest as long-range in haloes with rotational velocity $\bar{v} \gtrsim 10 \: {\rm km/ s}$ [c.f. Eq.~\eqref{eq:long-range}]. 
\newline
\emph{Right:} We consider $p_{_D}^+ - p_{_D}^+$ collisions in DM haloes, according to what described in Sec.~\ref{sec:self-inter}. For the blue (lower) line, we pick the minimum value of $\alpha_{_D}(m_{\bf p}, M_{_D})$ for which, under the assumption of single-component DM, there can be a significant effect on the dynamics of small haloes. In particular, we fix the momentum-transfer cross-section to $\sigma_{\bf pp} / m_{\bf p} = 0.5 \: {\rm cm^2/ g}$ at $\bar{v} = 10 \:  {\rm km/ s}$.  This ensures that the ellipticity of larger haloes is retained, since $\sigma_{\bf pp}/m_{\bf p}$ decreases with  $\bar{v}$. (Note though, that the value of $\alpha_{_D}$ specified in this way may be smaller than $\alpha_{_{D, \rm min}}$.)  
For the green (upper) line, we pick $\alpha_{_D}$ by setting $\sigma_{\bf pp} / m_{\bf p} = 1 \: {\rm cm^2/ g}$ at $\bar{v} = 220 \: {\rm km/s}$. This yields the maximum value of $\alpha_{_D}(m_{\bf p}, M_{_D})$ that is currently considered compatible with the observed ellipticity of haloes. For this choice of $\alpha_{_D}$, to the left of the green dot-dashed line, $\alpha_{_D} < \alpha_{_{D,\rm min}}$ and the scenario does not appear viable. That is to say, if for the $m_{\bf p}, \, M_{_D}$ values to the left of the green dot-dashed line, we set $\alpha_{_D} \gtrsim \alpha_{_{D, \rm min}}$, then the $p_{_D}^+  - p_{_D}^+$ interaction in haloes is too strong. However, when the formation of dark atoms in the early universe is taken into account, the DM self-scattering is suppressed, $\alpha_{_D}$ can be larger while respecting the ellipticity bound, and this part of the $m_{\bf p} - M_{_D}$ plane can produce viable scenarios, as we show in Sec.~\ref{sec:self-inter}.
In the grey-shaded regions, the perturbativity limit is exceeded, $\alpha_{_D} > 4\pi$, for the two choices of $\alpha_{_D}$ (lower and upper region respectively). 
}
\end{figure}

The condition \eqref{eq:MDasym} is satisfied in much of the parameter space of interest, namely that in which $\pD-\pD$ collisions in haloes are significant and compatible with  observations. As mentioned in the introduction, the ellipticity constraints from large haloes are most comfortably compatible with the requirement of significant interaction in smaller haloes if the DM self-scattering is long-range. The $\pD-\pD$ interaction manifests itself as long-range when the momentum transfer dominates over the mass of the mediator, i.e.~on average when $(\mpD/2) v \gtrsim \MD$. For dwarf-galaxy-size haloes with $v \sim 10 \km/\snd$ and larger haloes, this implies
\beq   
\MD \lesssim 2\times 10^{-5} \, \mpD \ . 
\label{eq:long-range}   
\eeq
The parameter space that fulfils the condition \eqref{eq:MDasym}, contains the long-range scattering regime of \eqref{eq:long-range} 
when $\aD \gtrsim q_\f^{-2} \times 7\times 10^{-9}$.
Taking into account the minimum value of $\aD$ required for efficient annihilation of DM in the early universe (c.f. \eqs{eq:aD sym} and \eqref{eq:alpha ann}), this corresponds to $\mpD \gtrsim q_\f^{-2} \times  45\keV$, which covers all of the range of interest (we assume $q_\f \sim {\cal O}(1)$). 
In fact, irrespectively of the long- or short-range nature of the $\pD-\pD$ interaction, the condition \eqref{eq:MDasym} encompasses much of the parameter space where the $\pD-\pD$ interaction can have a sizeable effect on the dynamics of the smaller haloes. (The rates for the $\pD - \pD$ scattering will be presented in Sec.~\ref{sec:self-inter}.) We illustrate these comparisons in Fig.~\ref{fig:conditions}.

\subsubsection{Dark atoms from a Yukawa potential}

If both $\pD^+$ and $\eD^-$ ions remain abundant at late times, they can potentially form U(1)$_{_D}$-neutral bound states. In the case of a massive dark photon, and in the non-relativistic regime, the interaction between $\pD^+$ and $\eD^-$ is described by the Yukawa potential 
\beq
V_Y = -\frac{\aD}{r} \, e^{-\MD r}  \ .
\label{eq:Yukawa}
\eeq
For small enough screening mass $\MD$, the attractive Yukawa potential has bound-state solutions. They can be found by solving Schroedinger's equation using instead the Hulth\'{e}n potential~\cite{HulthenA}, 
$V_H = -\a \tilde{M}_{_D} \exp(-\tilde{M}_{_D} r)/[1-\exp(-\tilde{M}_{_D} r)]$, with $\tilde{M}_{_D} = (\p^2/6)\MD$ being an appropriate approximation for the Yukawa potential~\cite{Cassel:2009wt}.  The binding energy of the ground state is estimated to be~\footnote{\label{foot:Dirac eq} For $\aD$ close to unity, relativistic effects in the bound-state dynamics become important.  In fact, for a Coulomb potential, the ground-state energy eigenvalues of a Klein-Gordon or a Dirac field become complex  at $\aD > 1/2$ and $\aD > 1$ respectively. Moreover, Gribov has shown that there is a critical coupling, $\a_{\rm crit} = \p (1-\sqrt{2/3} )  \simeq 0.58$, above which the Coulomb interaction between light fermions causes rearrangement of the perturbative vacuum~\cite{Gribov:1998kb,Gribov:1999ui,Dokshitzer:2004ie}.  Here, we shall thus consider only  $\aD < 0.5$.}
\beq
\D \approx \frac{1}{2} \aD^2 \muD \pare{1-\frac{\MD}{\aD \muD}}^2  \  .
\label{eq:Delta} 
\eeq
Dark atoms exist as long as 
\beq
\MD < \muD \aD \ ,
\label{eq:atom cond}
\eeq
i.e.~as long as the screening of the $\pD^+ - \eD^-$ interaction due to the mass of the dark photon occurs at larger length scales than the size of the atoms, which is determined by the Bohr radius $a_B = (\aD \muD)^{-1}$. The consistency condition of \eq{eq:muD upper} must still hold.  Dark atoms can form via the process
\beq
\pD + \eD \to \HD + \gammaD (\w)  \  ,
\label{eq:recomb}
\eeq
where $\w$ is the energy of the emitted dark photon $\gammaD$. Conservation of energy implies 
\beq
\w + \frac{\w^2-\MD^2}{2 \mH} = \D + \frac{1}{2} \muD \vrel^2 \ ,  \nn
\eeq
where $\vrel$ is the $\pD - \eD$ relative velocity in their centre-of-mass frame. For $\D + \muD \vrel^2/2 \ll \mH$, $\w \simeq \D + \muD \vrel^2/2$. Of course, the process \eqref{eq:recomb} is possible provided that $\w > \MD $.\footnote{Due to the kinetic mixing of \eq{eq:Lkin}, it is possible that dark atoms can still form with emission of an ordinary photon, even if $\MD > \D$. However, in this case, the cross-section for the formation of dark atoms is suppressed by $\e^2$, and is unlikely to be significant (for cosmological bounds on $\e$, see Sec.~\ref{sec:cosmo}).
}
In the early universe, bound states begin to form after the temperature of the dark plasma has dropped below the binding energy, $\TD \sim \muD \vrel^2/2 < \D$ (see Sec.~\ref{sec:recomb}), which means that the condition for their formation in the early universe is 
\beq
\MD < \frac{1}{2} \aD^2 \muD  \  .
\label{eq:MD<Delta}
\eeq
This condition is stronger than \eq{eq:atom cond}, as well as \eq{eq:MDasym} for the entire range of $\aD$ values we are considering, $\aD \leqslant 1/2$  (see footnote~\ref{foot:Dirac eq}).

%%%%%%%%%%%%%%%%%%%%%%%%%%%%%%%%%%%%%%%
%%%%%%%%%%%%%%%%%%%%%%%%%%%%%%%%%%%%%%%
\section{Cosmology}
\label{sec:cosmo}

We now examine in detail the cosmological history of the scenario under consideration, beyond the time of asymmetry generation. 
We sketch one possible sequence of events in Table~\ref{tab:cosmo sequence}.

\subsection{Kinetic equilibrium between SM particles and the dark sector}
\label{sec:kin equil}

Much of the cosmology and the low-energy phenomenology of DM in the scenario under consideration depends on the temperature of the dark plasma with respect to that of the SM particles at various important epochs in the cosmological evolution.   Assuming that the two sectors had once been thermally coupled,\footnote{As mentioned in footnote~\ref{foot:Tasym}, this is expected if the dark and the ordinary asymmetries were dynamically related in the early universe, but it is not a necessary assumption for the phenomenological aspects we are examining.}    
and that they decoupled at a common temperature $T_{\rm dec}$, the ratio of their temperatures at later times is determined by the number of d.o.f.s coupled in each sector at $T_{\rm dec}$ and their subsequent decoupling from the thermal bath of each sector. At visible and dark sector temperatures $\TV, \TD < T_{\rm dec}$, the comoving entropy is conserved separately in each sector, which implies 
\beq
\frac{\gD \TD^3}{\gV \TV^3} = \frac{ g_{_{\rm D, \, dec}} }{ g_{_{\rm V, \, dec}} }
\label{eq:rel entropy cons}
\eeq
where $\gV, \gD$ are the effective relativistic d.o.f.s in the visible and the dark sectors respectively, and the subscript ``dec" always refers to the last time the two sectors were coupled. Thus, depending on how many d.o.f.s  each sector contains, and the order in which they decouple, the ratio of temperatures
\beq
\ks \equiv \frac{\TD }{ \TV } = 
\pare{ \frac{ g_{_{\rm D, \, dec}} }{ g_{_{\rm V, \, dec}} }  \: \frac{\gV}{\gD} }^{1/3}
\label{eq:ksi}
\eeq
can vary and take values $\ks \gtrless 1$.  For example, if the mass spectrum in the dark sector is in general higher than that of the ordinary sector, then it is reasonable to expect that the dark d.o.f.s will decouple first, rendering $\ks > 1$ after their decoupling. Subsequent decoupling of the d.o.f.s of the ordinary sector will bring $\ks$ back to lower values. 
In the following, we will also find useful to define
\beq
\tilde{\ks} \equiv \left\{
\bal{5}
&\ks \: ,& \qquad &\text{if } \r_{_{\rm V}} > \r_{_{\rm D}}& 
\\
&1   \: ,& \qquad &\text{if } \r_{_{\rm V}} < \r_{_{\rm D}} \: ,& 
\eal
\right.
\label{eq:ksi tilde}
\eeq
where $\r_{_{\rm V}}, \, \r_{_{\rm D}}$ are the energy densities of the visible sector and the dark sector respectively, with 
$\r_{_{\rm U}} = \r_{_{\rm V}} + \r_{_{\rm D}}$
being the energy density of the universe.

Of course, if the dark photon is stable or sufficiently long-lived, it may eventually contribute significantly to the relativistic or non-relativistic energy density of the universe. In those cases, $\ks$ is typically constrained to be less than unity at late times.  Here, however, we consider a large parameter space, in a portion of which the cosmological abundance of dark photons decays early into SM particles.  Both $\ks <1$ and $\ks > 1$ remain thus viable, depending on the rest of the parameters.   We discuss constraints on $\ks$ from cosmology, and the fate of dark photons in Sec.~\ref{sec:dark photons}.

\medskip

Considering $\ks \neq 1$ at important times, such as the epoch of dark recombination (see Sec.~\ref{sec:recomb}), is valid provided that the kinetic mixing introduced in \eq{eq:Lkin} does not bring the dark and the ordinary sectors in equilibrium. Following Ref.~\cite{Carlson:1987si}, we estimate the upper bound on $\e$ which allows the two sectors to not equilibrate. The energy transfer between the two sectors occurs predominantly via scattering of ordinary electrons on the lightest charged species of the dark sector, either $\eD$ or $\fD$, with rate
\beq
\frac{d \r}{dt} \approx  n_e n_d \int d\W \, \frac{d\s}{d\W}  \, v \, \d E_1   \  ,
\eeq
where $\d E_1 \sim k (1-\cos \th)$ is the longitudinal momentum (and energy) transfer per collision, with $k \sim \sqrt{2} \, 3T$ being the relative average momentum of relativistic particles in a thermal bath of temperature $T$. $n_e$ and $ n_d$ are the number densities of the ordinary electrons and the relativistic dark-sector charged particles. The momentum transfer cross-section is
\beq
\int d\W \, (1-\cos \th) \, \frac{d\s}{d\W}  \approx
\frac{4 \p \e^2 \aD \a_{_{\rm EM}} \mu_{ed}^2 }{k^4} \
\ln [ \csc (\th_{\rm min} /2) ]
\ ,
\label{eq:sigma V-D}
\eeq
where $\a_{_{\rm EM}} \simeq 1/137$ is the fine-structure constant, and  $\mu_{ed} = \tilde{m}_e m_d/(\tilde{m}_e m_d)$ is the reduced mass of the scattering particles, with $\tilde{m}_e = 511 \keV$ being the ordinary electron mass and $m_d = \min(\meD, m_\f)$. Here, $m_\f$ is the (temperature-dependent) mass of the complex dark Higgs field $\fD$ before the dark phase transition; after the dark phase transition, the scattering of the physical scalar $\fvarD$ off ordinary electrons is loop-suppressed, as seen from~\eq{eq:L int}.  In \eq{eq:sigma V-D}, $\th_{\rm min}$ is the minimum scattering angle, which can be estimated as 
$\csc^2 (\th_{\rm min}/2) = 1 + (2 \l^{\rm Debye} k)^2/(\e^2 \a_{_{\rm EM}} \aD)$, 
where $\l^{\rm Debye} = \min(\l_{_V}, \l_{_D})$ is the smallest of the Debye screening lengths of the ordinary and the dark plasma $\l_{_V}$ and $\l_{_D}$ respectively, with $\l_{_D}^2 = T/[4\p \max(n_e \a_{_{\rm EM}}, n_d \aD)]$.

Equilibrium between the two sectors is not established so long as 
\beq
\frac{1}{\r} \frac{d \r}{dt} < H \ ,
\label{eq:non-equil}
\eeq
where $H$ is the Hubble parameter and $\r \simeq  (\p^2/30) g_* T^4$ is the energy density of either sector, with $g_*$ the corresponding relativistic d.o.f.s. 
For relativistic number densities of charged particles in the two sectors, $(1/\r)d\r/dt \propto 1/T$, i.e.~the energy exchange rate becomes more significant as the temperature drops. This is, of course, a manifestation of the long-range nature of the interaction. 
The condition \eqref{eq:non-equil} should thus be evaluated at the latest time when both sectors have significant (relativistic) number densities of charged particles, namely at $T \sim \max[\tilde{m}_e/3, \min(\meD/3,\vD)] \sim \max[\tilde{m}_e, m_d]/3$, where we used the estimated temperature of the dark phase transition, $T_{_{\rm PT}} \sim \vD$, as the latest time the $\fD$ particles can participate in the energy exchange between the two sectors. After one of the species becomes non-relativistic, the energy transfer is further suppressed.  This yields roughly the condition
\beq 
\e^2 \aD \lesssim 10^{-20} \: \sqpare{ \frac{\max(\tilde{m}_e, m_d)^3}{\tilde{m}_e \min(\tilde{m}_e, m_d)^2 } }  \ ,
\label{eq:equil bound}
\eeq
where we took $\ln [\csc (\th_{\rm min}/2)] \sim 20$.
Obviously, the term in the brackets is greater than 1 and the bound becomes more relaxed the heavier the dark-sector charged particles are.

We emphasise that the inequality~\eqref{eq:equil bound} is not necessarily a constraint. If satisfied, it allows for $\ks \neq 1$. However, if the dark photons decay, or redshift sufficiently due to entropy release in the ordinary sector, or become non-relativistic sufficiently early, cosmological considerations do allow for the dark and the ordinary sectors to have been in equilibrium before that time (see Sec.~\ref{sec:dark photons} and Fig.~\ref{fig:cosmo}). In this case, the condition \eqref{eq:equil bound} need not hold true. Note also that $\ks$ varies in general with time, albeit typically fairly mildly due to the small exponent in \eq{eq:ksi}.  In the following, we shall distinguish among the values of $\ks$ and $\tilde{\ks}$ at different epochs, using appropriate subscripts.

\subsection{Efficient annihilation of DM species}
\label{sec:annihilation}

The cross-section for the annihilation processes 
$\pD \bar{p}_{_{D}} \to \gammaD \gammaD$ and $\eD \bar{e}_{_{D}} \to \gammaD \gammaD$
is
\beq
(\s v)_{\rm ann} = \frac{\p \aD^2}{m_{\bf i}^2} \, S
\label{eq:sigma ann}
\eeq
where $m_{\bf i} = \mpD, \meD$ and $S$ is the Sommerfeld enhancement factor;  for a massless dark photon, $S = 2\p\z/(1-e^{-2\p\z})$, where $\z = \aD/\vrel$~\cite{Cassel:2009wt}. 
Dark protons, being heavier, can also annihilate into dark electrons via a dark photon, $\pD \bar{p}_{_{D}} \to \eD \bar{e}_{_{D}}$, with cross-section about equal to that of \eq{eq:sigma ann}.

For non-self-conjugate symmetric thermal-relic DM, and for $s$-wave annihilation, the annihilation cross-section has to be $(\s v)_{\rm sym} \simeq \tilde{\ks}_{\rm ann} \times 6 \times 10^{-26} \cm^3/\snd$~\cite{Feng:2008mu}.
This fixes $\aD$ with respect to the DM mass, $m\DM$
\beq
\a_{_{D, \rm sym}}(m\DM) = 4 \times 10^{-3} \pare{\frac{\tilde{\ks}_{\rm ann}}{S_{\rm sym}}}^{1/2} \pare{\frac{m\DM}{10^2 \GeV}} \ .
\label{eq:aD sym}
\eeq
The effective Sommerfeld enhancement, $S_{\rm sym}$, represents the effect of the thermal average of $S$ on the DM freeze-out, and is significant 
only for large $\aD$. In the symmetric DM limit, this corresponds to large DM masses,  $m\DM \gtrsim 800 \GeV$, with $S_{\rm sym}^{1/2} \sim 2$ at $m\DM \sim  10 \TeV$ (see e.g.~\cite{Feng:2010zp}). 
In the asymmetric regime, the efficient annihilation of the symmetric part of DM in the early universe necessitates $(\s v)_{\rm ann} \gtrsim f \times (\s v)_{\rm sym}$, where for $f>1.4~(2.4)$ the dark antiparticles make up less that 10\%~(1\%) of the DM density~\cite{Graesser:2011wi}.  This implies a minimum value of $\aD$, which in our scenario is set by the dark proton mass. Taking into account the two $\pD$ annihilation channels, and setting $f = 1.4$, the condition for efficient annihilation becomes
\beq 
\aD > \a_{_{D, \rm min}} \equiv  3.4 \times 10^{-3} \pare{\frac{\tilde{\ks}_{\rm ann}}{S_{\rm sym}}}^{1/2} \pare{\frac{\mpD}{10^2\GeV}}
\label{eq:alpha ann}
\eeq
The constraint of \eq{eq:alpha ann} may be relaxed if more annihilation channels exist, though this would imply a less minimal model.

The annihilation of the dark fermions freezes-out around $T_{_{\rm D, \, FO}} \sim m_{\bf i} / x_{_{\rm FO}}$, with~\cite{Kolb:1990vq,Graesser:2011wi,Feng:2008mu} 
\beq x_{_{\rm FO}} \approx 30 + \ln \pare{ \frac{m_{\bf i}}{100 \GeV} \, \frac{\tilde{\ks}_{\rm ann}^2 (\s v)_{\rm ann}}{10^{-24}\cm^3/\snd} } \ , \label{eq:xFO} \eeq 
typically before the onset of the dark recombination.

\subsection{Dark recombination and residual ionisation fraction}
\label{sec:recomb}

After the temperature of the dark sector drops below the binding energy of the dark atoms, and assuming a large $\eD^-$ density has survived until that time, it becomes favourable for the dark ions to form atoms via the process shown in \eqref{eq:recomb}. Depending on the parameters, we may discern three regimes describing dark recombination~\cite{CyrRacine:2012fz}.
For large $\aD$ or small masses (i.e.~large number densities of the dark species), the recombination process is quite efficient and occurs mostly in thermodynamic equilibrium. It can be described well by the Saha equation, until the recombination rate falls below the expansion rate of the universe and the ionisation fraction freezes-out. For intermediate couplings and/or masses, recombination happens in quasi-equilibrium and the details of the atomic transitions become important. For small $\aD$ or very large masses, recombination is very weak and most DM remains ionised. Dark matter remains ionised also if the inequality \eqref{eq:MD<Delta} is not satisfied.

The residual ionisation fraction can be approximated by~\cite{CyrRacine:2012fz}
\beq
\xD \approx  \left\{
\bal{5}
&\min \sqpare{  1, \  10^{-10} \    \frac{ \tilde{\ks}_{_{\rm DR}} }{\aD^4}   \pare{\frac{\mH \muD}{\GeV^2}} },&  
\qquad
&\MD < \frac{1}{2} \aD^2 \muD&  
\\
&1,& 
\qquad
&\MD > \frac{1}{2} \aD^2 \muD \ .&
\eal
\right.
\label{eq:xD resid}
\eeq
This approximation describes well the regime in which recombination is completed in equilibrium ($\xD \ll 1$) and the regime in which DM is mostly ions ($\xD \simeq 1$). It is less satisfactory in the regime where $\xD \lesssim 1$. However, because of the strong dependence of the $\xD<1$ branch of \eq{eq:xD resid} on $\aD$, this regime spans a fairly small portion of the parameter space (along the $\aD$ direction).

Using thermodynamic equilibrium equations, we may also estimate the dark-sector temperature, $T_{_{\rm D, \, rec \: fo}} \equiv \D / x_{\rm rec \: fo}$, at which the recombination process freezes out,
\beq
x_{\rm rec \: fo} \approx 53 + \ln \sqpare{ \tilde{\ks}_{_{\rm DR}} \pare{\frac{\aD}{0.1}}^5 \pare{\frac{10^3 \GeV^3}{\mH \muD^2}}}  \  .
\label{eq:rec FO}
\eeq
Depending on the parameters, dark recombination may be completed before or after dark photons acquire mass. 
In the former case, recombination proceeds as described. In the latter case, we still expect that \eq{eq:xD resid} is about as good approximation for $\MD >0$ as for $\MD = 0$.
This is because \eq{eq:xD resid} has been derived from equilibrium thermodynamics, taking into account only the ionised state and the ground state of the species involved.\footnote{The condition $\MD < \D$ implies to a good approximation that $\D(\MD \neq 0) \simeq \D(\MD = 0)$, since $\aD<0.5$ (cf. \eq{eq:Delta}). Thus, the estimated values for the various quantities of interest ($\xD, \: x_{\rm rec \: fo}$) which depend directly on $\D$, do not change significantly.}

\medskip
It is possible that dark Hydrogen atoms bind partially into molecules, $H_{_{D, 2}}$. While the formation of molecular Hydrogen from neutral Hydrogen atoms, $\HD+ \HD \to H_{_{D,2}} + \gammaD$, is rather slow, it could be catalysed by the presence of a small ionised DM component and proceed via the processes $\HD + \pD^+ \to H_{_{D,2}}^+ + \gammaD$ and $H_{_{D,2}}^+ + \HD \to H_{_{D,2}} + \pD^+$.  The smaller binding energy of the dark molecules in comparison to that of the dark atoms implies that in the case of a massive dark photon, a (somewhat) stronger conditions than \eqref{eq:MD<Delta} has to be satisfied for dark molecules to form. 
In the following we shall ignore the possibility of dark molecule formation, which merits a dedicated study. For a discussion on dark molecules and their scattering properties, see Ref.~\cite{Cline:2013pca}.

\subsection{The dark phase transition and the late-time dark-electron asymmetry}
\label{sec:eD asym}

We now return to the issue of the survival of a significant $\eD$ density at late times. The condition \eqref{eq:MDasym} ensures the generation of an $\eD$ asymmetry along with the $\pD$ asymmetry. However, the subsequent breaking of U(1)$_{_D}$ may potentially allow for the $\eD$ asymmetry to be washed out, and the $\eD$ abundance to be diminished by the recoupling of $\eD - \bar{e}_{_D}$ annihilations.  As we now discuss, the survival of a large $\eD$ abundance depends on how massive the dark photon is and on the charge $q_\f$ of the scalar field $\f_{_{D}}$ which breaks U(1)$_{_D}$.

For specific values of $q_\f$, dark electrons may acquire a Majorana mass, or Majorana-type mass mixing with other species, after U(1)$_{_D}$ breaking. For example, if $q_\f = 2$, dark electrons may couple to $\fD$ via the operators
\beq
{\cal L}_{\rm M} = 
- \frac{y_{_L}}{2} \, \fD \overline{\eDL^c} \eDL
- \frac{y_{_R}}{2} \, \fD \overline{\eDR^c} \eDR + \hc \ ,
\label{eq:Majorana}
\eeq
where $\eDL, \: \eDR$ are the left- and right-chirality components of $\eD$, and $y_{_L}, \: y_{_R}$ are dimensionless Yukawa couplings.\footnote{
It is important to keep in mind that the ingredients of the model we are invoking in our analysis, namely $\pD, \: \eD, \: \gammaD$ and $\fD$, make up only the low-energy effective theory of a more involved dark sector. This is essential in understanding how the breaking of U(1)$_{_D}$ is compatible with the assumption of asymmetric DM:  Values of $q_\f$ which generate Majorana-type mass terms for $\eD$ do not do the same for $\pD$, even though $|q_{\bf p}| = |q_{\bf e}|$ under U(1)$_{_D}$, provided that $\pD$ carries additional gauge charges (possibly broken at different scales than U(1)$_{_D}$, see e.g.~\cite{Petraki:2011mv,vonHarling:2012yn}). Alternatively, $\pD$ may be composed by particles which carry different gauge charges, with the ordinary matter providing here again an example of such a structure. One of these two features is, in most cases, necessary in order for the working hypothesis of the asymmetric DM scenario -- that there is a good low-energy dark baryon number symmetry -- to hold true.   We note though that an internal structure of $\pD$ and/or $\eD$ could have a variety of implications whose study is beyond the scope of this work. We only mention in passing that a composite $\pD$ appears e.g.\ in mirror DM~\cite{Foot:2014mia}, whose phenomenological complexity cannot be captured in its entirety by the present or previous studies of atomic DM consisting of fundamental particle species. 
}    
(We note that the $L$ and $R$ indices do not connote any SM gauge charges;  $\eD$ is a singlet under the SM gauge group, and the Dirac mass term $\meD \bar{e}_{_D} \eD$ is gauge invariant.)  
In this case, after U(1)$_{_D}$ breaking, dark electrons and anti-electrons acquire small Majorana masses and form a pseudo-Dirac pair.  
For $|y_{_{L}} - y_{_{R}}| \vD / \meD \ll 1$, the mass eigenstates are approximately the self-conjugate fields 
\beq
{e_{_{D,1}}}   \simeq \ -\frac{i}{\sqrt{2}} \,  \pare{ \eDL + \eDL^c - \eDR - \eDR^c } 
\quad {\rm and} \quad
{e_{_{D,2}}}  \simeq \  \frac{1}{\sqrt{2}} \,  \pare{ \eDL + \eDL^c + \eDR + \eDR^c }  \  ,
\label{eq:e mass eigen}
\eeq 
with masses $m_{1,2} \simeq \meD \mp y \vD$, where $y \equiv (y_{_L} + y_{_R})/2$. In terms of ${e_{_{D,1}}}$ and ${e_{_{D,2}}}$, the interactions to which the dark electrons participate are described by the Lagrangian 
\bea
\d {\cal L}_{\bf e} 
&=&  \frac{1}{2} \bar{e}_{_{D,1}} i \slashed{\partial} {e_{_{D,1}}}  +  \frac{1}{2} \bar{e}_{_{D,2}} i \slashed{\partial} {e_{_{D,2}}} 
- \frac{1}{2} (\meD - y \vD) \bar{e}_{_{D,1}} {e_{_{D,1}}} 
- \frac{1}{2} (\meD + y \vD) \bar{e}_{_{D,2}} {e_{_{D,2}}}   
\nn \\
&-& \frac{y}{2\sqrt{2}} \, \fvarD (\bar{e}_{_{D,2}} {e_{_{D,2}}}  - \bar{e}_{_{D,1}} {e_{_{D,1}}}) 
+ \pare{ \frac{i}{2} q_{\bf e}g A_{_D}^\m \bar{e}_{_{D,1}} \g_\m \, {e_{_{D,2}}}  + \hc}
\ ,
\label{eq:Le}
\eea
where we set $\fD = \vD + \fvarD/\sqrt{2}$, as before.  The Majorana masses can induce  $\eD^- - \eD^+$ oscillations, with frequency $\w_{\rm osc} = 2y\vD$. The oscillations can potentially erase the $\eD$ asymmetry if the expansion rate of the universe is lower than the oscillation frequence, $H \lesssim \w_{\rm osc}$.

However, the formation of $\pD^+ - \eD^-$ bound states can severely hinder the $\eD^- - \eD^+$ oscillations. The oscillation of a dark electron bound in a dark atom, into a dark positron, is energetically forbidden, if any energy gain from the oscillation does not suffice to render the dark electron unbound.  The energy difference between the two mass eigenstates of \eq{eq:e mass eigen} is $2 y \vD$. On the other hand, the expectation values of the kinetic and potential energies of a $\pD^+ - \eD^-$ bound state are $\ang{E_K} = \D, \ \ang{E_P} = -2\D$, with the total energy being $-\D$; for a $\pD^+ - \eD^+$ state characterised by the same wavefunction (or equivalently, the same superposition of plane waves), $\ang{E_K} = \D, \ \ang{E_P} = 2\D$, with the total energy summing to $3\D$. Thus if $2 y \vD < 4\D$, dark atoms are energetically stable. This sets an upper bound on the Yukawa coupling 
\beq   y \lesssim 20 \aD^{1/2} (\D/\MD) \ , \label{eq:atom stability} \eeq  
where we took $q_\f = 2$. Note that $\D/\MD > 1$ as per \eq{eq:MD<Delta}, and a lower limit on $\aD$ applies as well from requiring sufficient annihilation in the early universe, as described in Sec.~\ref{sec:annihilation}.
If the condition \eqref{eq:atom stability} is satisfied and dark atoms form cosmologically before the dark phase transition which generates the $\eD$ Majorana masses, i.e.~if $T_{_{\rm D, PT}} < \D/ x_{\rm rec \, fo}$, or
\beq
\MD < (32\p \aD)^{1/2} \, \D/ x_{\rm rec \, fo}  \ ,
\label{eq:MDmax}
\eeq
then DM today remains (partially) atomic, as described in Sec.~\ref{sec:recomb}.  $x_{\rm rec \, fo}$ given by \eq{eq:rec FO}. Note that the condition \eqref{eq:MDmax} is stronger than \eqref{eq:MD<Delta}. Wherever satisfied, the bound \eqref{eq:atom stability} on the Yukawa coupling becomes $y \lesssim 2 x_{\rm rec \, fo}$. This encompasses all the perturbative $y$ range.

If dark atoms have not formed before $\eD-\bar{e}_{_D}$ oscillations can begin, then the latter may erase the U(1)$_{_D}$ charge asymmetry carried by the dark electrons. Nevertheless,  $\eD-\bar{e}_{_D}$ oscillations cannot change the total abundance of dark electrons and anti-electrons (or ${e_{_{D,1}}}$ and ${e_{_{D,2}}}$),  if the $\eD - \bar{e}_{_D}$ annihilations are inefficient when the regeneration of the $\bar{e}_{_D}$ population occurs.   In terms of gauge eigenstates, this ensures that the abundance of $\eD^-$ ions can change only by a factor of at most 2.  The abundance of dark electrons after the freeze-out of annihilations and before oscillations occur, is  $Y_{{\bf e}, \rm FO} \equiv n_{{\bf e}, \rm FO}/s =  n_{{\bf p}, \rm FO}/s = \W\DM \r_c/(\mpD s_0)$,  where the subscript ``FO'' denotes the frozen-out or relic value.  Annihilations are inefficient if $\G_{\rm ann} < H_{_{\rm PT}}$, where  $\G_{\rm ann} = s_{_{\rm PT}} \, Y_{{\bf e}, \rm FO} \, (\s v)_{{\bf e}, \rm ann}$ and 
$H_{_{\rm PT}} \simeq 1.66 \sqrt{g_{_{\rm PT}}} \, T_{_{\rm PT}}^2/\mpl$ are the annihilation rate and 
the expansion rate respectively, at the time of the dark phase transition.   Here, 
$g_{_{\rm PT}}$, $T_{_{\rm PT}}$ and $s_{_{\rm PT}} \simeq (2\p^2/45)g_{_{\rm PT}} T_{_{\rm PT}}^3$ 
are the number of (entropic) effective relativistic d.o.f.s, the temperature and the entropy density of the universe 
at the time of the dark phase transition, with $T_{_{\rm PT}} \simeq T_{_{\rm D,PT}}/ \tilde{\ks}_{_{\rm PT}}$, with $T_{_{\rm D,PT}}$ given in \eq{eq:T_PT}.  
The dark-electron annihilation cross-section times relative velocity $(\s v)_{{\bf e}, \rm ann}$, is given in \eq{eq:sigma ann}.
Setting $q_\f = 2$, we conclude that if
\beq 
\MD \lesssim 10^{-11} \, \tilde{\ks}_{_{\rm PT}} (32\p \aD)^{1/2} \, \aD^{-2} \, \mpD \, \meD^2/\GeV^2 \ , 
\label{eq:MDmax osc} 
\eeq 
the relic $\eD$ abundance remains significant for cosmological and astrophysical considerations.  The above estimation does not take into account the effect of $\eD, \bar{e}_{_D}$ scatterings on particles of the thermal bath. Scatterings decohere the $\eD - \bar{e}_{_D}$ oscillations and can only relax the condition \eqref{eq:MDmax osc}. For a more detailed treatment of the coupled effect of oscillations, annihilations and scatterings, see Ref.~\cite{Cirelli:2011ac}.  The condition \eqref{eq:MDmax osc} is more stringent than \eqref{eq:MDmax} in most of the mass range of interest; however, only one of them has to hold to ensure that DM today contains a significant dark electron component.

Although the conditions \eqref{eq:MDasym} and \eqref{eq:MDmax} (or \eqref{eq:MDmax osc})  are sufficient for the DM in this scenario to be multi-component, they are not necessary.  It is possible that the breaking of U(1)$_{_D}$ does not generate a Majorana-type mass term for $\eD$; whether this occurs depends on the value of $q_\f$.   The generation of a Majorana-type mass term for $\eD$, as considered above, can occur only for very specific choices for $q_{\f}$, while the possibility of no Majorana-type mass term is realised for an infinitude of $q_{\f}$ values.  For example, $q_\f = 2$ allows the interactions of \eq{eq:Majorana}, which yield Majorana mass terms for $\eD$ after the U(1)$_{_D}$ breaking. Other values of $q_\f$ could result in Majorana-type mass-mixing of $\eD$ with other particles; this depends of course on the particle content of the theory. With the exception of this finite set of values, all other possible $q_\f$ assignments do not generate a Majorana-type mass term for $\eD$ after U(1)$_{_{D}}$ breaking. 

The absence of such Majorana terms amounts to the conservation of a global U(1) remnant symmetry, under which $\eD$ is charged; we shall call this symmetry the dark lepton number, $\LD$.  The global symmetries of a theory can of course always be redefined by linear transformations, and in this case it is convenient to (re)define the global low-energy symmetries of the model, $\BD$ and $\LD$, such that the charges of $\pD$ and $\eD$ are $\BD(\pD) = 1, \ \BD(\eD) = 0$ and $\LD(\pD) = - \LD(\eD) = 1$. This makes the analogy to the case of an unbroken U(1)$_{_D}$ obvious: The $\BD$ asymmetry generation presumes the breaking of $\BD$ by high-energy processes, which however conserve $\LD$. Equal asymmetries in $\pD$ and $\eD$ are thus generated. The dark electron can be thought as the lightest particle charged under $\LD$; its asymmetry is conserved and its stability is ensured.  In this setup, DM is multi-component independently of the mass of the dark photon.

Obviously, the range of dark-photon masses encompassed by \eq{eq:MDmax}/\eqref{eq:MDmax osc} is limited if $\meD$ is very small. However, dark electrons cannot be arbitrarily light without implications. If $\meD < \MD$, then the annihilation of non-relativistic dark electrons into dark photons is kinematically forbidden. In this case, dark electrons decouple while relativistic. This again implies that their relic number density is significant (perhaps even more so than if they decouple when non-relativistic). Furthermore, their large number density would likely imply cosmological bounds on the ratio of the temperatures of the dark and the ordinary plasma; these bounds depend on how light dark electrons are. 
It is possible, of course, that additional annihilation channels of $\eD$ into lighter d.o.f.s exist. In addition to having the disadvantage of being a less minimal model, it is likely that similar reasoning would constrain the cosmological abundances of these lighter d.o.f.s.  In Sec.~\ref{sec:dark photons}, we will discuss such bounds in the context that dark photons are the relativistic thermal relics; it is straightforward to generalise these bounds to the case in which the dark electrons (or some other species) are the lightest d.o.f.s into which the energy and entropy of the dark sector are eventually deposited.

If the conditions \eqref{eq:MDasym} and \eqref{eq:MDmax} or \eqref{eq:MDmax osc} are satisfied in the parametric regimes of interest, or if the $\eD$ asymmetry is not washed-out due to a global remnant symmetry of U(1)$_{_D}$, then considering simply the $\pD-\pD$ collisions in haloes does not capture the dynamics of DM self-interaction properly. A proper treatment should take into account the formation of dark atoms in the early universe, and incorporate atom-atom, atom-ion and ion-ion collisions, with both species of ions, $\pD^+$ and $\eD^-$, included (and possibly also $\eD^+$ if they are regenerated via oscillations).  We do so in Sec.~\ref{sec:self-inter}.

\subsection{The fate of the dark photons and the dark-to-ordinary temperature ratio}
\label{sec:dark photons}

The dark protons and the dark electrons decouple from the dark photons at the end of the dark recombination. 
The dark photons and the physical scalar field $\fvarD$ remain chemically coupled via the annihilations $\fvarD \fvarD \leftrightarrow \gammaD \gammaD$ until after the heaviest of them becomes non-relativistic. In the following we shall assume that $\fvarD$ is heavier than $\gammaD$ and derive constraints from considering the abundance of the dark photons after their chemical decoupling. 
This is the case if the coupling $\l$ introduced in \eq{eq:Lf}, is $\l > 2\p q_\f^2 \aD$. It is straightforward to reverse this assumption. 

Immediately after the U(1)$_{_D}$-breaking phase transition, $\gammaD$ and the physical scalar $\fvarD$ are still relativistic or quasi-relativistic. The $\fvarD$ bosons become non-relativistic at $\TD \lesssim m_\fvar /3 \sim \vD/3$ (where we assumed self-coupling of ${\cal O}(1)$). In the non-relativistic regime, the $\fvarD\fvarD \rightarrow \gammaD \gammaD$ annihilation cross-section is 
\beq 
(\s v)_{\fvarD\fvarD \rightarrow \gammaD \gammaD} \simeq \frac{44 \p q_\f^4 \aD^2}{m_\fvar^2} 
\pare{1 - \frac{20}{11} \frac{\MD^2}{m_\fvar^2} + \frac{12}{11}\frac{\MD^4}{m_\fvar^4} } \pare{1-\frac{\MD^2}{m_\fvar^2}}^{1/2} \ . 
\label{eq:phi annihilation}
\eeq
For the case of interest, $m_\fvar \ll \mpD$, and the $\fvarD$ annihilation cross-section is evidently very large,
$(\s v)_{\fvarD\fvarD \rightarrow \gammaD \gammaD} \ggg (\s v)_{\bar{p}_{_D}\pD \rightarrow \gammaD \gammaD} \gtrsim (\s v)_{\rm sym}$, rendering the frozen-out abundance of $\fvarD$ cosmologically insignificant.\footnote{Moreover, if $m_\fvar > 2\MD$, the $\fvarD$ bosons decay promptly after they decouple, into dark photons, $\fvarD \to \gammaD \gammaD$; if $\MD + 1.022 \MeV< m_\fvar < 2\MD$, they may decay into $\fvarD \to \gammaD e^+ e^-$ via a virtual dark photon and its kinetic mixing with hypercharge. For the decay rates, see Ref.~\cite{Batell:2009yf}.}
The $\fvarD$ bosons freeze-out when $m_\fvar / \TD = x_\fvar  \sim 41 +\ln [ q_\f^4 \aD^2 \tilde{\ks}^2 \, ({\rm GeV} /m_\fvar)]$.

The $\fvarD - \gammaD$ chemical decoupling sets the abundance of the dark photons. At that time, the temperature is $T_{_{\rm D, \gammaD}} = m_\fvar  / x_\fvar \sim \vD/x_\fvar$, thus
\beq 
x_{\gammaD} \equiv \MD/T_{_{\rm D, \gammaD}} = x_\fvar (\MD/m_\fvar) \sim x_\fvar \, (8 \p q_\f^2 \aD)^{1/2} \ .  \label{eq:MD/T} 
\eeq
We see that, as a result of their coupling to the scalar field $\fD$, the dark photons decouple while non-relativistic in most of the parameter range of interest: 
$x_{\gammaD}  \gtrsim 3$ for $q_\f^2 \aD \gtrsim 2 \times 10^{-4}$ (with the exact value depending on $m_\fvar$ and $\tilde{\ks}$).  The relic dark photons may decay via the kinetic mixing of \eq{eq:Lkin} into SM charged fermions, provided that $\MD > 1.022 \MeV$.  However, if $\e$ is very small, or $\MD < 1.022 \MeV$, dark photons may be very long-lived or even cosmologically stable.   The decay of the cosmological abundance of dark photons into SM charged particles injects relativistic energy density, which can affect BBN and CMB. On the other hand, a significant relic abundance surviving until very late, or even today, could affect the time of matter-radiation equality, or contribute to the matter density of the universe.  We shall thus require that dark photons either (i) decay before BBN, or (ii) their energy density is sufficiently small, as specified below. 

\benu[(i)]

\item Decay before BBN.

The dark photons must acquire mass before BBN, when the ordinary sector is at temperature
$T_{_{\rm V, \, PT}} \! =  T_{_{\rm D, \, PT}} / \ks_{_{\rm PT}} > T_{_{\rm V, \, BBN}} \!\sim 1 \MeV$. This happens if
\beq
\ks_{_{\rm PT}} < 20 \: \pare{\frac{10^{-2}}{ q_\f^2 \, \aD}}^{1/2}  \pare{\frac{\MD}{10\MeV}} \ .
\label{eq:ksi PT}
\eeq
After the dark photons acquire mass, they may decay into SM charged fermions with rate given in \eq{eq:gammaD to f+f-}. Requiring that the dark photons decay before BBN, $T_{_{\rm V, \, decay}} > T_{_{\rm V, \, BBN}} \sim 1 \MeV$, yields
\beq
\e > 10^{-10} \frac{1}{f_{_{\rm EM}}^{1/2}}\pare{\frac{10 \MeV}{\MD}}^{1/2} \ .
\label{eq:decay bf BBN}
\eeq
Both conditions \eqref{eq:ksi PT} and \eqref{eq:decay bf BBN}, as well as $\MD>1.022 \MeV$, must hold in order for the dark photons to decay before BBN.

\item
Survive through BBN.

If either of the inequalities \eqref{eq:ksi PT}, \eqref{eq:decay bf BBN} is not satisfied, or $\MD <1.022 \MeV$, dark photons may decay after BBN, or even survive until today. If dark photons are relativistic at the time of BBN, i.e. if 
\beq 
\MD < 3 \ks_{_{\rm BBN}} T_{_{\rm V, \: BBN}} \simeq \ks_{_{\rm BBN}} \times 3\MeV \ , 
\label{eq:gammaD rel BBN}
\eeq  
we must require 
\beq \ks_{_{\rm BBN}} \lesssim 0.6 \ . \label{eq:BBN bound} \eeq 
This upper limit corresponds to the relativistic energy density of one extra neutrino species, as allowed by current data~\cite{Mangano:2011ar}.  If   $\MD > \ks_{_{\rm BBN}} \times 3\MeV$, dark photons are non-relativistic at BBN, and there is no constraint on $\ks_{_{\rm BBN}}$. 

Independently of whether the inequality \eqref{eq:gammaD rel BBN} is satisfied, we must require that the abundance of dark photons does not alter the time of matter-radiation equality. Moreover, if dark photons are stable, we must require that they comprise only a subdominant component of the DM of the universe. 
Their number density $n(\gammaD)$, normalised with the entropy density of the universe $s$, is 
\[ Y(\gammaD) \equiv \frac{ n(\gammaD) }{s}  \: \approx \:  \frac{\ks_{\gammaD}^3}{ g_{*,S} (t_{\gammaD}) } \, h(x_{\gammaD}) \ , \] 
where $g_{_{*, S}} (t_{\gammaD})$ is the number of effective relativistic d.o.f.s at the time of the dark photon chemical decoupling, $x_{\gammaD}$ is given in \eq{eq:MD/T} and
\[
h(x) 
\equiv \frac{135}{4\p^4} \int_{x}^\infty dy \: \frac{y \sqrt{y^2 - x^2}}{e^x - 1}
\sim \left\{
\bal{6}
&0.8,& \qquad &x < 3 &
\\
&0.4 \, x^{3/2} \: e^{-x},& \qquad &x \gtrsim 3 \ .&
\eal
\right. \]
We discern the following cases:

\benu[a.]
\item
Dark photons would alter the time of  matter-radiation equality if they became non-relativistic and dominated the energy density of the universe at some temperature $T_{_{\rm V, dom}} \geqslant T_{_{\rm V, eq}}$, where $ T_{_{\rm V, eq}} \sim 5 \eV$ is the temperature at matter-radiation equality. In this case, at $\TV = T_{_{\rm V, dom}}$, 
$s Y(\gammaD) \MD \approx \r_{_U} = (\p^2 / 90)g_* \TV^4$, where $\r_{_U}$ is the energy density of the universe and $g_*$ are the relativistic d.o.f.s. This gives $T_{_{\rm V, dom}} \approx 4 Y(\gammaD) \MD $. If $\MD / (\ks_{_{\rm dom}} T_{_{\rm V, dom}}) >3$ and $T_{_{\rm V, dom}}  \geqslant T_{_{\rm V, eq}}$, that is if 
\beq 
\frac{1.25\eV}{\MD}
\ < \ \frac{\ks_{\gammaD}^3 h(x_{\gammaD})}{g_{*,S}(t_{\gammaD})} \ < \ \frac{0.08}{ \ks_{_{\rm dom}} }
\ , 
\label{eq:gammaD dom}
\eeq
we must require that dark photons decay at $T_{_{\rm V, decay}} > T_{_{\rm V, dom}}$. This necessitates
\beq
\e > 6 \times 10^{-9} \, f_{_{\rm EM}}^{-1/2} \:
\frac{ \ks_{\gammaD}^3 }{ g_{_{*, S}} (t_{\gammaD}) }  \pare{\frac{\MD}{10\MeV}}^{1/2}
\:  h(x_{\gammaD})  \ .
\label{eq:decay af BBN}
\eeq
as well as $\MD > 1.022 \MeV$.

\item
If the condition \eqref{eq:gammaD dom} is not satisfied, then the dark photons do not dominate the energy density of the universe before matter-radiation equality and the bound of \eq{eq:decay af BBN} does not apply. However, we still have to require that, if the dark photons have become non-relativistic today, their relic abundance is a subdominant component of the DM. Their contribution to the matter density of the universe is $\W(\gammaD) = s_0 Y(\gammaD) \MD /\r_c $. Requiring $\W(\gammaD) < 0.01$ implies
\beq
\ks_{\gammaD} < 0.02
\pare{\frac{ g_{_{*, S}} (t_{\gammaD}) }{10}}^{1/3} \pare{\frac{100 \keV}{\MD}}^{1/3} 
\: h(x_{\gammaD})^{-1/3}
\ .
\label{eq:beyond MR eq NR}
\eeq 
Note that this condition also ensures that $T_{_{\rm V, dom}} < T_{_{\rm V, eq}} $.
The condition \eqref{eq:beyond MR eq NR} applies provided that $\MD / (\ks_0 T_{_{\rm V, 0}}) > 3$, or
\beq
\MD > 3 \ks_0 T_{_{\rm V, 0}} \simeq \ks_0 \times 7 \times 10^{-4} \eV \ ,
\label{eq:NR today}
\eeq
where the subscript $_0$ refers to the present epoch. The condition \eqref{eq:beyond MR eq NR} is the equivalent of the Cowsik-McClelland bound and the Lee-Weinberg bound for $x_{\gammaD} < 3$ and $x_{\gammaD} > 3$ respectively, adapted to our scenario.

\item
If the inequality~\eqref{eq:NR today} is not satisfied, that is if
\beq
\MD < \ks_0 \times 7 \times 10^{-4} \eV \ ,
\label{eq:R today}
\eeq
dark photons are still relativistic today, and the only applicable bound is \eqref{eq:BBN bound}. 
This case includes the limit of a massless dark photon.

\begin{figure}[t]
\centering
\includegraphics[width=0.5\textwidth]{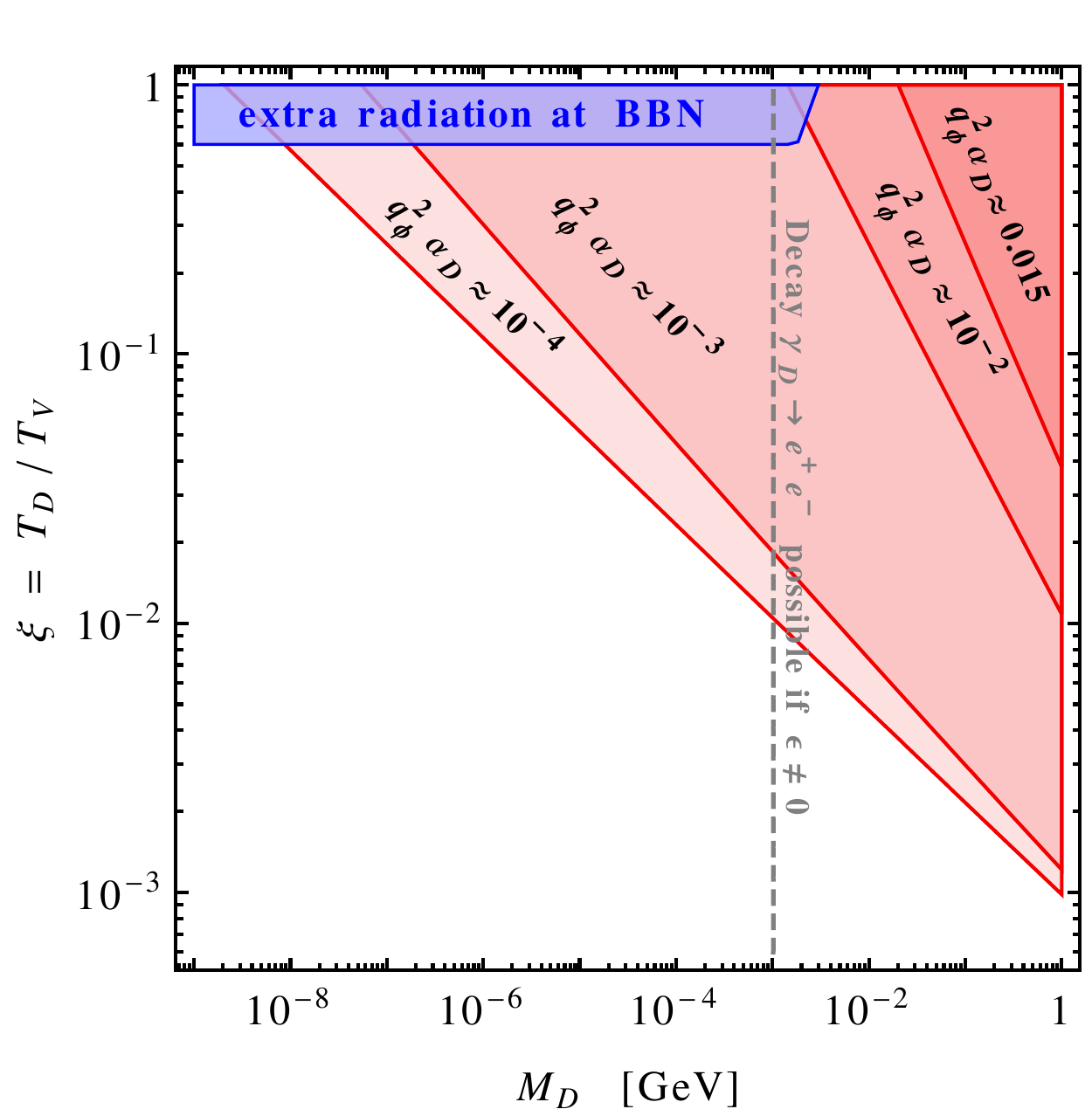}
\caption{\footnotesize
Bounds on the dark-to-ordinary temperature ratio, $\xi = T_{_{\rm D}} / T_{_{\rm V}}$, vs the dark photon mass $M_{_D}$, assuming $\epsilon\to 0$. In the blue-shaded region on the top left, the extra radiation due to relativistic dark photons exceeds the BBN limit. This bound applies to $\xi_{_{\rm BBN}}$ (i.e.~evaluated at the time of BBN). In the red-shaded regions on the right, the relic abundance of the dark photons may alter the time of matter-radiation equality or dominate the DM density. Each region corresponds to the value of $q_\phi^2 \alpha_{_D}$ shown on the plot. These bounds apply to $\xi_{\gamma_{_D}}$ (i.e.~evaluated at the time of the dark photon chemical decoupling). We have assumed that $m_\f \approx \vD$.
To the right of the grey dashed line, the cosmological abundance of the dark photons may decay into SM charged fermions if the dark force mixes kinetically with hypercharge. If the decay is sufficiently fast, the bounds on $\xi$ may be relaxed or eliminated. 
}
\label{fig:cosmo}
\end{figure}

\eenu
\eenu
We illustrate the above constraints in Fig.~\ref{fig:cosmo}, for the case of $\e \to 0$. 

If $\e \neq 0$, these constraints can only be relaxed due to the possibility of dark photon decay. To assess the viability of a parameter set in the case of $\e \neq 0$, the conditions \eqref{eq:decay bf BBN} or \eqref{eq:decay af BBN}, if and whichever applicable, should be compared to \eqref{eq:equil bound}. Of course, $\e$ is itself constrained by various experiments. (For a compilation of bounds on $\e$, see e.g.~Refs.~\cite{Jaeckel:2012yz,Lees:2014xha}, and for bounds on minicharged particles, see e.g.~Ref.~\cite{Goodsell:2009xc}. For recent stringent bounds on very light dark photons mixing with hypercharge, see Ref.~\cite{An:2013yua}.)
Moreover, the kinetic mixing of the dark force with hypercharge opens the possibility for direct and indirect DM detection, and it implies a number of other observational signatures~\cite{Batell:2009yf,Pospelov:2008zw}. 
Exploring the direct detection prospects of atomic DM involves taking into account all DM components and the different nature of their interactions with ordinary matter, including elastic and inelastic scattering~\cite{Cline:2012is,Cline:2013zca,Laha:2013gva}. The direct detection of atomic DM in the limit of a massless dark photon has been considered in Refs.~\cite{Cline:2012is,Cline:2013zca}. Related studies of direct detection of (multicomponent) DM with long-range interactions can be found in Refs.~\cite{Foot:2003iv,Foot:2010hu,Foot:2013msa,Fornengo:2011sz}.
The present scenario of atomic DM with a massive dark photon may also produce indirect detection signatures. The ionised component of DM may form bound states in the dense environment of the haloes today~\cite{Pospelov:2008jd,Shepherd:2009sa,Pearce:2013ola}. The formation of bound states is invariably accompanied by emission of a mediator, here a dark photon, whose subsequent decay into SM particles may yield observable signals~\cite{Pearce:2013ola}. 
We leave the study of direct and indirect signatures of atomic DM with a massive dark photon, as well as a more detailed discussion of the bounds on $\e$, for future work.

\subsection{Dark-matter kinetic decoupling and large-scale structure}

The coupling of DM to a dark radiation bath, if it persists until late times, may affect the matter power spectrum and gravitational clustering. The acoustic oscillations of the coupled DM and dark-radiation system can imprint a new characteristic scale on the matter power spectrum, which in turn may give rise to novel features on the CMB temperature and polarisation spectra~\cite{Foot:2012ai,CyrRacine:2012fz,Feng:2009mn,Cyr-Racine:2013fsa}. Consequently, galaxy surveys and CMB constrain the coupling of DM to dark radiation.

In the scenario under consideration, DM and dark radiation remain coupled mostly via Compton scattering of dark photons on dark ions and Rayleigh scattering of dark photons on neutral dark atoms~\cite{CyrRacine:2012fz}. The scale of the dark acoustic oscillations is determined by the quantity~\cite{Cyr-Racine:2013fsa}
\beq
\S_{\rm DAO} \equiv \aD \pare{\frac{\eV}{\D}} \pare{\frac{\GeV}{\mH}}^{1/6} \ .
\eeq
Galaxy surveys and CMB, including low-multiple, high-multiple and lensing data, constrain this quantity to be $\S_{\rm DAO} < 10^{-4.5}$ if the inequality \eqref{eq:BBN bound} is saturated, $\ks \approx 0.6$. Lower values of $\ks$ relax this bound~\cite{Cyr-Racine:2013fsa}.  For a massive dark photon, these constraints apply only if the dark photons are still relativistic at CMB, i.e.~if $\MD/T_{_{\rm D, \,CMB}} \lesssim 3$, or $\MD \lesssim 3 \, \ks_{_{\rm CMB}} T_{_{\rm V, \, CMB}} \simeq \ks_{_{\rm CMB}} \times 15 \eV$.
While these bounds are very important, both their strength and the dark photon mass range to which they apply are fairly limited, with the constraints from the ellipticity of large haloes being by far more severe, as we discuss next.

%%%%%%%%%%%%%%%%%%%%%%%
\begin{table}[t!]
\centering

\newcolumntype{D}{>\raggedright p{0.6cm}}
\newcolumntype{L}{>{\varwidth[c]{\linewidth}}l<{\endvarwidth} }
\newcolumntype{F}{X}
\def\arraystretch{2.7}

%\fbox{
\begin{tabularx}{0.94\textwidth}{|D L |F|}
\hline
(i)   & \[ T_{\rm asym}\] 	&    Dark asymmetry generation \\  \hline 
(ii)  & \[ T_{\rm dec}\] 		&     Thermal decoupling of the dark and ordinary sectors  \\ \hline
(iii) & \[ T_{_{\rm D, \, {\bf p} - FO }} \approx \mpD/30 \] &  Freeze-out of  $\pD-\bar{p}_{_D}$ annihilations \\ \hline
(iv)  & \[ T_{_{\rm D, \, {\bf e} - FO }} \approx \meD/30 \] &  Freeze-out of  $\eD-\bar{e}_{_D}$ annihilations \\ \hline
(v)   & \[ \D \gtrsim T_{_{\rm D, DR}} \gtrsim \D/50 \] &   Dark recombination \\ \hline
(vi)  & \[ T_{_{\rm D, \, PT }}  \sim \MD / (8\p q_\f^2 \aD)^{1/2} \] &  U(1)$_{_D}$-breaking  phase transition  \\ \hline
(vii) & \[ T_{_{\rm D, \gammaD}} \sim \MD / [41 (8\p q_\f^2 \aD)^{1/2}]  \] &   
Dark photon chemical decoupling \\ \hline
(viii)& \[ T_{_{\rm V, \, BBN}} \approx 1 \MeV \] &   Big Bang Nucleosynthesis \\ 
\hline
\end{tabularx}
%}
\caption{\footnotesize
One possible cosmological sequence in the scenario of atomic DM with a massive dark photon. For other cases and further discussion, see text. In (i) and (ii), the dark and the ordinary sectors have common temperature. (iii) -- (vii) refer to the dark-sector temperature, while (viii) refers to the temperature of the ordinary sector.
}
\label{tab:cosmo sequence}
\end{table}
%%%%%%%%%%%%%%%%%%%%%%

\section{Dark-matter self-interaction in haloes}
\label{sec:self-inter}

In this section, we explore the effect of the DM self-interactions in haloes. For definiteness, we shall assume that in the entire parameter space, DM is made up of equal amounts of $\pD^+$ and $\eD^-$, with no relic $\pD^-$ and $\eD^+$ present, and that the DM ions are bound in atoms as described by \eq{eq:xD resid}.  Of course, the multi-component nature of DM is strictly inevitable under the minimal assumptions of the model (asymmetric DM coupled to gauge vector boson) only when the conditions \eqref{eq:MDasym} and  \eqref{eq:MDmax} are satisfied. However, these conditions are sufficient, but not necessary; asymmetric DM coupled to a massive gauge vector boson may be multi-component and atomic even outside the validity of \eqref{eq:MDasym} and \eqref{eq:MDmax}, depending on the specifics of the model, as explained in Sec.~\ref{sec:massive} and Sec.~\ref{sec:eD asym}.  Here, we choose to focus on this multi-component DM case. For a study of the DM self-interactions in haloes in the case of single-component DM coupled to a vector boson, see Refs.~\cite{Tulin:2012wi,Tulin:2013teo}.

\subsection{DM scattering rates}

We assume that DM has a velocity distribution which is locally Maxwellian
\beq
f (v, \bar{v}) = \pare{\frac{3}{2 \p \bar{v}^2}}^{3/2}  e^{-3 v^2/ 2 \bar{v}^2} \: ,
\label{eq:distributions}
\eeq
where $\bar{v}$ is the average rms velocity, and of course $\int d^3 v \: f (v, \bar{v}) = 1$.  
The average velocity is a function of the position inside the halo, $\bar{v} = \bar{v} (r)$.

Let $\G_{\bf p}$, $\G_{\bf e}$, and $\G_{\bf H}$ be the average rates of momentum-changing collisions for dark protons, dark electrons and dark Hydrogen atoms. Each of these rates includes the contributions from scattering with all other species, 
\beq 
\bal{5}
&\G_{\bf p}	& \ = \  & \G_{\bf pp} + \G_{\bf pe} + \G_{\bf pH}  \: ,	& \\
&\G_{\bf e}	& \ = \  & \G_{\bf ep} + \G_{\bf ee} + \G_{\bf eH}	  \: ,	& \\
&\G_{\bf H}	& \ = \  & \G_{\bf Hp} + \G_{\bf He} + \G_{\bf HH} \: .	& 
\eal
\label{eq:rates total}
\eeq
$\G_{\bf ij}$ is the average momentum-loss rate from species \textbf{i} to species \textbf{j}~\cite{CyrRacine:2012fz},
\beq
\G_{\bf ij} = \frac{ \dot{p}_{\bf ij} }{ \bar{p}_{\bf i} } \: ,
\label{eq:Gamma ij}
\eeq
where $\bar{p}_{\bf i} = m_{\bf i} \bar{v}$ is the average rms momentum of species \textbf{i}, and $\dot{p}_{\bf ij}$ is the average momentum-loss rate of species \textbf{i} due to collisions with species \textbf{j}, in a DM halo. We estimate it as
\bea
\dot{p}_{\bf ij}  
&=&  n_{\bf j}(r)  \int d^3 v_{\bf i} \, f (v_{\bf i}, \bar{v}) \int d^3 v_{\bf j} \, f (v_{\bf j}, \bar{v}) 
\: |{\bf v_i - v_j}| \int d\W \frac{d \s_{\bf ij}}{d\W} \: \d p_{\bf ij}  \nn \\
&=&   n_{\bf j}(r)  \int d^3 v f (v, \bar{v}_{\rm rel}) \, v \int d\W \frac{d \s_{\bf ij}}{d\W} \: \d p_{\bf ij}
\ ,
\label{eq:pdot ij}
\eea
where $n_{\bf j} (r)$ is the number density of the \textbf{j} species in the halo, $d\s_{\bf ij}/d\W$ is the ${\bf i-j}$ differential scattering cross-section, $\d p_{\bf ij}$ is the momentum transfer from \textbf{i} to \textbf{j}, and $\bar{v}_{\rm rel} \equiv \sqrt{2}\bar{v}$. 
Both $d\s_{\bf ij}/d\W$ and $\d p_{\bf ij}$ depend only on the relative velocity $v = |{\bf v_i - v_j}|$ and the scattering angle $\theta$, with
\beq
\d p_{\bf ij} =  \frac{m_{\bf i} m_{\bf j}}{m_{\bf i} + m_{\bf j}} v \times g (\th) \ .
\label{eq:delta p}
\eeq
The angular function $g(\theta)$ depends on whether the total, the longitudinal, or the transverse momentum transfer is considered,
\bea
g_{\rm tot}(\th) &=& 
\left \{
\bal{5}
&2 \sin (\th/2),& \qquad &\text{if }{\bf i \neq j} \\
&2 \min \sqpare{\sin (\th/2), \: \cos (\th/2) },& \qquad &\text{if }{\bf i = j}
\eal
\right.
\label{eq:gtot(theta)}
\\
g_{\rm l}(\th) &=& 
\left \{
\bal{5}
& 1-\cos \theta,& \qquad &\text{if }{\bf i \neq j} \\
& \min \sqpare{1-\cos \th, \: 1+ \cos \th },& \qquad &\text{if }{\bf i = j}
\eal
\right.
\label{eq:gl(theta)}
\\
g_{\rm tr}(\th) &=& | \sin \th |  \:  .
\label{eq:gtr(theta)}
\eea
The case \textbf{i = j} takes into account that the forward and backward scattering of identical particles are equivalent. The transverse momentum transfer is in either case the same. Quite often, the energy transfer in the transverse direction is instead considered; this can be accounted by setting $g(\th) = \sin^2 \th$.
In recent self-interacting DM simulations, the momentum transfer is parametrised in terms of the longitudinal component and assuming distinguishable particles, with the momentum-transfer cross-section defined as 
\beq
\s_{\rm mt} \equiv \int d\W \, (1-\cos \theta) \, \frac{d\s}{d\W}  \ .
\label{eq:sigma mt}
\eeq

For the ion-ion collisions, governed by the  Yukawa potential of \eq{eq:Yukawa}, we use existing analytical formulae for $\s_{\rm mt}$. In the small coupling regime, $r_{\bf ij} \equiv 4\aD \mu_{\bf ij}/\MD \ll 1$, where $\mu_{\bf ij} = m_{\bf i} m_{\bf j}/(m_{\bf i} + m_{\bf j})$, the Born approximation is valid and the momentum transfer cross-section in a \textbf{i-j} collision is
\beq
\s_{\rm mt, \, \bf ij}^{\rm Born} = \frac{2 \p \, \b_{\bf ij}^2}{\MD^2}
\sqpare{\ln\pare{1+\frac{r_{\bf ij}}{\b_{\bf ij}}} - \frac{r_{\bf ij} }{\b_{\bf ij} + r_{\bf ij}}} \: ,
\label{eq:Born}
\eeq
where $\b_{\bf ij} = \aD \MD/(v_{\rm rel}^2 \m_{\bf ij})$, with $v_{\rm rel}$ being the relative velocity of the \textbf{i-j} pair in the DM halo. For larger couplings, the Born approximation breaks down, and the classical approximation becomes relevant.   For the attractive $\eD - \pD$ interaction, in the classical approximation~\cite{Khrapak:2004apr,Khrapak:2003jun,Khrapak:2004Nov,Tulin:2013teo}
\beq
\s_{\rm mt, \, \bf ep}^{\rm clas} \simeq \left \{
\bal{5}
&\frac{4\p}{\MD^2} \, \b_{\bf ep}^2 \, \ln \pare{1+ \frac{1}{\b_{\bf ep}}} \ , \qquad&
& \,\b_{\bf ep}\, \lesssim 10^{-1} &
\\
&\frac{8\p}{\MD^2} \frac{\b_{\bf ep}^2}{1+1.5 \b_{\bf ep}^{1.65}} \ ,  \qquad& 
&10^{-1} \lesssim  \,\b_{\bf ep}\,  \lesssim 10^3 &
\\
&\frac{0.81\p}{\MD^2} \sqpare{\ln^2 \b_{\bf ep} + 2 \ln \b_{\bf ep} +2.5 + \frac{4}{\ln \b_{\bf ep}}} \ , \qquad&
& 10^3 \lesssim \,\b_{\bf ep} \, .&
\eal
\right.
\label{eq:ep clas}
\eeq
The classical approximation for the $\eD^- - \pD^+$ scattering is valid outside the Bohr-approximation regime and for $r_{\bf ep}/\b_{\bf ep} >1$. For $1 < r_{\bf ep} < \b_{\bf ep}$, the $\eD^- - \pD^+$ scattering exhibits resonances due to the contribution of (virtual) bound states~\cite{Tulin:2012wi,Tulin:2013teo}. Here, for simplicity, we ignore the resonant structure, which does not affect the bulk of the parameter space. We shall adopt the classical approximation everywhere outside the Born approximation, i.e. for all $r_{\bf ep} >1$.
For the repulsive $\pD^+ - \pD^+$ and $\eD^- - \eD^-$ scattering~\cite{Khrapak:2004Nov,Feng:2009hw} 
\beq
\s_{\rm mt, \, \bf ii}^{\rm clas} \simeq \left \{
\bal{5}
&\frac{2\p}{\MD^2} \, \b_{\bf ii}^2 \, \ln \pare{1+ \frac{1}{\b_{\bf ii}^2} } \ , \qquad&
& \,\b_{\bf ii}\, \lesssim 1 &
\\
&\frac{\p}{\MD^2} \pare{\ln 2\b_{\bf ii} - \ln \ln 2\b_{\bf ii}}^2    \ ,  \qquad& 
&\b_{\bf ii} \,  \gtrsim 1 \, .&
\eal
\right.
\label{eq:ee pp class}
\eeq

In contrast to the ion-ion collisions, the atom-atom and atom-ion collisions are not expected to be significantly affected by a non-zero dark photon mass.  In the massless dark photon limit, the range of the interatomic potential is of the order of the Bohr radius $a_B = (\muD \aD)^{-1}$ (see Ref.~\cite{Cline:2013pca} and references therein). Significant modifications due to a non-zero dark photon mass are expected to appear only at distances $r \gtrsim 1/\MD$, i.e.~larger than the Bohr radius whenever the condition \eqref{eq:MD<Delta} is satisfied.  The atom-atom and atom-ion scattering rates estimated assuming a massless dark photon are thus a good approximation for the case of a massive dark photon, in the parameter space where dark atoms can form in the early universe.  Nevertheless, even in the limit of a massless dark photon, there is currently considerable uncertainty in the existing literature about the atom-atom and atom-ion collision rates. In the following, we adopt two different approaches, developed in Ref.~\cite{Cline:2013pca} and Ref.~\cite{CyrRacine:2012fz}. Below, we summarise their main results relevant to our analysis.

The authors of Ref.~\cite{Cline:2013pca} calculated the low-energy atom-atom scattering cross-section by direct computation of the phase shifts induced by the interatomic potentials. Because of the multiple states of the Hydrogen atoms, the atom-atom scattering cross-section exhibits a rich resonant structure, which we shall ignore, as in the case of $\pD^+ - \eD^-$ scattering.
Away from resonances, Ref.~\cite{Cline:2013pca} found that the energy dependance of the transverse energy-transfer cross-section can be fit for a wide energy range by the analytical expression
\beq
\s_t \approx  \pare{\muD \aD}^{-2} 
\sqpare{b_0 + b_1 \pare{\frac{\mH v^2}{4\muD \aD^2}} + b_2 \pare{\frac{\mH v^2}{4\muD \aD^2}}^2 }^{-1} \ ,
\label{eq:CL sigma HH}
\eeq
where $\s_t$ is defined as
\beq \s_t \equiv \int d\W \,\sin^2 \th \, \frac{d\s}{d\W} \ .
\label{eq:sigma t}
\eeq
The parameters $b_0, \: b_1, \: b_2$ are determined by numerical fits, and depend mildly on the ratio $R \equiv \mpD/\meD$.
As noted above, $\s_t$ is better suited than $\s_{\rm mt}$ for estimating the momentum and energy transfer in collisions between identical particles; however, the comparison of $\s_t$ with results from current simulations is more precarious. We use \eq{eq:CL sigma HH} and the numerical values for the fitting parameters provided in Ref.~\cite{Cline:2013pca} to estimate the efficiency of the atom-atom collisions in haloes.  To cover a continuum range of $R$, we interpolate $b_0, \: b_1, \: b_2$ between the values provided.  As an example, we give here the fitting parameters at $R = 10$: $b_0 = 0.012, \: b_1 = 0.197, \: b_2 = 0.053$~\cite{Cline:2013pca}. 
We present our results using this approach in 
Figs.~\ref{fig:CL_alphaVmH}, \ref{fig:CL_DeltaVmH}, \ref{fig:CL_alphaVmH 2}, \ref{fig:CL_continuous MD}, \ref{fig:CL_me=mp small MD} -- \ref{fig:CL_fixed xD}.

Reference~\cite{Cline:2013pca} does not provide any estimate for the atom-ion scattering cross-section; we thus ignore the atom-ion collisions when adopting their estimates for the atom-atom scattering.  This is justified because we expect that atom-ion collisions are either not dominant or not significant in the entire parameter space. As mentioned above, the screening scale for interactions involving atoms is the Bohr radius, while ion-ion collisions are screened by the dark photon mass; given the condition \eqref{eq:MD<Delta} for the formation of dark atoms, interactions involving atoms are always more strongly screened than ion-ion interactions. Thus, in the parameter space where $\xD \gtrsim 0.5$, ion-ion collisions dominate due to both a stronger cross-section and comparable or larger number densities of the colliding species. The atom-ion cross-section may be comparable or stronger that the atom-atom cross-section. However, collisions involving ions cannot play any significant role if $\xD \ll 1$. It is possible that the atom-ion collisions dominate or contribute significantly to the total momentum-transfer rate in the regime where $0.1 \lesssim \xD \lesssim 0.5$. However, in this (fairly limited) regime, the gauge coupling is typically not large enough to render the atom-ion or the atom-atom interactions significant for the dynamics of haloes.  We confirm this assertion when adopting the analysis of Ref.~\cite{CyrRacine:2012fz}, which includes an atom-ion scattering rate, and which we describe next.

The authors of Ref.~\cite{CyrRacine:2012fz} estimate the atom-atom and atom-ion collision rates by appropriate rescaling of the experimentally measured rates for ordinary atoms and ions. They consider the momentum-transfer cross-section as defined in \eq{eq:sigma mt}, and average over a Maxwellian velocity distribution.  They estimate the atom-atom and atom-ion momentum-transfer rates to be
\bea
\G_{\bf HH} &\simeq&  n_{_{\bf H}} 
\sqpare{15 \p \, (4/3)^{3/8} \, \G(19/8)} 
\pare{\frac{ \aD^2 \, \bar{v}^{3/4} }{\D^2}}
\pare{\frac{\tilde{m}_e \mH}{\muD \D}}^{-1/8}
\sqpare{1+\frac{\tilde{m}_e \mH}{\muD \D} \: \frac{ \bar{v}^2}{225}}^{-19/8} \! ,
\label{eq:Gamma HH}
\\
\G_{\bf pH} &\simeq& n_{_{\bf H}} \
\frac{ 30\sqrt{3 \p^3} \, \aD^2 \bar{v} }{\D^2} \:
\frac{\pare{\mH \mpD}^{1/2}}{\mH+\mpD}
\sqpare{1+\frac{\tilde{m}_e \mpD}{(\muD+\mpD) \D} \: \frac{ \bar{v}^2}{150}}^{-5/2}  ,
\label{eq:Gamma pH}
\\
\G_{\bf eH} &\simeq& n_{_{\bf H}} \
\frac{ 30\sqrt{3 \p^3} \, \aD^2 \bar{v}}{\D^2} \: 
\frac{\pare{\mH \meD}^{1/2}}{\mH+\meD}
\sqpare{1+\frac{\tilde{m}_e \meD}{(\muD+\meD) \D} \: \frac{ \bar{v}^2}{150}}^{-5/2}  ,
\label{eq:Gamma eH}
\\
\G_{\bf Hp}  &\simeq&  \frac{n_{\bf p} \mpD }{n_{_{\bf H}} \mH  } \, \G_{_{\bf pH}} \ , \label{eq:Gamma Hp}
\\
\G_{\bf He}  &\simeq&  \frac{n_{\bf e} \meD }{n_{_{\bf H}} \mH  } \, \G_{_{\bf eH}} \ , \label{eq:Gamma He}
\eea
where we remind that $\tilde{m}_e = 511 \keV$ is the ordinary electron mass. The range of validity of the above rates is considered to be the energy interval $10^{-3} \lesssim E_{\rm cm} / \D \lesssim 10$, where $E_{\rm cm} = \m_{\bf ij} v^2/2$ is the centre-of-mass energy of the colliding particles, with $\m_{\bf ij}$, $v$ being their reduced mass and relative velocity respectively.
As noted in Ref.~\cite{CyrRacine:2012fz}, \eq{eq:Gamma HH} is expected to over-estimate the atom-atom collision rate at low-energies. Indeed, the rate of \eq{eq:Gamma HH} diverges as $v \to 0$, in contrast to the result of Ref.~\cite{Cline:2013pca}, which finds that at low energies $s$-wave scattering dominates and the atom-atom cross-section becomes velocity-independent. Our numerical calculations show that, even within the energy range of validity, the atom-atom scattering rate of \eq{eq:Gamma HH} is typically significantly larger than that estimated using the the cross-section of \eq{eq:CL sigma HH}. This is, at least partly, due to the difference between the definitions of \eqs{eq:sigma mt} and \eqref{eq:sigma t}.
We present our results using the rates of \eqs{eq:Gamma HH} - \eqref{eq:Gamma He} in Figs.~\ref{fig:SI_alphaVmH}, \ref{fig:SI_DeltaVmH}, \ref{fig:SI_alphaVmH 2}, \ref{fig:SI_continuous MD}.

\subsection{Effect of DM self-interaction in haloes}

As we have established, the scenario we consider in this paper generically results in multi-component DM, with different types of inter- and intra-species interactions. Obviously, existing DM simulations of single-component DM, which have examined a limited number of interaction types and strengths, do not directly apply to this scenario. Nevertheless, here we shall use the insight from these simulations to devise reasonable conditions which will allow us to gauge the impact of the DM interactions in the scenario under consideration, on the dynamics of haloes.  Our goal is two-fold: 
(i) To place rough constraints which ensure that the DM scattering in Milky-Way-size haloes does not destroy their observed ellipticity.
(ii) To identify the regions of the parameter space which could affect the dynamics of dwarf-galaxy-size haloes,  and bring predictions in better agreement with observations.

We define an effective average momentum-transfer rate 
\beq
\G_{\rm eff} \equiv 
h_{\bf p} \min\pare{\G_{\bf p}, \G_{\rm crit} / h}  +  h_{\bf e} \min\pare{\G_{\bf e}, \G_{\rm crit} /h}  +  h_{_{\bf H}} \min\pare{\G_{\bf H}, \G_{\rm crit}/h}   
\label{eq:Gamma eff} \ ,
\eeq 
where $h_{\bf p}$, $h_{\bf e}$ and $h_{_{\bf H}}$ are the mass fractions carried by dark protons, dark elecrons, and dark Hydrogen atoms respectively,
\beq
\bal{10}
&h_{\bf p}	& \ \equiv \  &\frac{ \xD \mpD }{\xD (\mpD + \meD)  + (1-\xD)\mH }	
&  \ \simeq \ &\frac{\xD \mpD}{\mH}	& 
\ , \\
&h_{\bf e}	& \ \equiv \  &\frac{\xD \meD  }{\xD (\mpD  + \meD)  + (1-\xD)\mH}	
&  \ \simeq \ &\frac{\xD \meD}{\mH}	& 
\ , \\
&h_{_{\bf H}}	& \ \equiv \  &\frac{(1-\xD) \mH }{\xD (\mpD + \meD) + (1-\xD)\mH}
&  \ \simeq \ &1-\xD & 
\ .
\eal
\eeq
Note that $\G_{\rm eff}$ depends on the position in the DM halo through its dependence on the densities and the velocity dispersion of the DM species. The dependence on the velocity dispersion, in particular, arises mostly due to the strong velocity dependence of the scattering cross-sections [cf.~\eqs{eq:ep clas} -- \eqref{eq:CL sigma HH}].

$\G_{\rm crit}$ is an estimate (to be specified below) for the magnitude of the effective momentum-transfer rate above which there is a significant effect on the DM halo under consideration; it is what we will eventually compare $\G_{\rm eff}$ with. Since the various rates, $\G_{\bf p}, \ \G_{\bf e}, \ \G_{\bf H}$, depend on the position inside the DM halo, the estimate for $\G_{\rm crit}$ should also depend on the position at which these rates are evaluated. In \eq{eq:Gamma eff}, we weigh the contributions of the various species to $\G_{\rm eff}$ by the mass fraction they carry, but we also cap the contribution of each species at $\G_{\rm crit}/h$, with $h<1$. Indeed, if the momentum-transfer rate for a given species is very large, while this species carries only a tiny fraction of the mass of the halo, the effect of the momentum loss by this species on the halo dynamics is negligible. Then, the contribution of this species to $\G_{\rm eff}$ should not be allowed to drive $\G_{\rm eff}$ to or above the critical value.\footnote{Note from \eq{eq:Gamma ij} that, although the momentum loss by a species is weighted by the momentum carried by this species, the definition of $\G_{\bf ij}$ is \emph{not} such that $\int \G_{\bf ij} dt \leqslant 1$.}     
On the other hand, if a species carries a sufficiently large portion of the halo mass, its interactions are expected to largely determine the dynamics of the halo.  Capping the contribution of each species to $\G_{\rm eff}$ at  $\G_{\rm crit}/h$ encapsulates these considerations:  $h$ is the fraction of DM, which, if very strongly interacting, can drive $\G_{\rm eff}$ to its critical value. In the following, we choose (somewhat arbitrarily) $h = 50\%$. This choice is partly informed by the dynamics of the  dark matter and ordinary matter mixture in the haloes; while ordinary matter, which is quite self-interacting and dissipative, makes up about 15\% of the mass in the universe, it does not affect significantly the clustering of dark matter at most scales.

The strongest constraints on the DM self-interaction arise from the observed ellipticity of haloes of the size of the Milky Way or larger.\footnote{
For bounds on the DM self-interaction from colliding clusters, see Ref.~\cite{Kahlhoefer:2013dca}. For velocity independent cross-sections, these bounds are milder or comparable to the bounds from the ellipticity of Milky-Way-size haloes. For long-range interactions, bounds from cluster collisions are more easily satisfied, due to the larger velocity dispersion at cluster scales, $\sim 10^3 \km/\snd$.}  
The relevant observations correspond to distances $r \sim (4 -50) \kpc$ from the centre of the galactic haloes~\cite{Buote:2002wd}. We thus choose to evaluate $\G_{\rm eff}$ for the Milky Way  at $\r\DM = 1 \GeV/\cm^3$, which is estimated to occur at  $r \sim 4.5 \kpc$ for both an NFW and an isothermal profile.
We also set $\bar{v}=220 \km/\snd$. Then, in \eq{eq:Gamma eff}, we substitute $\G_{\rm crit} \to \G_{\rm crit}^{\rm MW}$, and require that
\beq
\G_{\rm eff}^{\rm MW} < \G_{\rm crit}^{\rm MW}
\label{eq:MW cond}
\eeq
where we determine $\G_{\rm crit}^{\rm MW}$ by the following consideration:
For the chosen values of the DM density and velocity dispersion, and at the limit of single-component DM of mass $m$ with $v$-independent scattering cross-section, the condition of \eq{eq:MW cond} reduces to $\s_{\rm mt}/m \lesssim 1\cm^2/\gr$~\cite{Peter:2012jh}.\footnote{Note that this bound does not include the possible effect of baryonic matter. If stars dominate the inner 5-10~kpc of a galaxy of the size of the Milky Way, then their non-spherical distribution may induce some ellipticity on the DM halo~\cite{Kaplinghat:2013xca}. This would relax the upper bound on the DM self-scattering cross-section.}
For single component DM, $\G_{\rm eff} = \r\DM (\s_{\rm mt}/m) \bar{v}$, thus we set
\beq 
\G_{\rm crit}^{\rm MW} =  (1 \GeV/\cm^3) (1 \cm^2/\gr) (220 \km/\snd) 
\simeq 1.2 \Gyr^{-1} \simeq 17 \, H_0  \  .
\label{eq:MW bound}
\eeq 
This is a reasonable upper bound on the average momentum transfer rate for preventing thermalisation and isotropisation of the halo.\footnote{For comparison, Ref.~\cite{CyrRacine:2012fz} uses $\G_{\rm crit} = 10 H_0$, although there are differences in their and our definition of $\G_{\rm eff}$.}

Moreover, we want to identify the parameter space which can affect the dynamics of smaller haloes. Since the dwarf spheroidal galaxies of the Milky Way are consistent with isothermal isotropic profiles, we only set a lower bound on $\G_{\rm eff}$. We evaluate the momentum-transfer rates at $\r\DM = 0.5 \GeV / \cm^3$ and $\bar{v} = 10 \km/\snd$ (for a review on the kinematics of dwarf spheroidal galaxies, see Ref.~\cite{Battaglia:2013wqa}), set $\G_{\rm crit} \to \G_{\rm crit}^{\rm DW}$ and require 
\beq 
\G_{\rm eff}^{\rm DW} > \G_{\rm crit}^{\rm DW}  \  .
\label{eq:DW cond}
\eeq
We choose $\G_{\rm crit}^{\rm DW}$ such that at the limit of single-component DM, the condition \eqref{eq:DW cond} reduces to $\s_{\rm mt} / m > 0.5 \cm^2/\gr$~\cite{Vogelsberger:2012sa,Vogelsberger:2012ku,Peter:2012jh,Rocha:2012jg}. Thus, we pick 
\beq 
\G_{\rm crit}^{\rm DW} = (0.5 \GeV / \cm^3) (0.5 \cm^2/\gr) (10\km/\snd) 
\simeq 0.014 \Gyr^{-1} \simeq 0.2 \: H_0  \  .
\label{eq:DW bound}
\eeq

Note that the above approach in choosing $\G_{\rm crit}^{\rm MW}$ and  $\G_{\rm crit}^{\rm DW}$ renders our bounds independent of the DM density at which the momentum-transfer rates are evaluated, and establishes a reasonable connection with estimated constraints from $N$-body simulations of benchmark DM models.
Moreover, in \eqs{eq:MW bound} and \eqref{eq:DW bound} we have picked somewhat different values for the critical cross-section over mass, in order to allow for the (fairly limited) range of values in the case of a $v$-independent cross-section, which can affect the small-halo dynamics while preserving the ellipticity of larger haloes.

To evaluate the various momentum-transfer rates, we need to know the spatial distributions of the various species in the DM halo. These, in turn, depend on the strength of the interactions among DM particles and the relative abundances of the species. Obviously, detailed simulations are needed to study the clustering of multi-component and self-interacting DM. Here, we shall make the simplifying assumption that all species follow the same density profile, 
\beq
n_{_{\bf H}}(r) \simeq (1-\xD)\r\DM(r)/\mH 
\qquad {\rm and} \qquad
n_{\bf p}(r)  = n_{\bf e}(r) \simeq \xD \r\DM(r)/\mH \ .
\label{eq:densities of species}
\eeq 
$n_{\bf p}(r) = n_{\bf e}(r)$ is indeed expected due to the $\pD-\eD$ attractive interaction, and the resulting screening of the intra-species repulsion. However, since the ion-ion interaction is rather strong, while the ion-atom and atom-atom interaction is typically significantly weaker, it is possible that atoms and ions settle in separate profiles, with the ionised component forming its own isothermal halo~\cite{Kaplan:2011yj}. Nevertheless, in the regimes where $\xD \approx 1$ or $\xD \ll 1$, we expect all of the DM particles to follow the same profile, determined mostly by the gravitational pull of the dominant species.

\subsection{Discussion}

According to the above, the DM self-scattering in haloes is described by five parameters, $\aD$, $\mpD$, $\meD$, $\MD$ and $\ks_{_{\rm DR}}$ (equivalently, $\mpD$ and $\meD$ can be exchanged for $\mH$ and $\muD$, or for $\mH$ and $\D$). Moreover, the efficient annihilation of DM in the early universe sets a lower bound on $\aD$ which depends on $\tilde{\ks}_{\rm ann}$. We shall take $\tilde{\ks}_{\rm ann} = \tilde{\ks}_{_{\rm DR}}$.  Figures~\ref{fig:CL_alphaVmH} to \ref{fig:CL_fixed xD} illustrate the effect of DM self-interactions in slices of the parameter space. For easy reference, in tables \ref{tab:symbols} and \ref{tab:conditions}, we summarise the meaning of the various symbols used and the conditions applied. In the following, we discuss some general features.

\bit

\item {\bf Non-monotonic dependence of $\G_{\rm eff}$ on $\aD$ and $\mH$.} 

Because of the possibility of formation of bound states in the early universe, the DM scattering rate in haloes varies non-monotonically with  $\aD$ and $\mH$. For small $\aD$, dark recombination is inefficient and DM today consists mostly of ions, $\xD \simeq 1$. 
Of course, even if DM is fully ionised, very low values of $\aD$ imply negligible DM self-interaction. 
Increasing $\aD$ increases the ion-ion scattering rate, which becomes sizeable for moderate values of the coupling. However, increasing $\aD$ also implies more efficient formation of dark atoms in the early universe. As a result, when the coupling becomes strong enough to drive $\xD$ to non-maximal values, the DM scattering rate becomes suppressed. Further increase of $\aD$, beyond the point where dark atoms are already the dominant component of DM today, enhances the atom-atom scattering cross-section and increases again the DM self-interaction in haloes. 
Similar considerations apply for the variation of $\G_{\rm eff}$ with $\mH$, which determines the DM number density. Large $\mH$, or small number density, suppresses both the recombination rate in the early universe and the scattering rate in haloes today. The variation of $\mH$ has thus the converse effect of the variation of $\aD$, on $\xD$ and $\G_{\rm eff}$.
These considerations explain the ``wedge'' feature which appears in Figs.~\ref{fig:CL_alphaVmH} -- \ref{fig:SI_DeltaVmH}.

This behaviour exemplifies the importance of considering carefully the cosmology of models in which DM couples to a light mediator.
Clearly, failing to properly account for the formation of bound states in the early universe would result in over-estimating the DM self-scattering in haloes, and would yield inaccurate upper bounds on $\aD$ and lower bounds on $m\DM$ and $\MD$.

\item {\bf The effect of the velocity dependence of the scattering cross-sections.}

Both the atom-atom and the ion-ion cross-sections decrease with increasing velocity.  For ion-ion scattering, $\s_{\bf ij} \propto 1/v^4$ at the $\MD \to 0$ limit with a milder dependence on $v$ for $\MD >0$, as seen from \eqs{eq:Born} -- \eqref{eq:ee pp class}.  The sensitivity of $\s_{_{\bf HH}}$ on $v$ varies: At very low energies atom-atom scattering is velocity-independent, while at higher energies it can be even as sensitive to $v$ as the ion-ion scattering, as seen from \eq{eq:CL sigma HH}~\cite{Cline:2013pca}.  The velocity dependence of $\s_{_{\bf HH}}$ and $\s_{\bf ij}$ results in sizeable parameter regions which satisfy both conditions \eqref{eq:MW cond} and \eqref{eq:DW cond}, as seen in Figs.~\ref{fig:CL_alphaVmH} to \ref{fig:CL_fixed xD}.  This feature is rather prominent both in the $\xD \simeq 1$ and $\xD < 1$ regimes.  As suggested in the introduction, it is a major motivation for considering the present scenario.

\item {\bf Ionisation fraction vs DM annihilation}

In much of the parameter range where $\aD$ provides sufficient annihilation in the early universe, DM has efficiently recombined in atoms. Large ionisation fraction, $\xD > 0.5$, and efficient annihilation occur for [c.f. \eqs{eq:alpha ann}, \eqref{eq:xD resid}]
\beq
2.4 \times 10^{-5} \pare{\frac{\mpD}{\GeV}} \pare{\frac{\tilde{\ks}_{\rm ann}/S_{\rm sym}}{0.5}}^{1/2}
\ \lesssim \ \aD \ \lesssim \
2.3 \times 10^{-3} \pare{\frac{\mH}{\GeV}}^{1/2} \pare{\frac{4 \muD}{\mH}}^{1/4} \pare{\frac{\tilde{\ks}_{_{\rm DR}}}{0.5}}^{1/4} \ ,
\label{eq:ann-ion aD}
\eeq
which necessitates
\beq
\mpD \lesssim 20 \TeV
\pare{\frac{4 \muD}{\mpD + \meD}}^{1/2} \pare{\frac{\tilde{\ks}_{_{\rm DR}}}{0.5}}^{1/2} \pare{\frac{0.5}{\tilde{\ks}_{\rm ann}}} S_{\rm sym}\ .
\label{eq:ann-ion mp}
\eeq
The range of this regime is maximised for $\meD = \mpD$. In Figs.~\ref{fig:CL_me=mp small MD} -- \ref{fig:CL_fixed xD}, we set $\meD = \mpD$ and explore the effect of varying $\MD$ and $\ks_{\rm ann}, \: \ks_{_{\rm DR}}$, as we describe below.

\smallskip

The $\meD = \mpD$ and $\aD \approx \a_{_{D, \rm min}}$ limit resembles most closely the case of single-component symmetric DM coupled to a light or massless dark photon (recall that  $\a_{_{D, \rm min}} \approx \a_{_{D, \rm sym}}$).

\item {\bf The effect of the dark photon mass}

A non-zero dark photon mass screens the ion-ion interactions, and is thus important only in the parameter regions where ions are the dominant component of DM. Significant screening occurs for $\b_{\bf pp} \gtrsim 1$, or $\MD \gtrsim \mpD v^2/2\aD$, albeit the efficacy of the screening depends also on the DM number density, i.e.~on the DM mass. As seen in Fig.~\ref{fig:CL_fixed xD}, for $\xD \sim 0.9$, the screening by $\MD$ can reconcile the DM self-interaction with current bounds, if $\MD \gtrsim 40 \MeV$ for $\mpD, \, \meD \sim 100 \GeV$, with a smaller $\MD$ needed for larger DM masses and/or smaller ionisation fractions. In fact, even for $\xD \sim 0.9$, a very small or zero $\MD$ is viable if $\mpD, \, \meD \gtrsim \TeV$.

For lighter DM, arbitrarily small dark photon masses, including a zero mass, also produce viable and interesting scenarios, due to the formation of dark atoms in the early universe, which suppresses the DM self-scattering rate. Figures~\ref{fig:CL_continuous MD}, \ref{fig:SI_continuous MD} show that a continuum of values for the dark photon mass can produce scenarios of either effectively collisionless or self-interacting DM.  A small dark photon mass is, of course, confluent with the existence and formation of dark atoms. For $\MD \gtrsim \a^2 \muD/2$, dark atoms are kinematically forbidden to form, while for $\MD \gtrsim \aD \muD$ bound states do not exist. In this regime, $\xD = 1$; however, in this case, the sizeable value of $\MD$ screens the ion-ion scatterings and yields again viable scenarios. This regime is depicted in Figs.~\ref{fig:CL_alphaVmH 2} -- \ref{fig:CL_me=mp large MD}.

\item {\bf Dark photon mass vs dark photon relic abundance}

If ions are the dominant component of DM, then a sizable dark-photon mass may be necessary to screen the DM self-interaction (see e.g. Fig.~\ref{fig:CL_fixed xD}). For cosmologically stable dark photons, large $\MD$ implies a stronger upper limit on the dark-to-ordinary temperature ratio $\ks$, as seen from Fig.~\ref{fig:cosmo}.  This bound becomes weak for moderate or large values of $q_\f^2 \aD$; however, to retain a large ionisation fraction, $\aD$ cannot be too large. 
If U(1)$_{_D}$ mixes kinetically with hypercharge, then the cosmological abundance of dark photons can be reduced via decay, consequently relaxing the bounds on $\ks$ independently of $\aD$. 
As already discussed, the kinetic mixing can induce decay of dark photons into SM charged fermions, provided that $\MD > 1.022 \MeV$. In Fig.~\ref{fig:CL_me=mp large MD}, we consider large dark photon masses and depict the effect of a larger $\ks$ value. 
A non-zero $\e$ implies also channels for direct and indirect detection, thereby potentially probing this part of the parameter space.

\item {\bf The limit(s) of collisionless CDM}

As can be seen in Figs.~\ref{fig:CL_alphaVmH}--\ref{fig:CL_fixed xD}, there is ample parameter space in which DM in the scenario under consideration behaves as collisionless CDM. In fact, there are more than one ways to approach this limit, as evident from the previous discussion.  Large $\mH$ implies small number density and small DM scattering rate. Moderate or large values of $\aD$ imply tightly bound dark atoms which can be rather weakly interacting. Small $\aD$ suppresses all kinds of interactions (but is constrained by the requirement of efficient annihilation in the early universe).  Large $\MD$ suppresses the ion-ion scattering rate. Small $\MD$ ensures that dark atoms can form, which in turn neutralises DM and suppresses the DM self-interactions.

\eit

%%%%%%%%%%%%%%%%%%%%%%%%%%%%%%%%%%%%%%%
%%%%%%%%%%%%%%%%%%%%%%%%%%%%%%%%%%%%%%%
\section{Conclusion}	
\label{sec:conc}

Dark matter self-interacting via a light mediator is motivated by the observed galactic structure. It can be well accommodated within the asymmetric DM scenario, which allows for arbitrarily large DM annihilation cross-sections and thus for sizeable direct couplings of DM to light species. Due to the DM long-range self-interactions and the particle-antiparticle-asymmetric relic abundance, the cosmology of DM in such scenarios can be quite complex, with important implications for the phenomenology of DM in today's universe. 

In this work, we explored the scenario of asymmetric DM coupled to a light but not necessarily massless gauge vector boson. This is one of the most minimal scenarios in which DM self-interactions may manifest as long-range in haloes today, as well as one of the most minimal asymmetric DM scenarios. Yet, its cosmology is rather involved. We showed that in much of the parameter space where the DM self-interactions can have an important effect on the gravitational clustering, DM is necessarily multi-component and can combine into bound states in the early universe. The multi-component and atomic character of DM are features which appear not only in the limit of a massless mediator, but also in the case of a light but massive dark photon and a mildly broken gauge symmetry.

The formation of bound states in the early universe changes dramatically the DM self-interactions in haloes today, which can therefore be correctly estimated only by consistently taking into account the preceding cosmology.  We did so for the scenario under consideration; we placed constraints based on the observed ellipticity of large haloes, and we identified parameter regions where the DM self-scattering can affect the DM clustering patterns in smaller haloes, bringing theory in better agreement with observations. We showed that viable and interesting scenarios exist for a continuum of dark-photon masses, from zero to sufficiently large such that the DM self-interaction is effectively short-range.

Of course, the precise determination of the effect of DM self-interactions in haloes requires high-resolution simulations. The added complexity of this endeavour in the context of the model under consideration, and of models with similar features, is two-fold: the multi-component nature of DM, and the variety of intra- and inter-species interactions. Here we devised and applied conditions on the DM self-scattering rate, based on reasonable considerations which, among else, ensured that these conditions reduce to established constraints and estimates in the limit of single-component DM. Such methods do not certainly circumvent the need for detailed simulations; rather, studies such as the present showcase the features and the parameter space that should be investigated in future numerical works.

%%%%%%%%%%%%%%%%%%%%%%%%%%%%%%%%%%%%%%%
%%%%%%%%%%%%%%%%%%%%%%%%%%%%%%%%%%%%%%%
\section*{Acknowledgements}

A.K. and L.P. were supported by DOE Grant DE-SC0009937. A.K. was supported by the World Premier International Research Center Initiative (WPI Initiative), MEXT, Japan.  K.P. was supported by the Netherlands Foundation for Fundamental Research of Matter (FOM) and the Netherlands Organisation for Scientific Research (NWO). We thank Dan Butter, Ian Shoemaker, Ray Volkas and Hai-Bo Yu for useful discussions and comments.

\clearpage

%%%%%%%%%%%%%%%%%%%%%%%
\begin{table}[h!]
\centering

\newcolumntype{D}{>\raggedright p{4.7cm}}
\newcolumntype{L}{>{\varwidth[c]{\linewidth}}l<{\endvarwidth} }
\newcolumntype{F}{X}
\def\arraystretch{1.7}

%\fbox{
\begin{tabularx}{0.52\textwidth}{|D| L |F|}
\hline
Particle						& Symbol				&   Mass \\  \hline 
\hline
Dark proton				& $ \pD $		&  $\mpD $ \\  \hline 
Dark electron 			& $ \eD$				&  $ \meD $  \\ \hline
Dark hydrogen atom 	& $ \HD $ 			&  $ \mH $   \\ \hline
Dark photon				& $ \gammaD $ 	&  $ \MD $  \\ \hline
Dark Higgs (physical d.o.f.) 	& $ \fvarD $ 			&  $ m_\fvar $ \\ 
\hline
\end{tabularx}

\bigskip\medskip

\begin{tabularx}{0.83\textwidth}{|D | F | }
\hline
Quantity &  Symbol      \\ 
\hline\hline
Dark fine structure constant & $ \aD $     \\ \hline
$\pD - \eD$ reduced mass   & $ \muD = \mpD \meD /(\mpD + \meD) $     \\ \hline
Binding energy & $ \D \simeq (1/2) \a^2 \muD [1-\MD/(\aD\muD)]^2 $     \\ \hline
Residual ionisation fraction & $ \xD  $     \\ \hline
Dark-to-ordinary sector temperature ratio & $ \ks = \TD / \TV $    (subscripts on $\ks$ denote the epoch,   ``DR": dark recombination) \\ 
\hline
\end{tabularx}
%}
\caption{Summary of particles and symbols.  }
\label{tab:symbols}
\end{table}
%%%%%%%%%%%%%%%%%%%%%%

\begin{table}[h!]
\centering

\newcolumntype{D}{>\raggedright p{3.6cm}}
\newcolumntype{G}{>\raggedright p{5.2cm}}
\newcolumntype{L}{>{\varwidth[c]{\linewidth}}l<{\endvarwidth} }
\newcolumntype{F}{X}
\def\arraystretch{1.7}

\begin{tabularx}{\textwidth}{|D |G| L| F | }
\hline
Regions & Meaning  & Condition &  Relevant equation(s)     \\ 
\hline\hline
Red/pink-shaded  	  & Disfavoured by ellipticity of large haloes. &  $\G_{\rm eff}^{\rm MW} > \G_{\rm crit}^{\rm MW}$ & \eqref{eq:MW cond}, \eqref{eq:MW bound} \\ \hline
Enclosed by blue line  & Favoured by galactic substructure. & $\G_{\rm eff}^{\rm DW} > \G_{\rm crit}^{\rm DW}$ & \eqref{eq:DW cond}, \eqref{eq:DW bound} \\ \hline
Hashed & Insufficient annihilation in the early universe. & $\aD < \a_{_{D, \rm min}} $  & 
\eqref{eq:alpha ann}   \\ \hline
Grey-shaded & Unphysical parameter space & $4 \muD > \mH + \D$   &  \eqref{eq:muD upper} \\ \hline
Dashed grey lines &\multicolumn{2}{l|}{ Contours of constant ionisation fraction,  $\xD$.} & \eqref{eq:xD resid} \\  \hline
Dotted green lines & Maximum $\MD$ allowing radiative formation of dark atoms. & $\MD = (1/2) \aD^2 \muD$ & \eqref{eq:MD<Delta} \\  \hline
Dot-dashed yellow lines & Maximum $\MD$  allowing existence of bound states (dark atoms). & $\MD = \muD \aD$ & \eqref{eq:atom cond} \\ 
\hline
\end{tabularx}
%}
\caption{Conditions sketched in Figs.~\ref{fig:CL_alphaVmH} -- \ref{fig:CL_fixed xD}.  }
\label{tab:conditions}
\end{table}
%%%%%%%%%%%%%%%%%%%%%%

%%%%%%%%%%%%%%%%%%%%%%%%%%%%%%%%%%%%%%%
%  FIGURES
%%%%%%%%%%%%%%%%%%%%%%%%%%%%%%%%%%%%%%%

%%%%%%%%%%%%%%%%%%%%%%%%%%%%%%%%%%%%%%%%%%%%%%%%%%%%%%%%%%%%%%%%%%%%%%%%%%%%%%%%%%%%%%%%%
% 1st set
%%%%%%%%%%%%%%%%%%%%%%%
\begin{figure}[h!]
\centering
\includegraphics[width=0.45\textwidth]{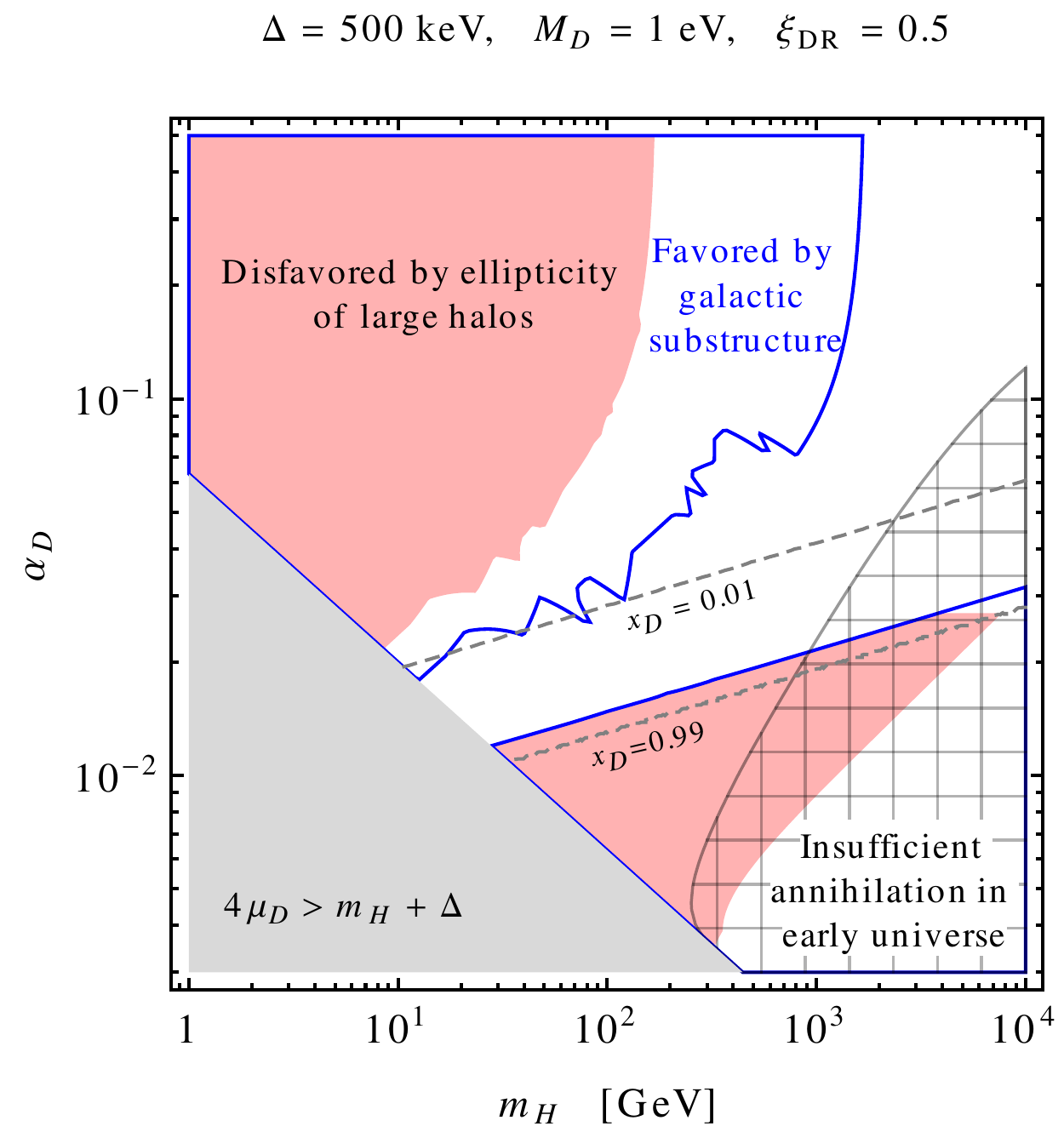}
\includegraphics[width=0.45\textwidth]{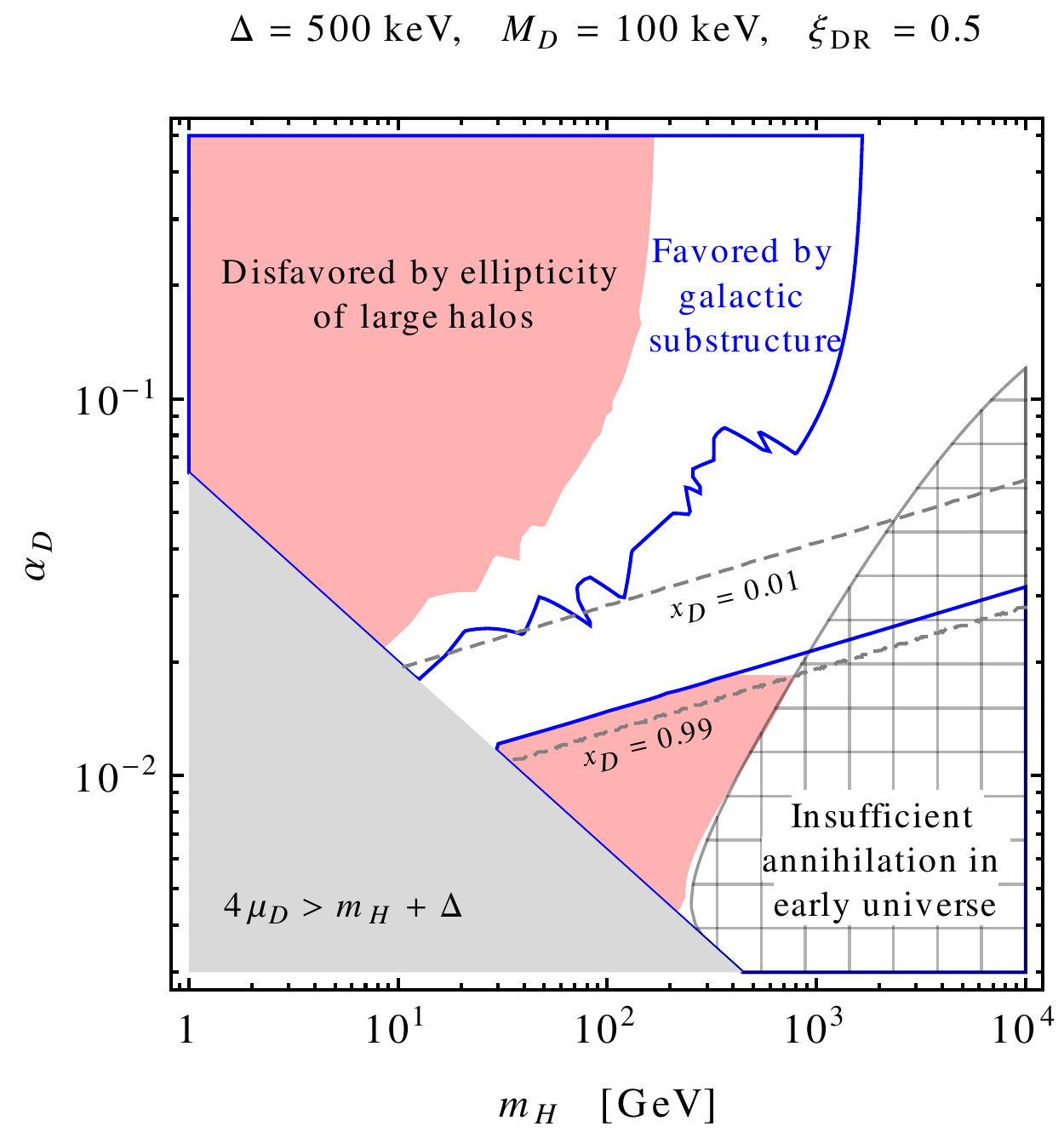}

\bigskip\bigskip
\includegraphics[width=0.45\textwidth]{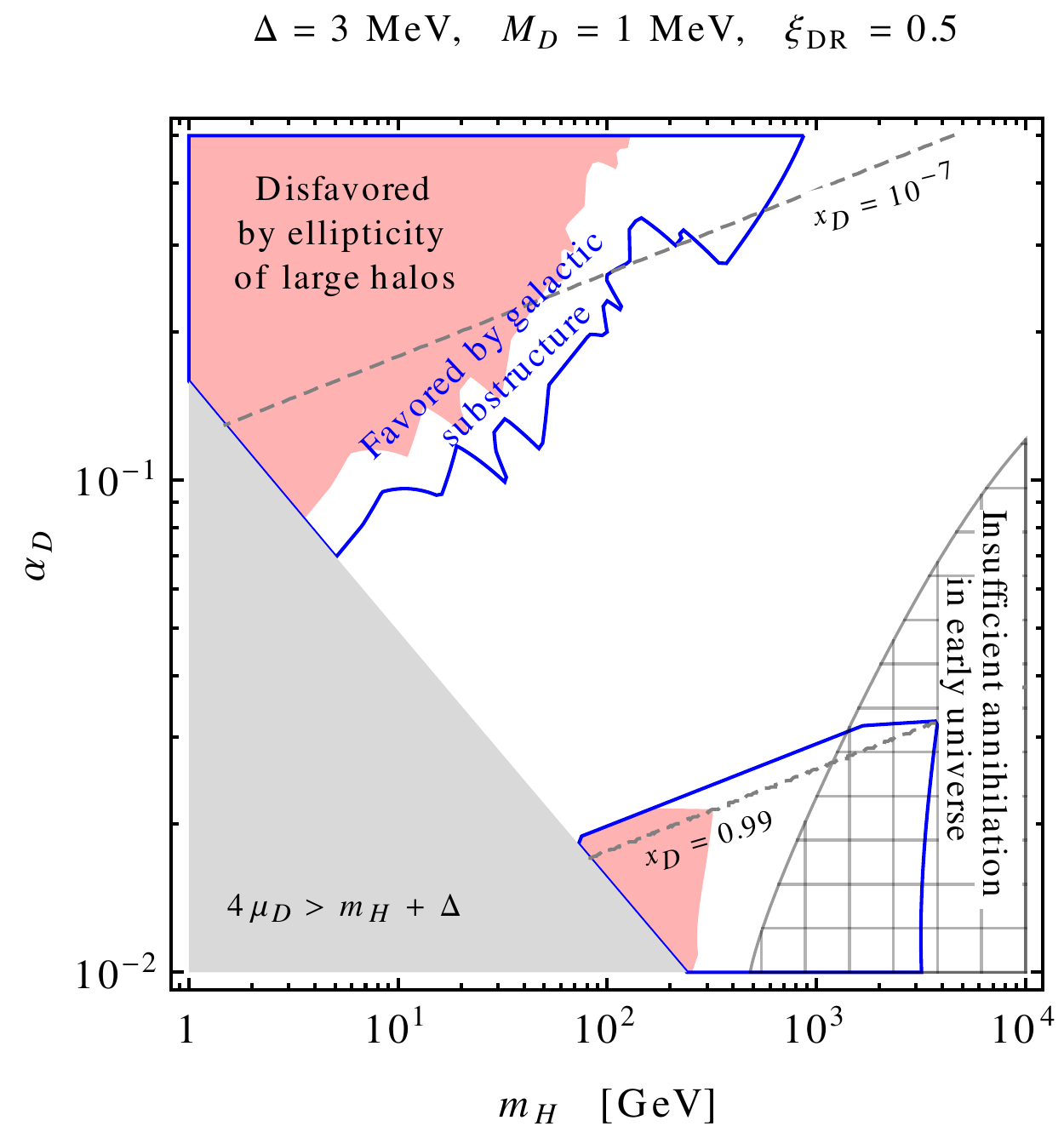}
\includegraphics[width=0.45\textwidth]{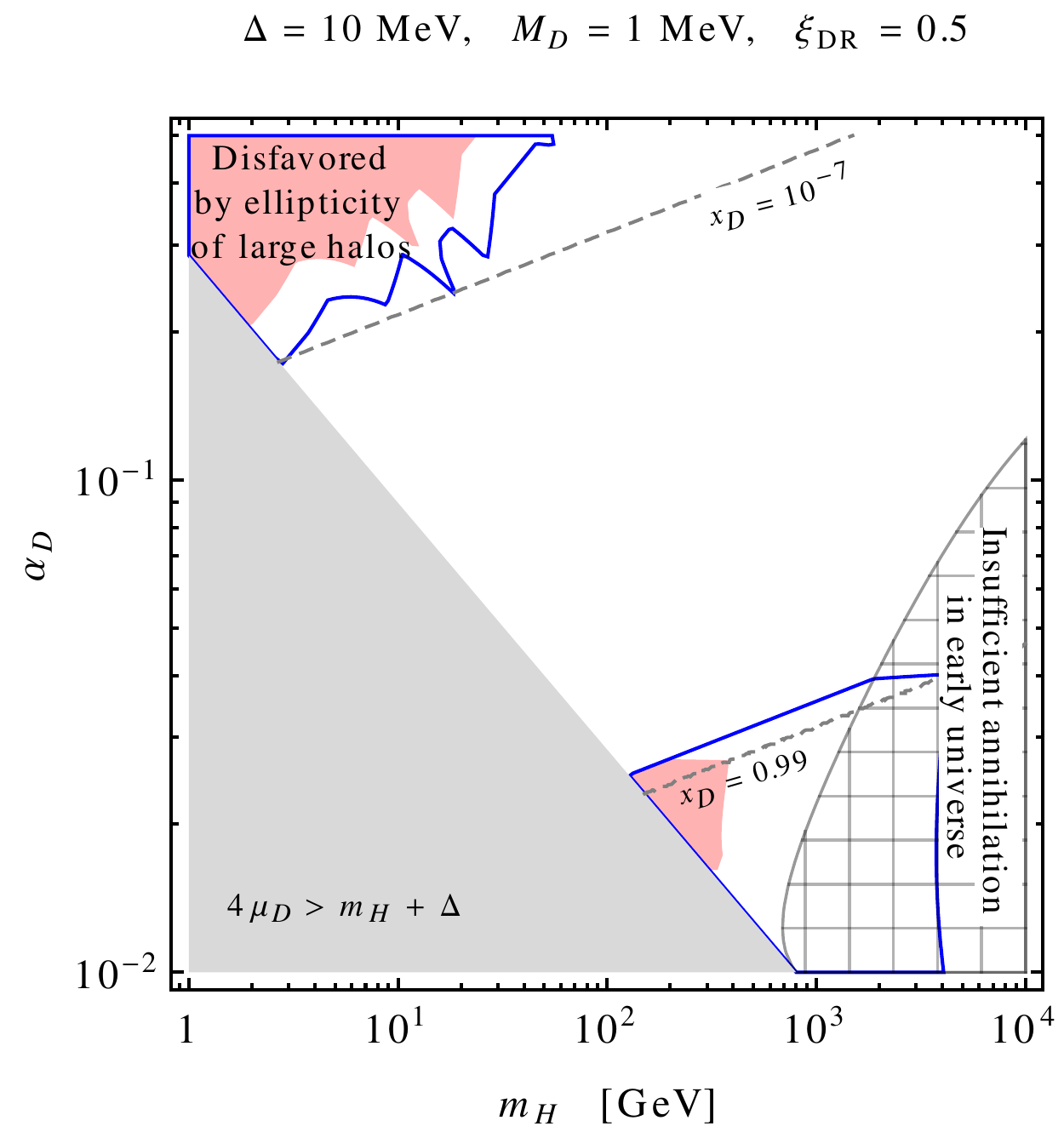}
\caption{\footnotesize
In the red-shaded region, DM self-interaction rate violates the condition \eqref{eq:MW cond}; this region is disfavoured by ellipticity of Milky-Way-size and larger haloes. In the region enclosed by the blue solid line, the DM self-scattering satisfies the condition \eqref{eq:DW cond} and can affect the dynamics of dwarf-galaxy-size haloes. In the cross-hatched region, the DM annihilation in the early universe is insufficient, under minimal assumptions; this bound can be relaxed if more annihilation channels exist. In the grey-shaded region, the consistency condition of Eq.~\eqref{eq:muD upper} is not satisfied; this region does not correspond to any meaningful parameter space. The dashed grey lines denote fixed values of the residual ionisation fraction $x_{_D}$. For each of the plots in this set, the binding energy $\Delta$, the dark photon mass $M_{_{D}}$ and the dark-to-ordinary temperature ratio at the time of dark recombination $\xi_{_{\rm DR}}$ are fixed to the values mentioned in the plot labels. For the annihilation bound, we take $\xi_{\rm ann} = \xi_{_{\rm DR}}$. In this set of plots, the atom-atom scattering was estimated according to Ref.~\cite{Cline:2013pca} (c.f. \eq{eq:CL sigma HH}) and atom-ion collisions were ignored.}
\label{fig:CL_alphaVmH}
\end{figure}
%%%%%%%%%%%%%%%%%%%%%%
%%%%%%%%%%%%%%%%%%%%%%
\begin{figure}[t]
\centering
\includegraphics[width=0.45\textwidth]{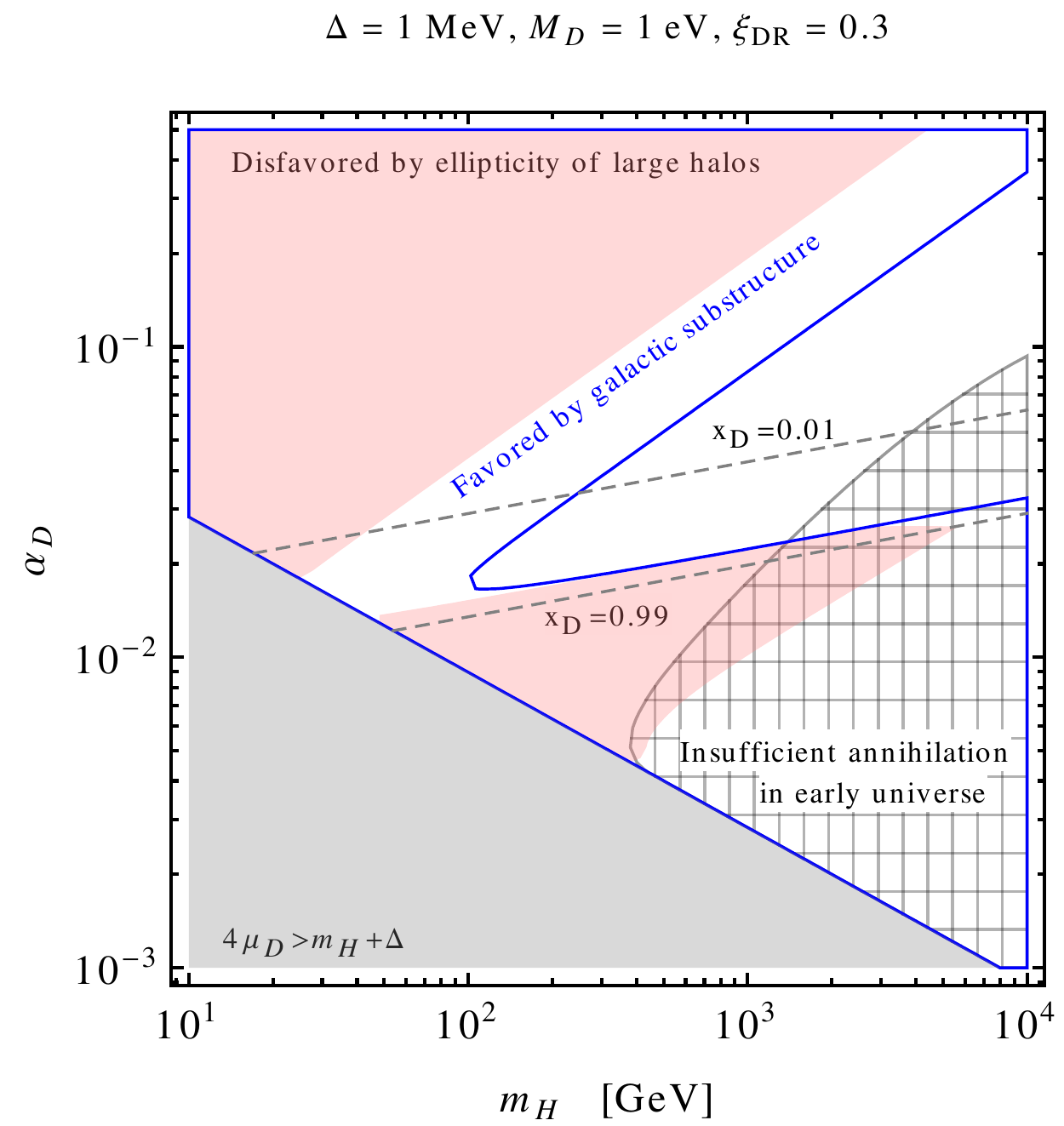}
\includegraphics[width=0.45\textwidth]{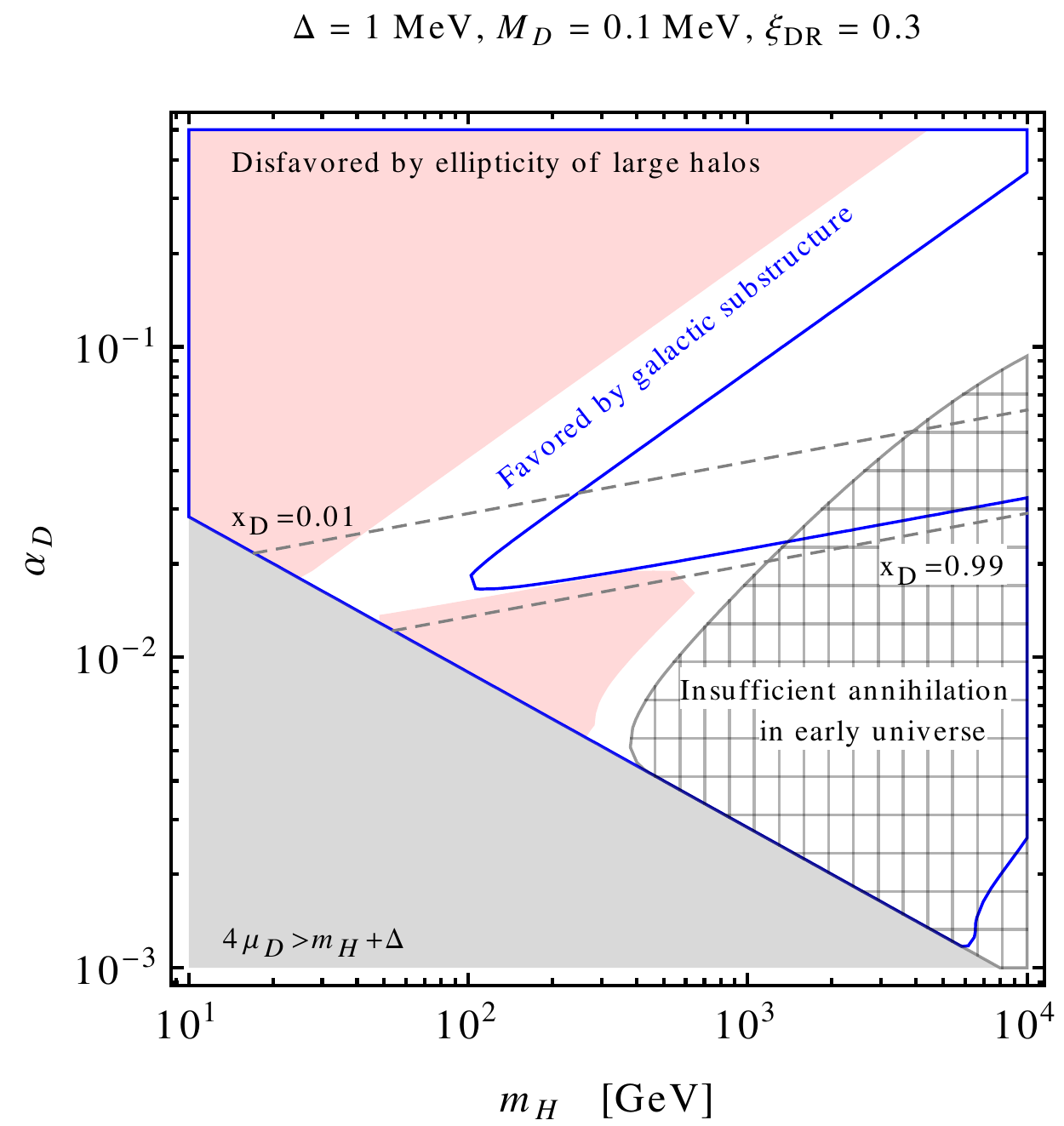}

\bigskip\bigskip
\includegraphics[width=0.45\textwidth]{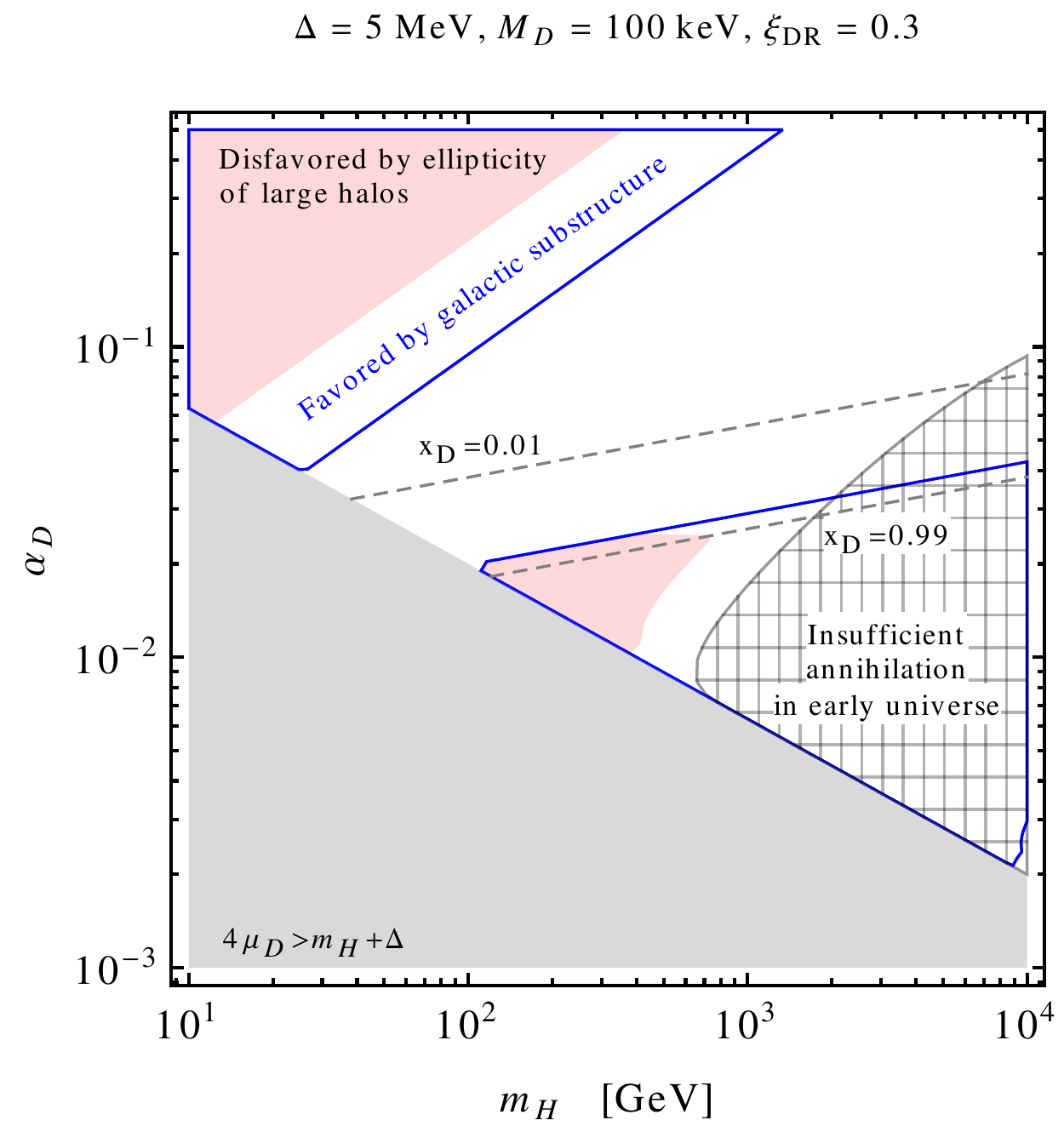}
\includegraphics[width=0.45\textwidth]{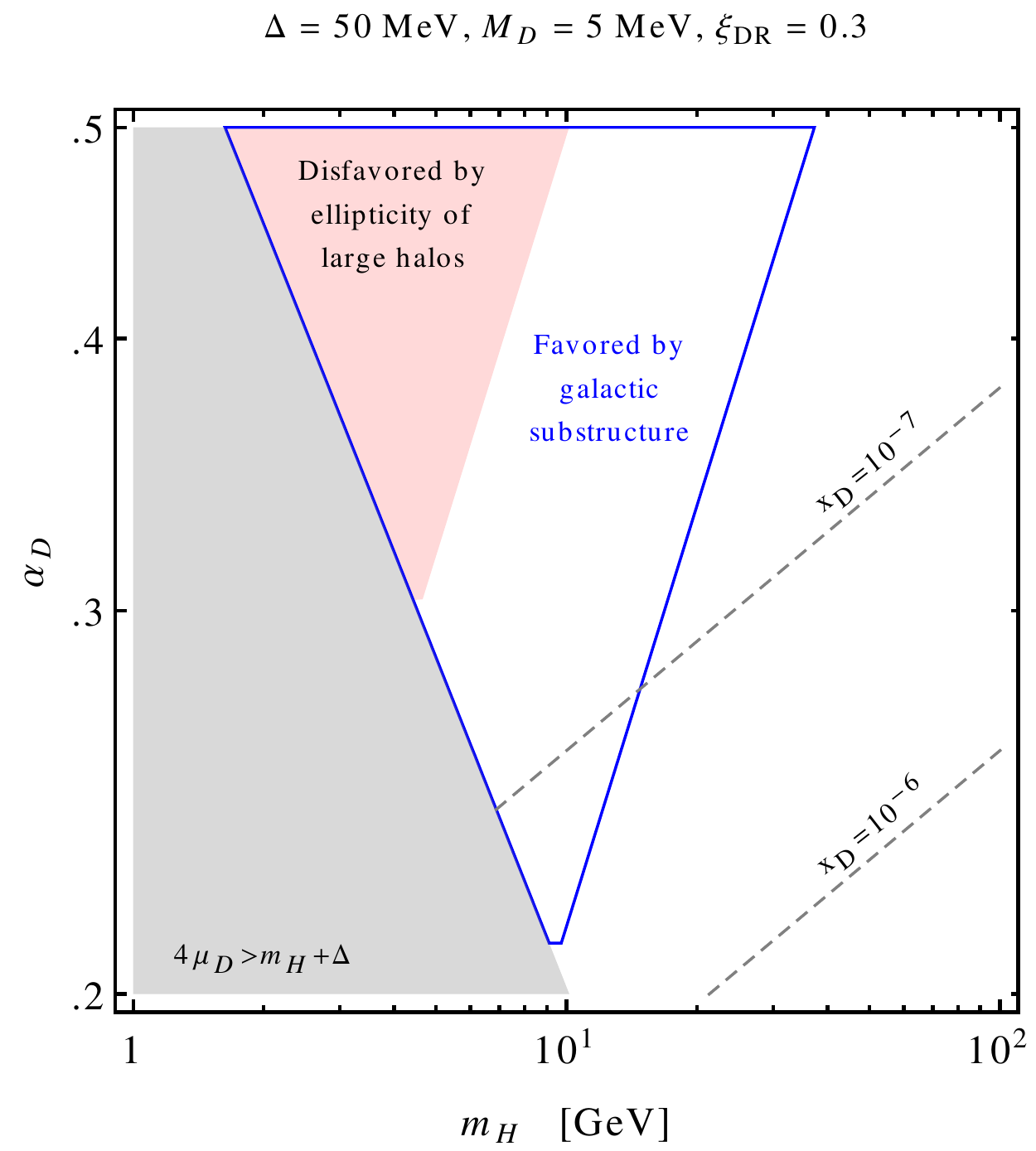}
\caption{\footnotesize 
Same as Fig.~\ref{fig:CL_alphaVmH}, but with the atom-atom and atom-ion scattering estimated according to Ref.~\cite{CyrRacine:2012fz} (c.f. Eqs.~\eqref{eq:Gamma HH} - \eqref{eq:Gamma He}). To facilitate the visual identification of the method used, in this and subsequent sets of plots using the approach of Ref.~\cite{CyrRacine:2012fz}, red shading has been switched to pink.}
\label{fig:SI_alphaVmH}
\end{figure}
%%%%%%%%%%%%%%%%%%%%%%

%%%%%%%%%%%%%%%%%%%%%%%%%%%%%%%%%%%%%%%%%%%%%%%%%%%%%%%%%%%%%%%%%%%%%%%%%%%%%%%%%%%%%%%%%
% 2nd set
%%%%%%%%%%%%%%%%%%%%%%%
\begin{figure}[h!]
\centering
\includegraphics[width=0.45\textwidth]{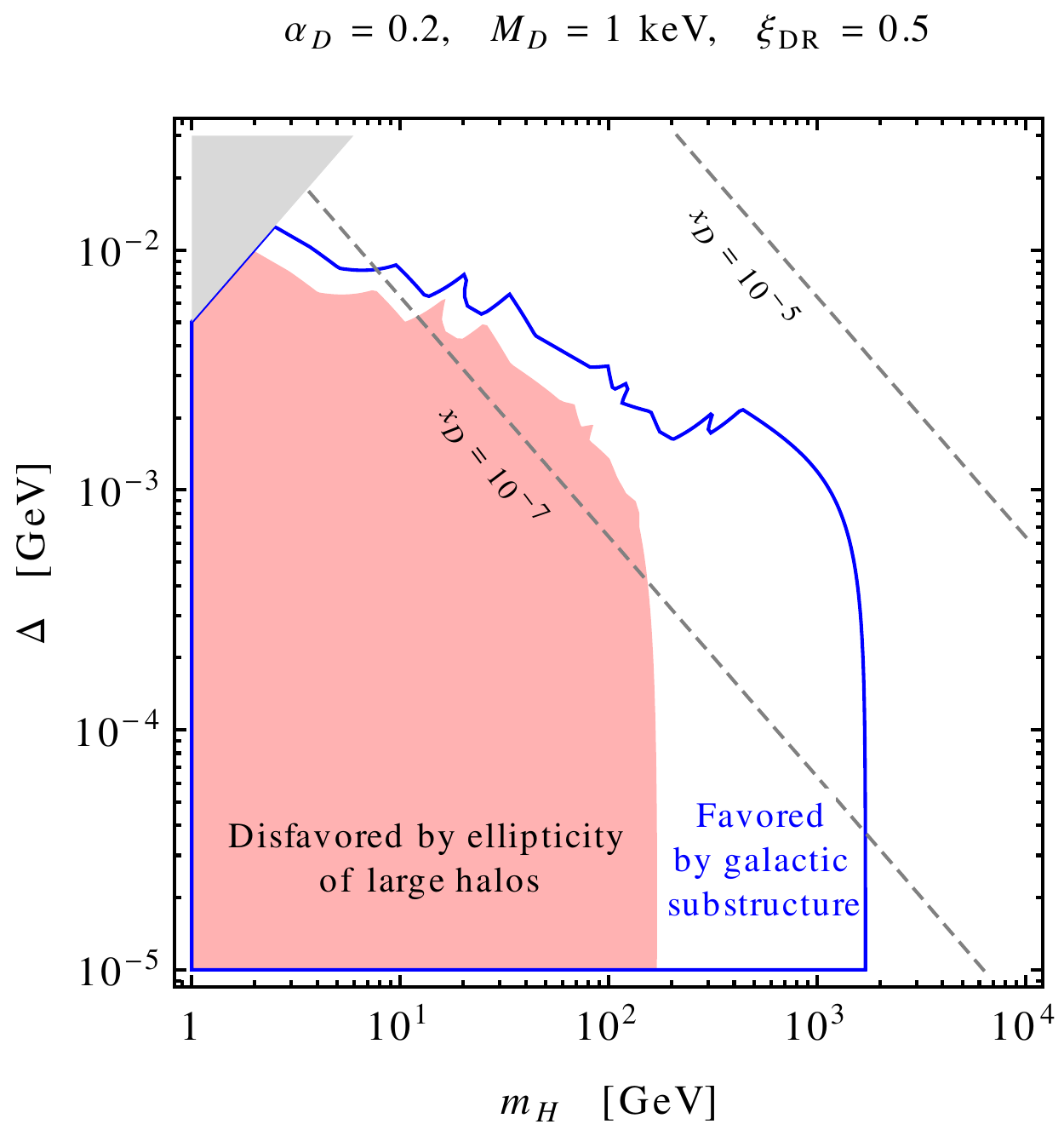}
\includegraphics[width=0.45\textwidth]{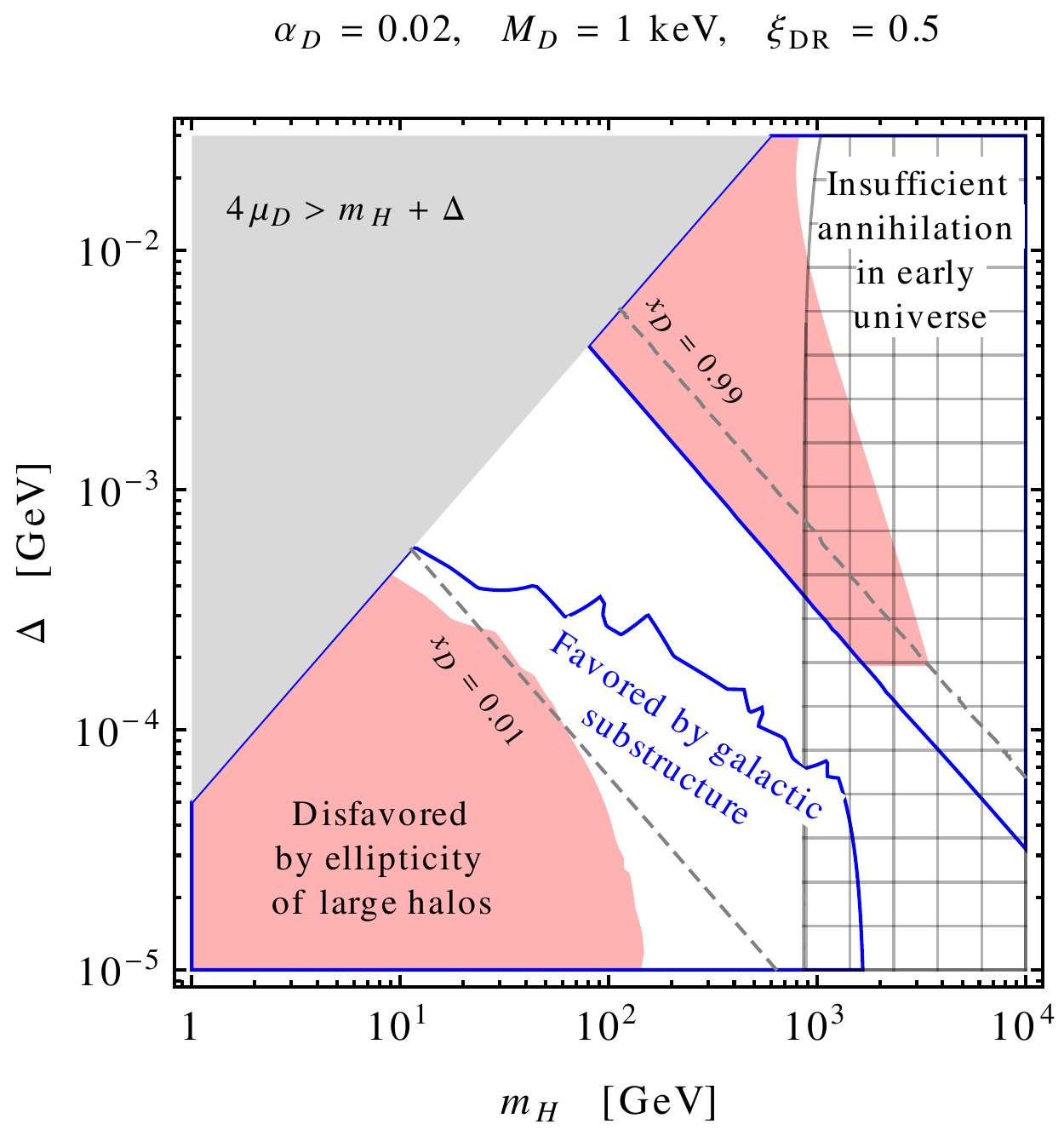}
\caption{\footnotesize 
Same as in Fig.~\ref{fig:CL_alphaVmH}, for fixed values of $\alpha_{_D}$, $M_{_D}$ 
and $\xi_{_{\rm DR}}$. We have used the approach of Ref.~\cite{Cline:2013pca} for atom-atom scattering.}
\label{fig:CL_DeltaVmH}
\end{figure}
%%%%%%%%%%%%%%%%%%%%%%%
%%%%%%%%%%%%%%%%%%%%%%%
\begin{figure}[ht]
\centering
\includegraphics[width=0.45\textwidth]{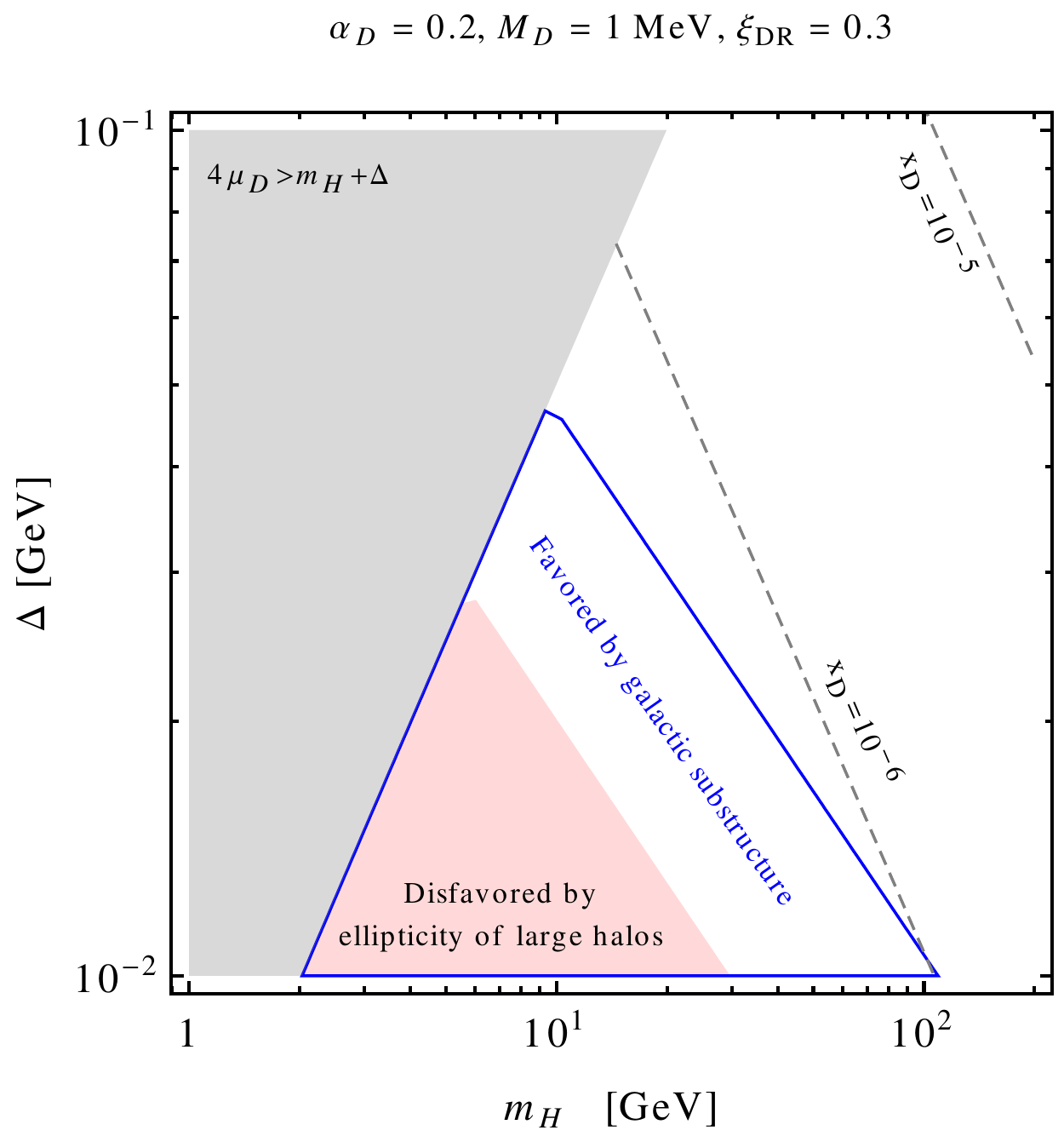}
\includegraphics[width=0.45\textwidth]{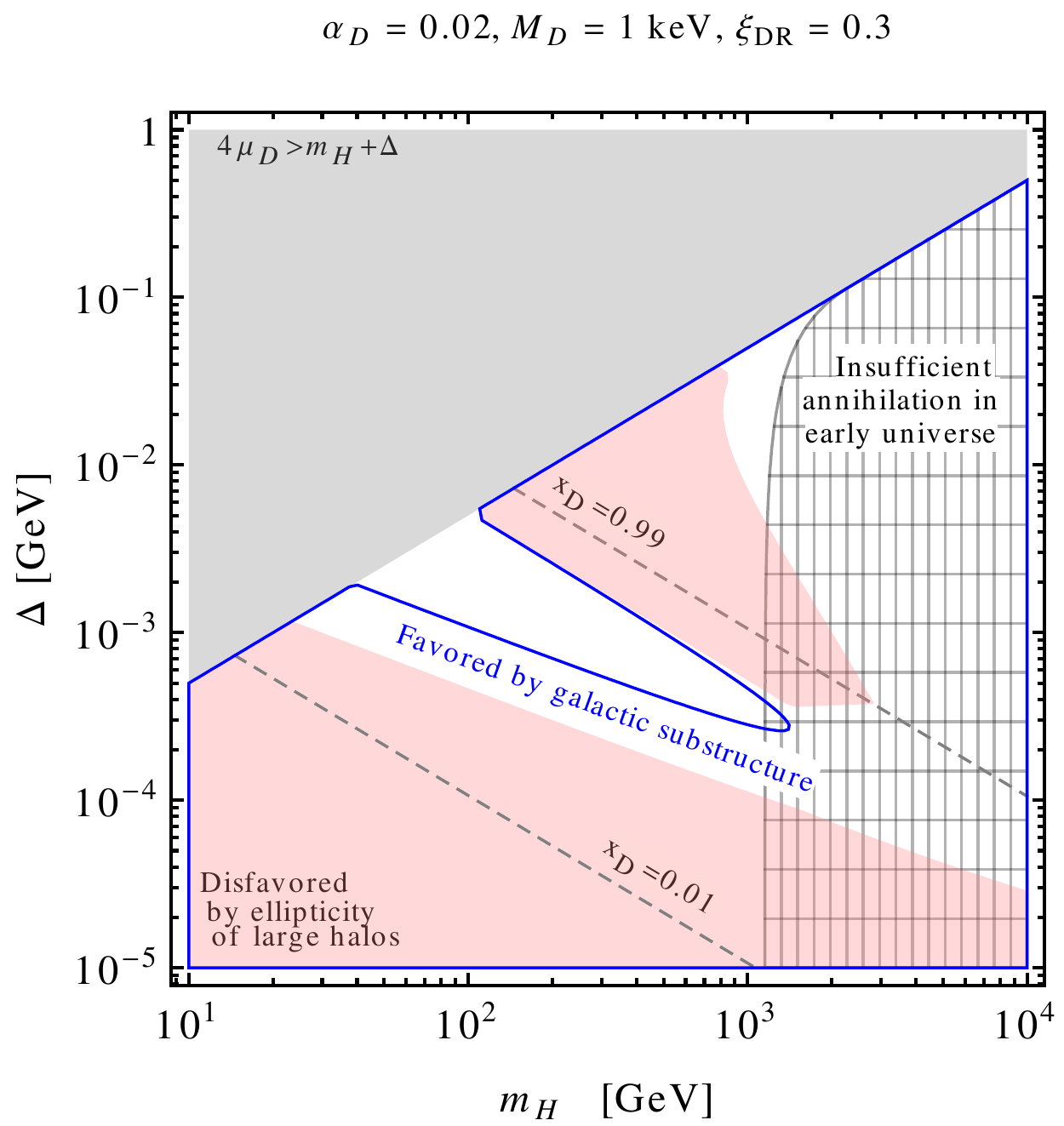}
\caption{\footnotesize  
Same as in Fig.~\ref{fig:CL_DeltaVmH}, using the approach of 
Ref.~\cite{CyrRacine:2012fz} for collisions involving atoms.}
\label{fig:SI_DeltaVmH}
\end{figure}
%%%%%%%%%%%%%%%%%%%%%%%
%%%%%%%%%%%%%%%%%%%%%%%%%%%%%%%%%%%%%%%%%%%%%%%%%%%%%%%%%%%%%%%%%%%%%%%%%%%%%%%%%%%%%%%%%

%%%%%%%%%%%%%%%%%%%%%%%%%%%%%%%%%%%%%%%%%%%%%%%%%%%%%%%%%%%%%%%%%%%%%%%%%%%%%%%%%%%%%%%%%
% 3rd set
%%%%%%%%%%%%%%%%%%%%%%
\begin{figure}[h!]
\centering
\includegraphics[width=0.45\textwidth]{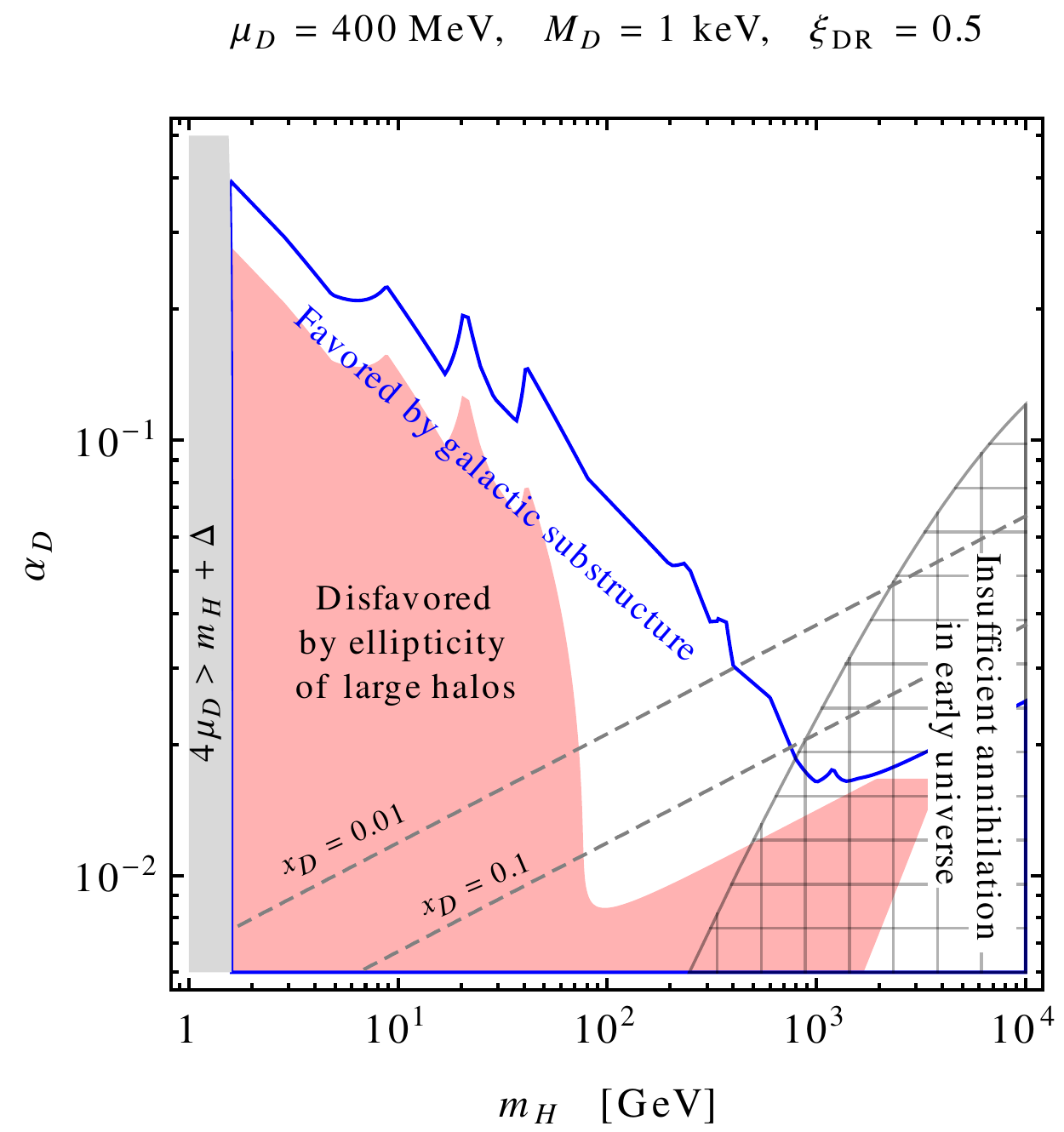}
\includegraphics[width=0.45\textwidth]{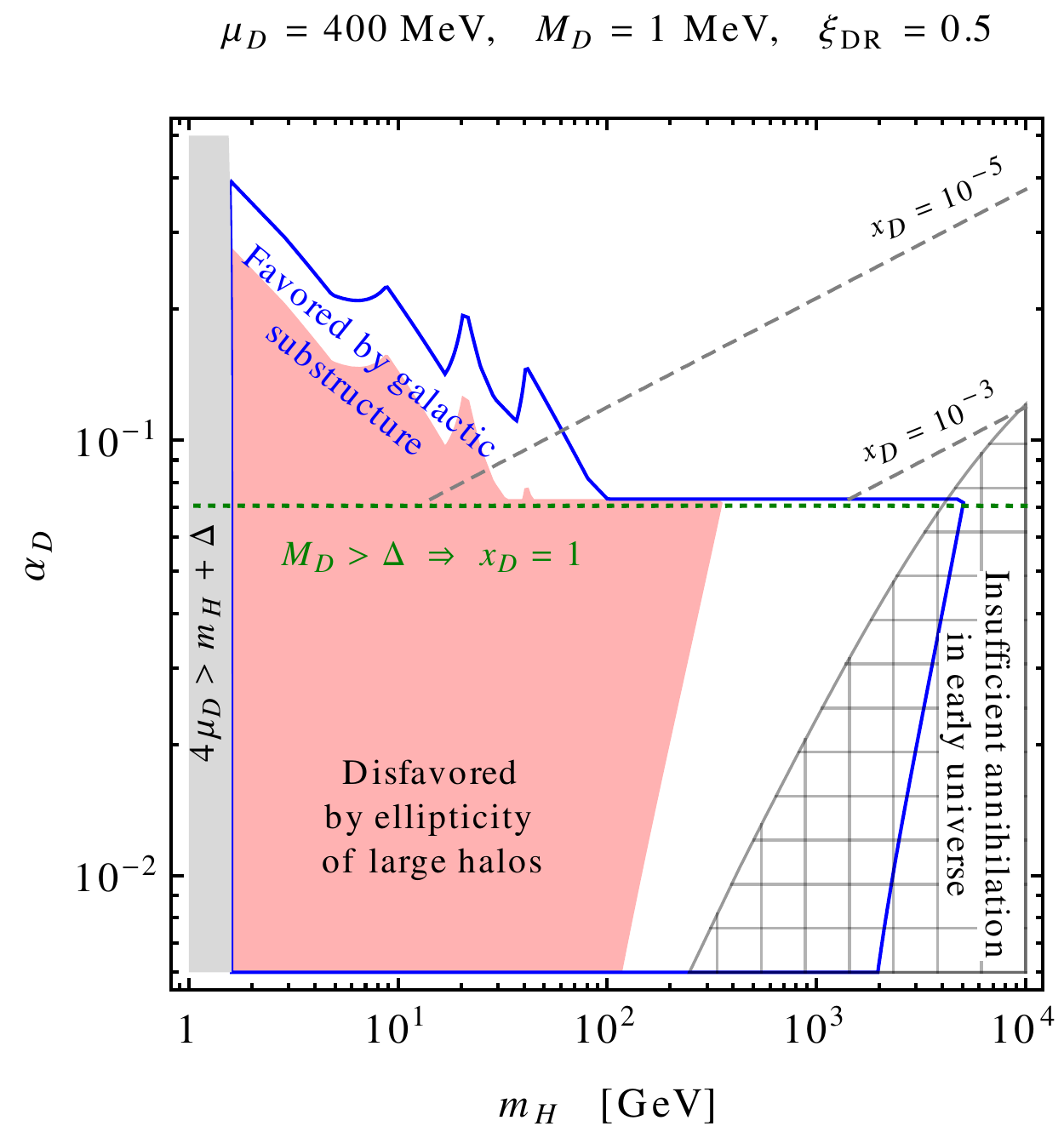}
\caption{\footnotesize   
Same as in Fig.~\ref{fig:CL_alphaVmH}, for fixed values of 
$\mu_{_D}$, $M_{_D}$ and $\xi_{_{\rm DR}}$. Below the dotted green line, the 
formation of dark atoms via emission of a dark photon is not kinematically 
possible, and DM remains fully ionised, $x_{_D} = 1$. 
We have used the approach of Ref.~\cite{Cline:2013pca} for atom-atom scattering.}
\label{fig:CL_alphaVmH 2} 
\end{figure}
%%%%%%%%%%%%%%%%%%%%%%
%%%%%%%%%%%%%%%%%%%%%%
\begin{figure}[ht]
\centering
\includegraphics[width=0.45\textwidth]{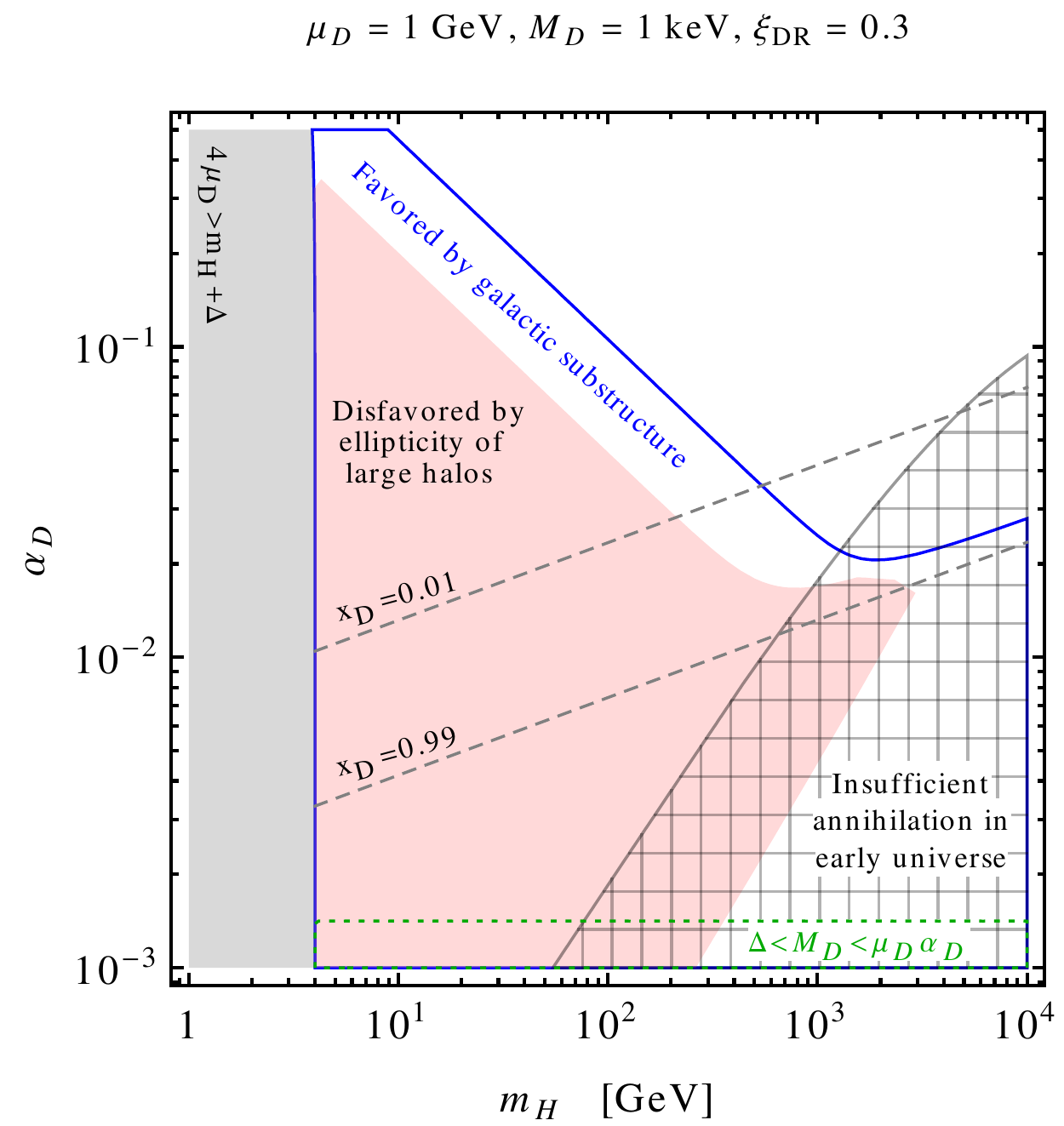}
\includegraphics[width=0.45\textwidth]{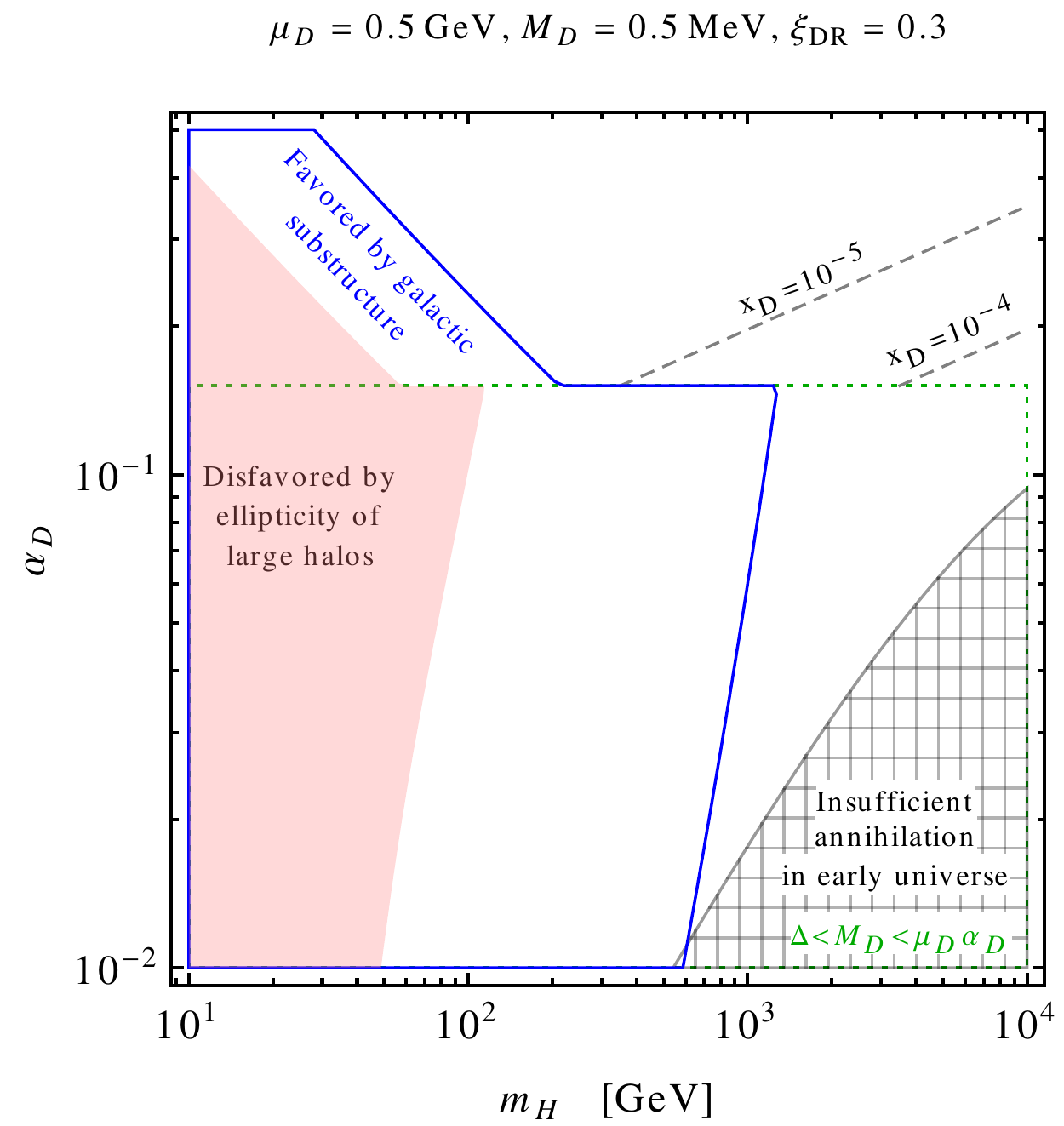}
\caption{\footnotesize   
Same as in Fig.~\ref{fig:CL_alphaVmH 2}, using the approach of 
Ref.~\cite{CyrRacine:2012fz} for collisions involving atoms. }
\label{fig:SI_alphaVmH 2} 
\end{figure}
%%%%%%%%%%%%%%%%%%%%%%
%%%%%%%%%%%%%%%%%%%%%%%%%%%%%%%%%%%%%%%%%%%%%%%%%%%%%%%%%%%%%%%%%%%%%%%%%%%%%%%%%%%%%%%%%

%%%%%%%%%%%%%%%%%%%%%%%%%%%%%%%%%%%%%%%%%%%%%%%%%%%%%%%%%%%%%%%%%%%%%%%%%%%%%%%%%%%%%%%%%
% 4th set
%%%%%%%%%%%%%%%%%%%%%%%
\begin{figure}[ht]
\centering
\includegraphics[width=0.4\textwidth]{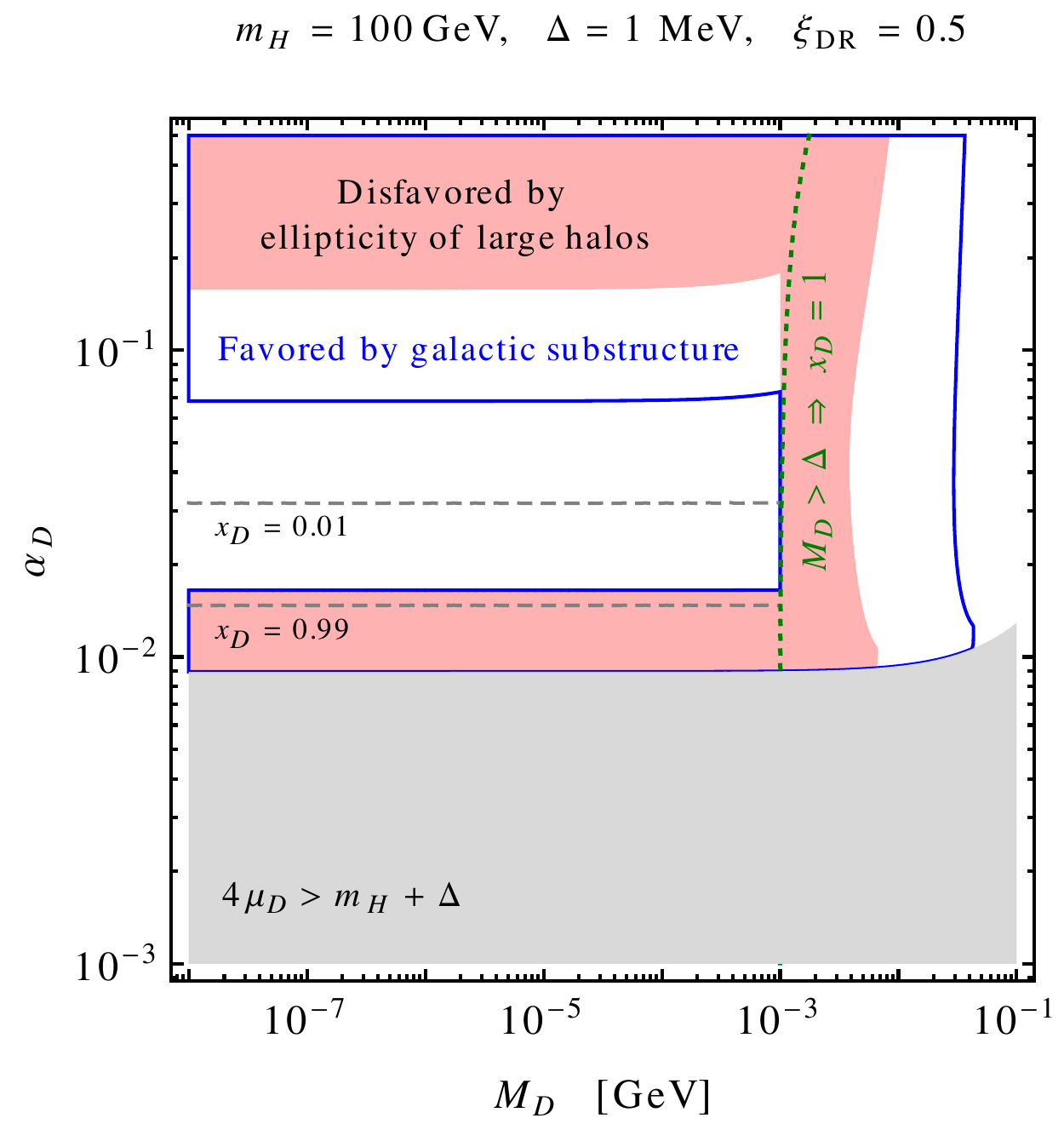}
\includegraphics[width=0.4\textwidth]{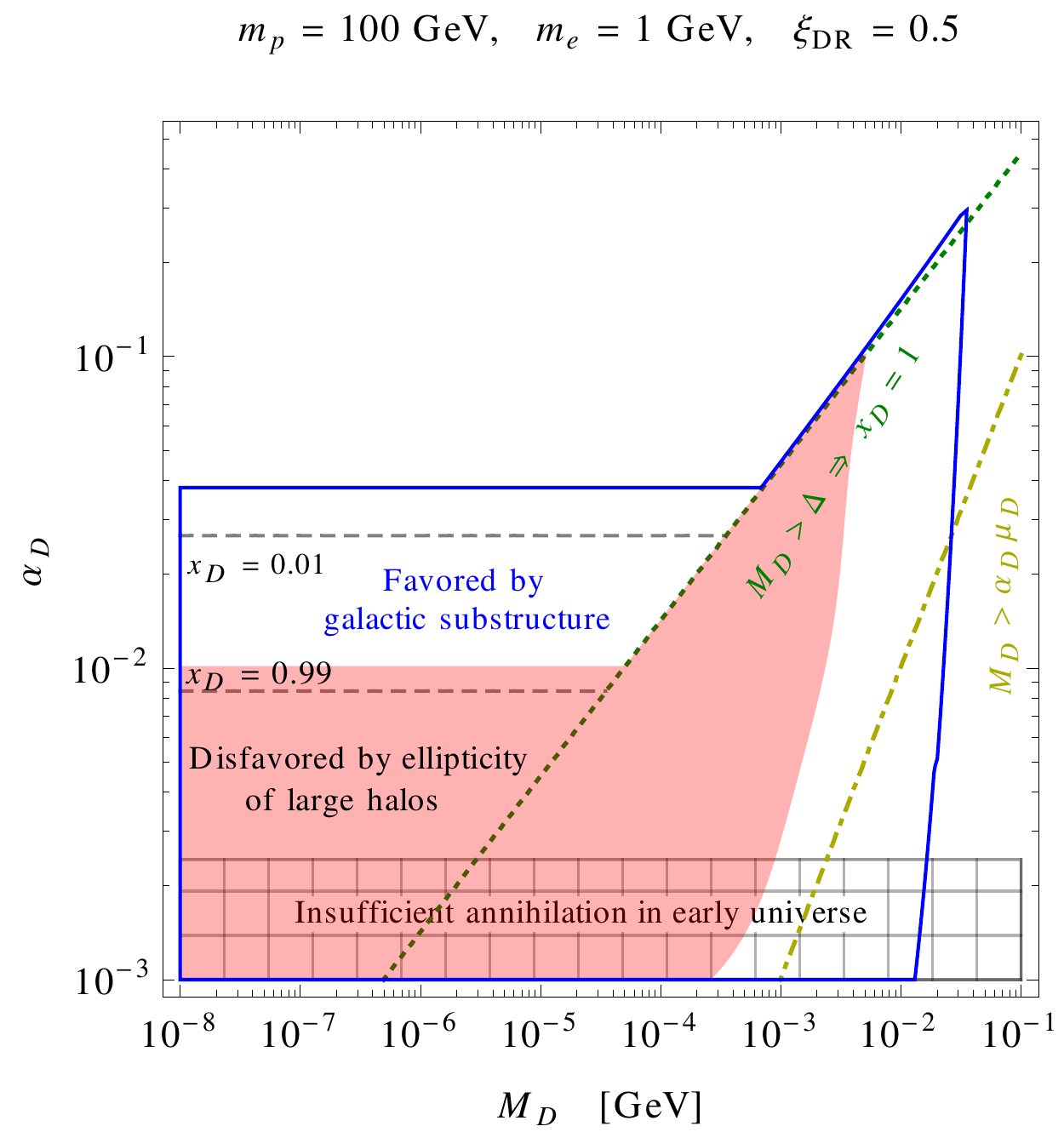}

\bigskip
\includegraphics[width=0.4\textwidth]{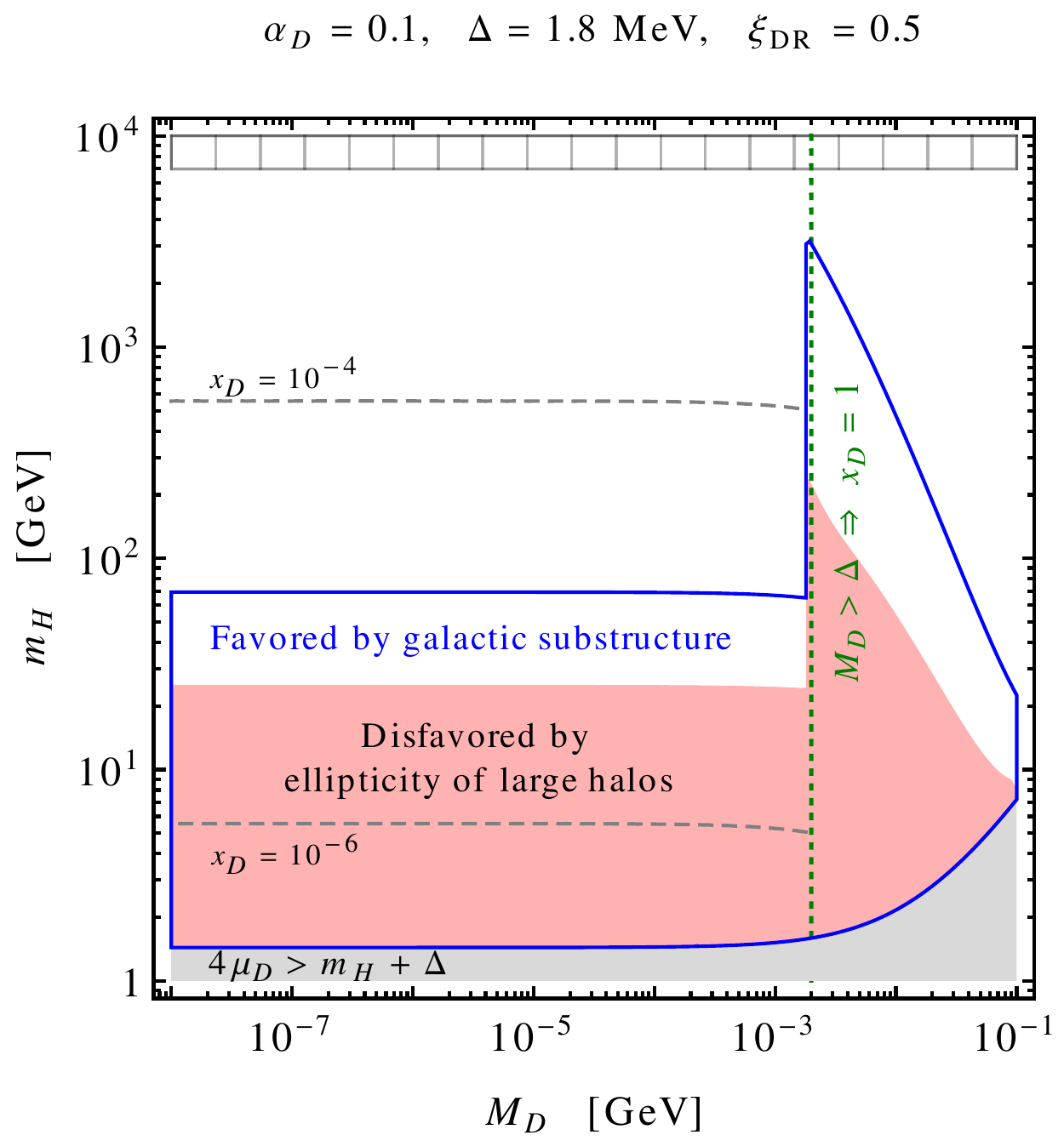}
\includegraphics[width=0.4\textwidth]{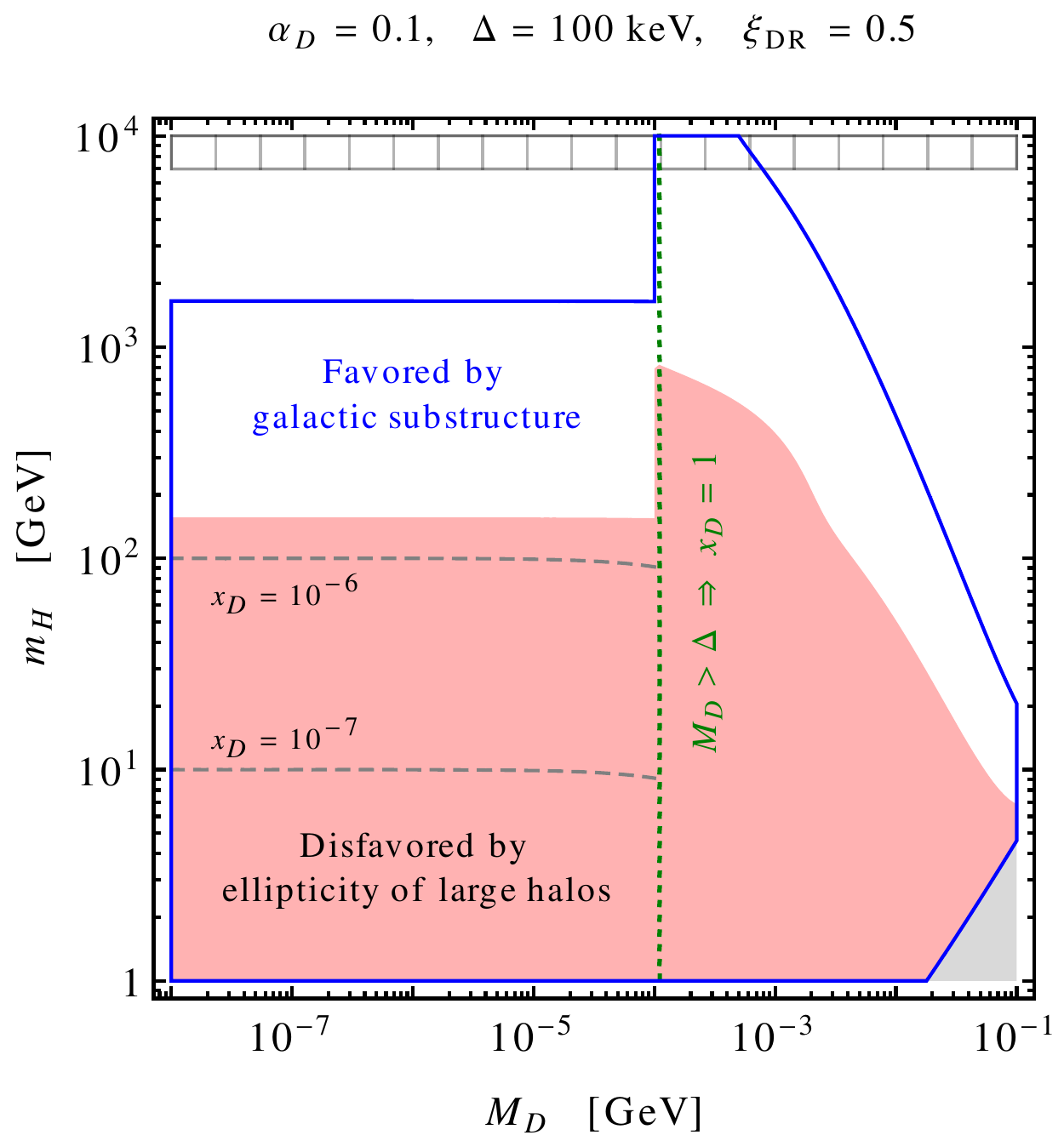}

\bigskip
\includegraphics[width=0.4\textwidth]{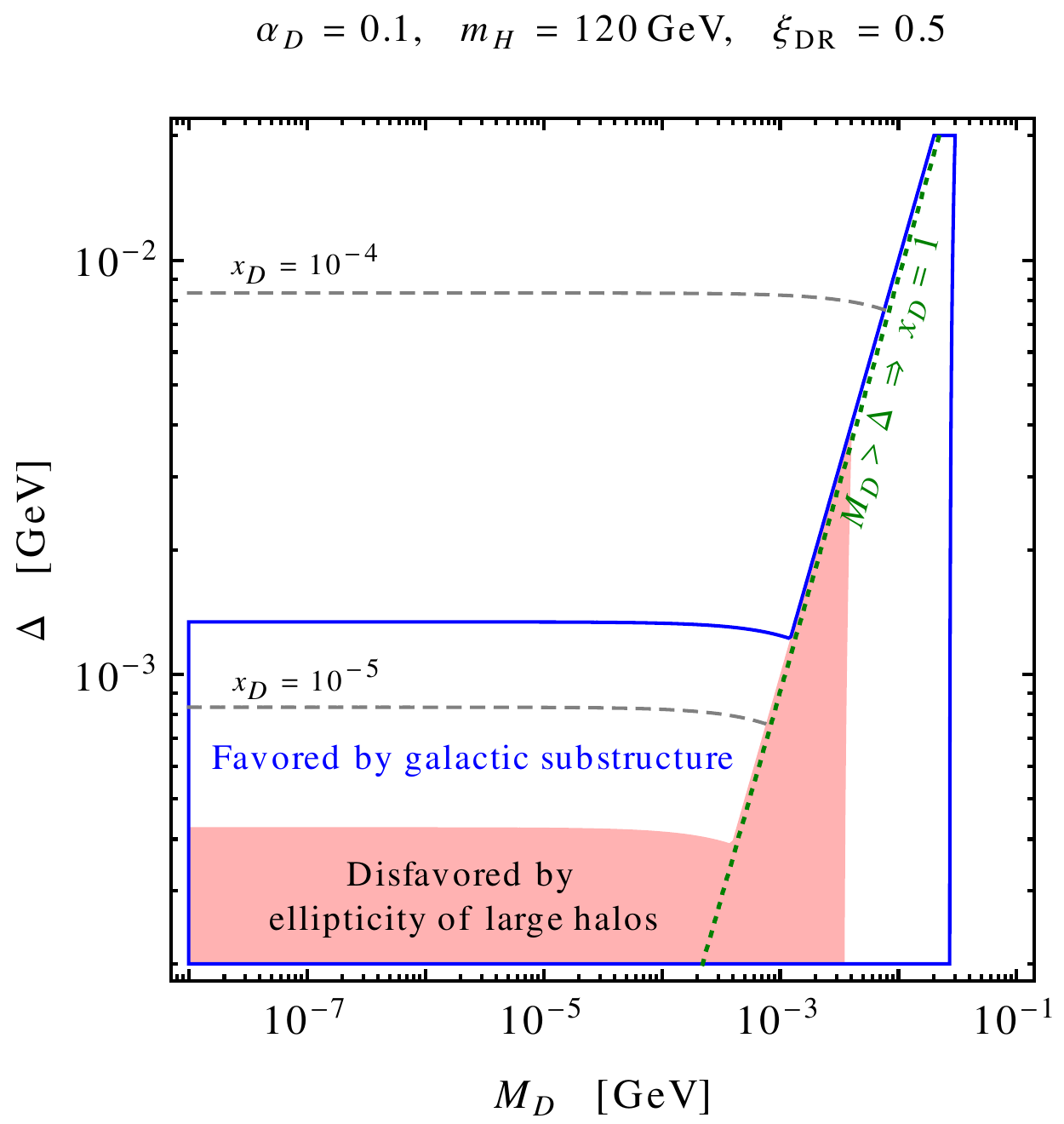}
\includegraphics[width=0.4\textwidth]{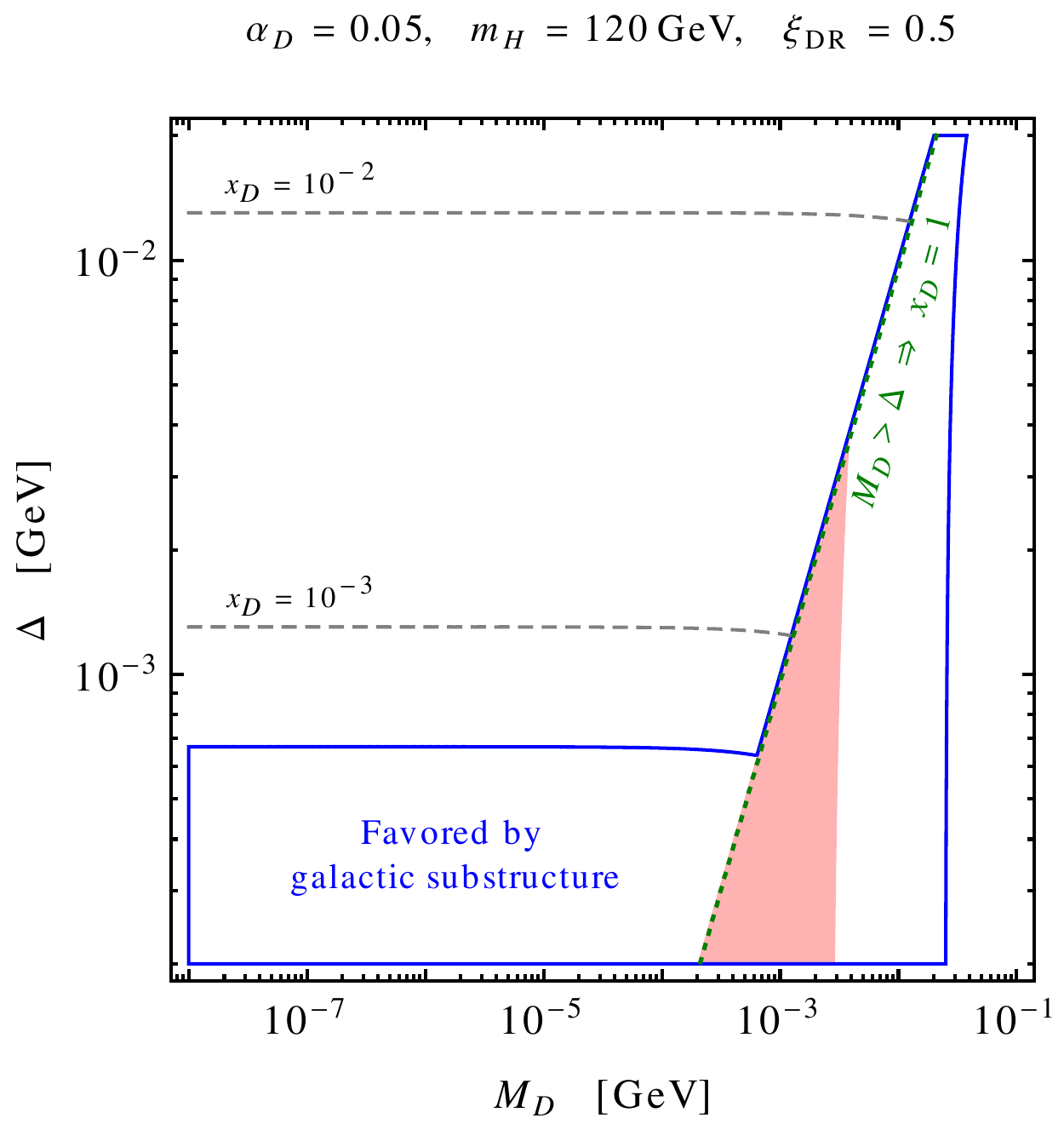}
\caption{\footnotesize  
Same as in Fig.~\ref{fig:CL_alphaVmH}, for fixed values of the 
parameters mentioned on the plot labels. In the plots of the right column, and to the right 
of the dotted green line, $M_{_D} > \alpha_{_D}^2 \mu_{_D}/2$ and dark atoms cannot form. 
In these regions, $x_{_{D}} = 1$. Further to the right, in the grey-shaded regions, 
$M_{_D} > \alpha_{_D} \mu_{_D}$ and bound states do not exist.}
\label{fig:CL_continuous MD}
\end{figure}
%%%%%%%%%%%%%%%%%%%%%%
%%%%%%%%%%%%%%%%%%%%%%%
\begin{figure}[ht]
\centering
\includegraphics[width=0.4\textwidth]{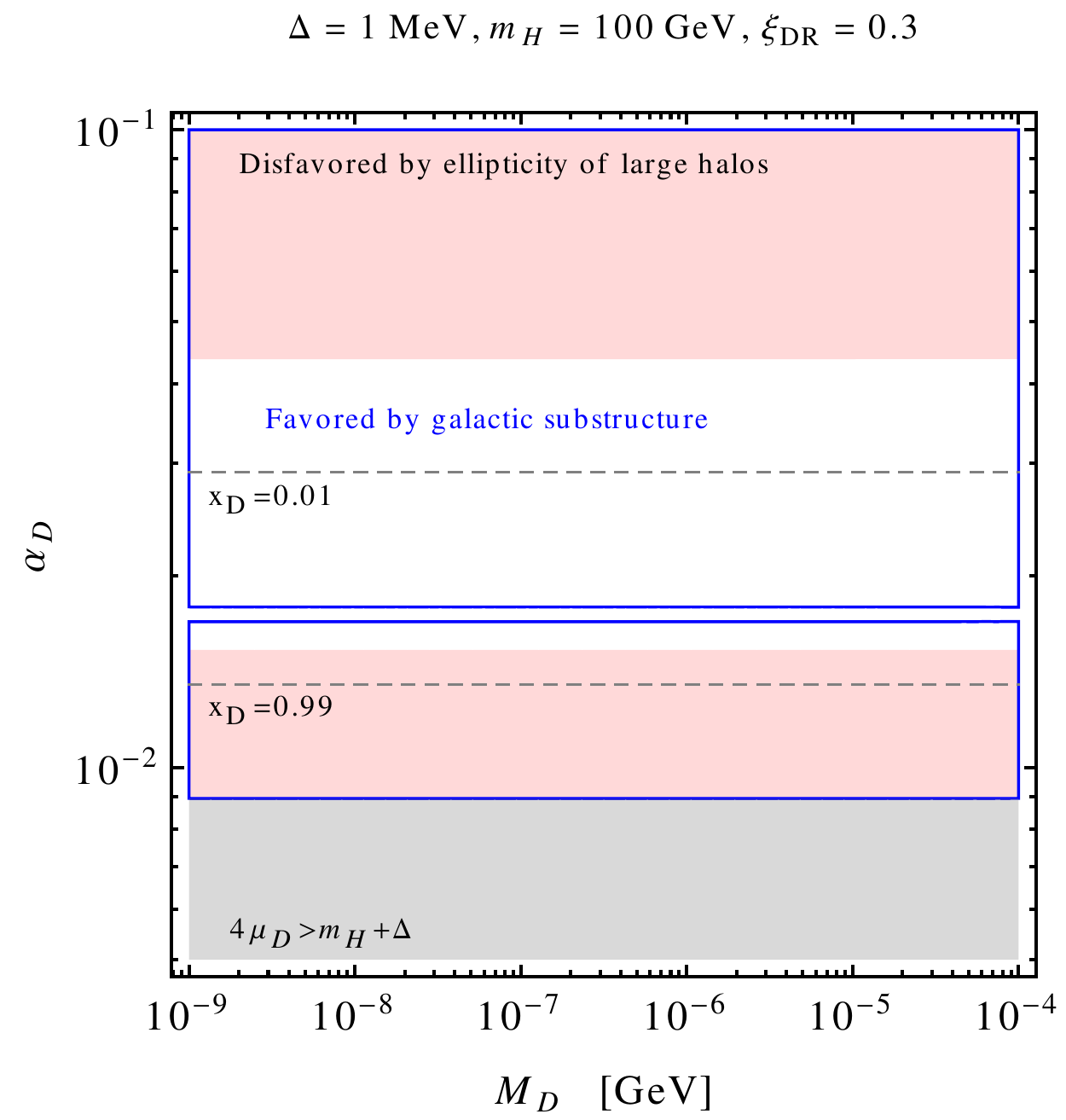}
\includegraphics[width=0.4\textwidth]{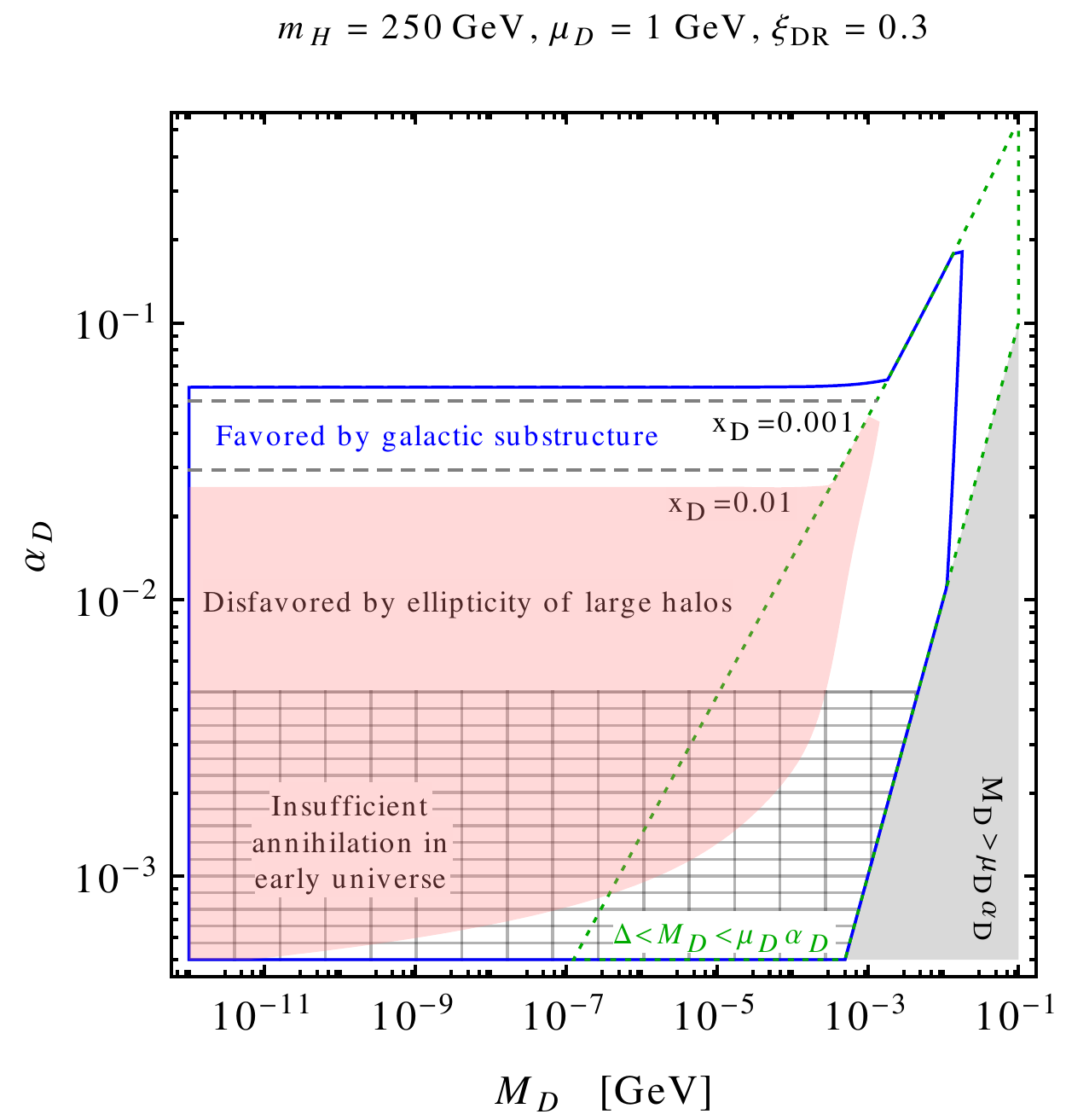}

\bigskip
\includegraphics[width=0.4\textwidth]{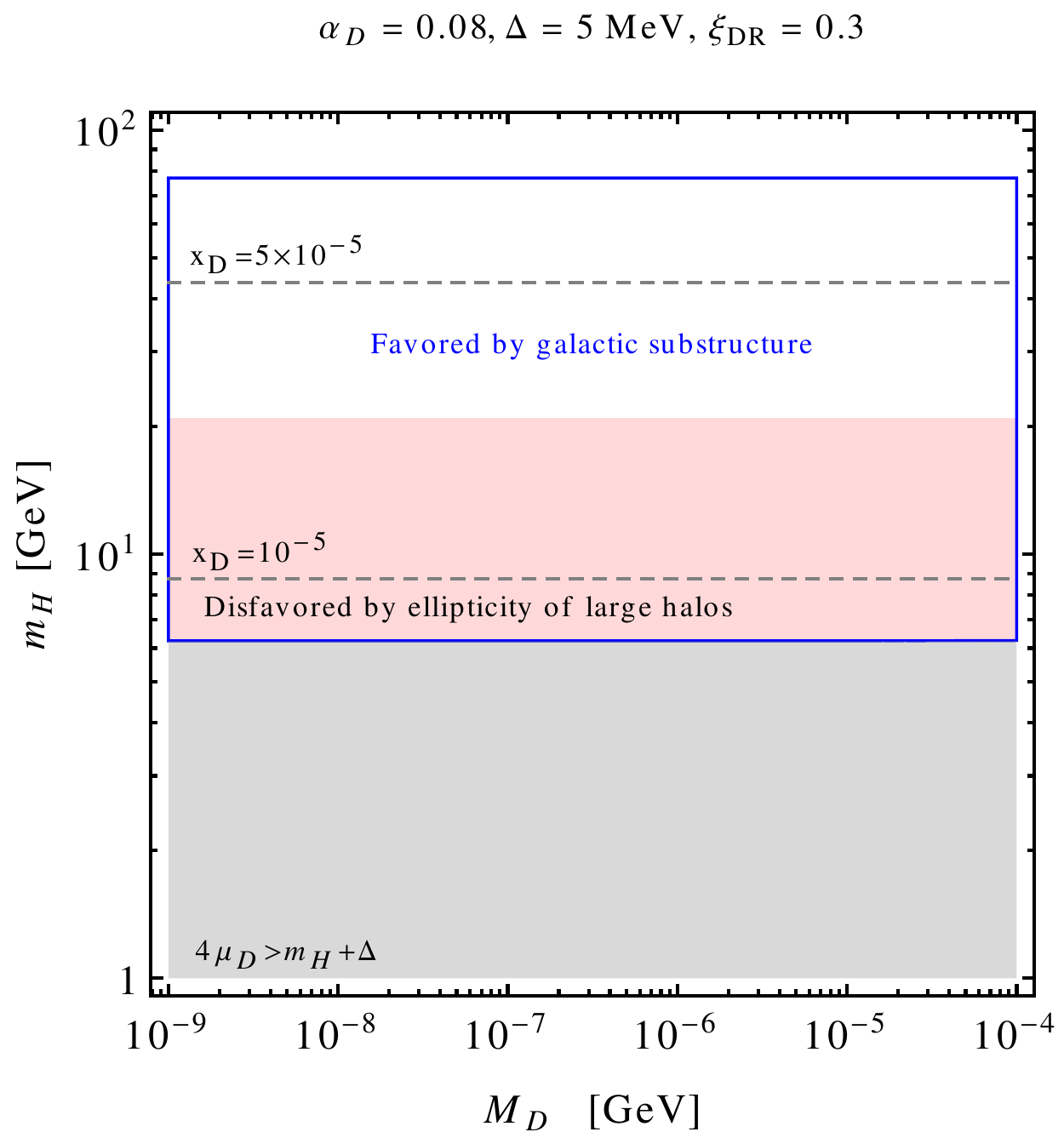}
\includegraphics[width=0.4\textwidth]{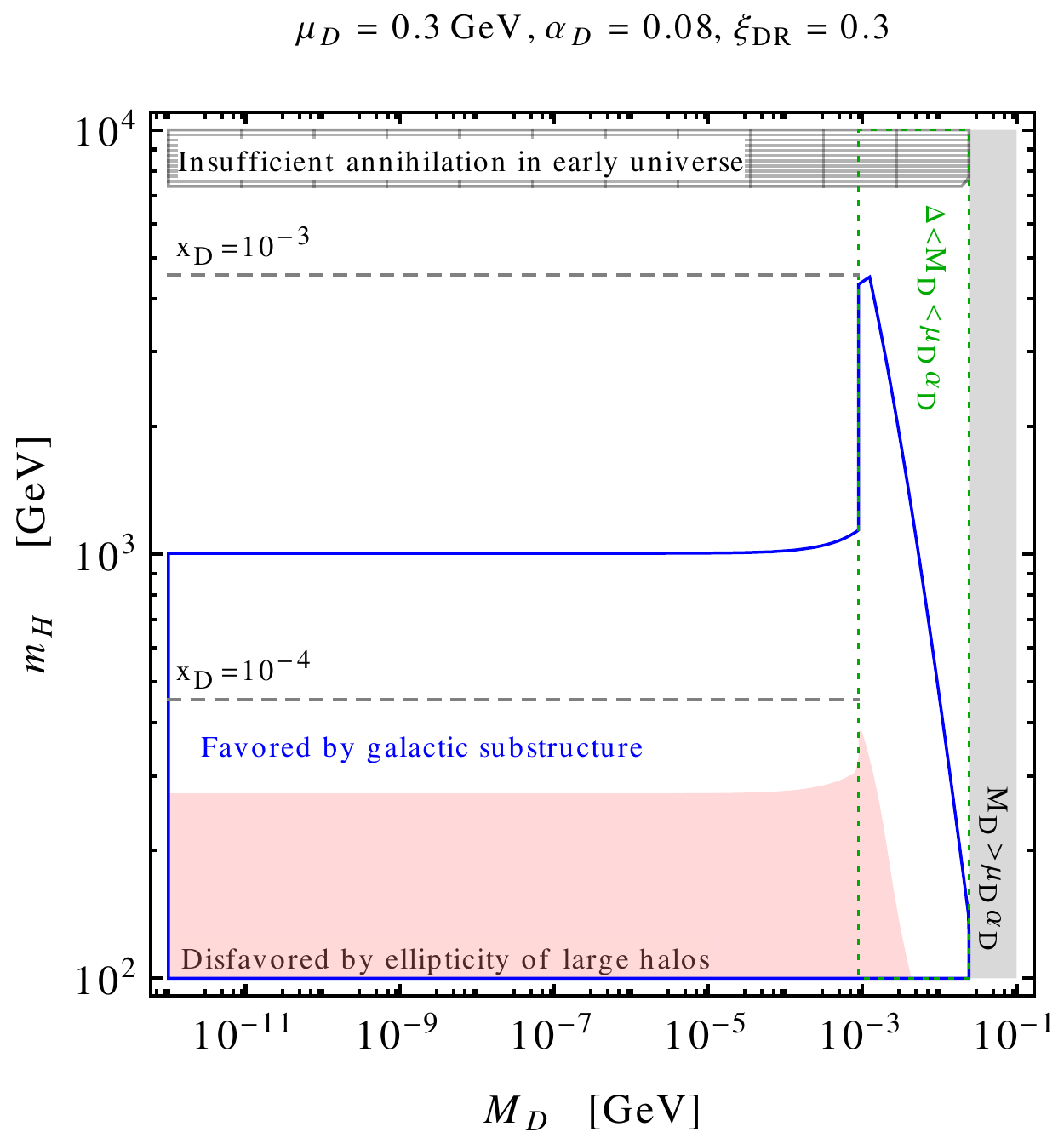}

\bigskip
\includegraphics[width=0.4\textwidth]{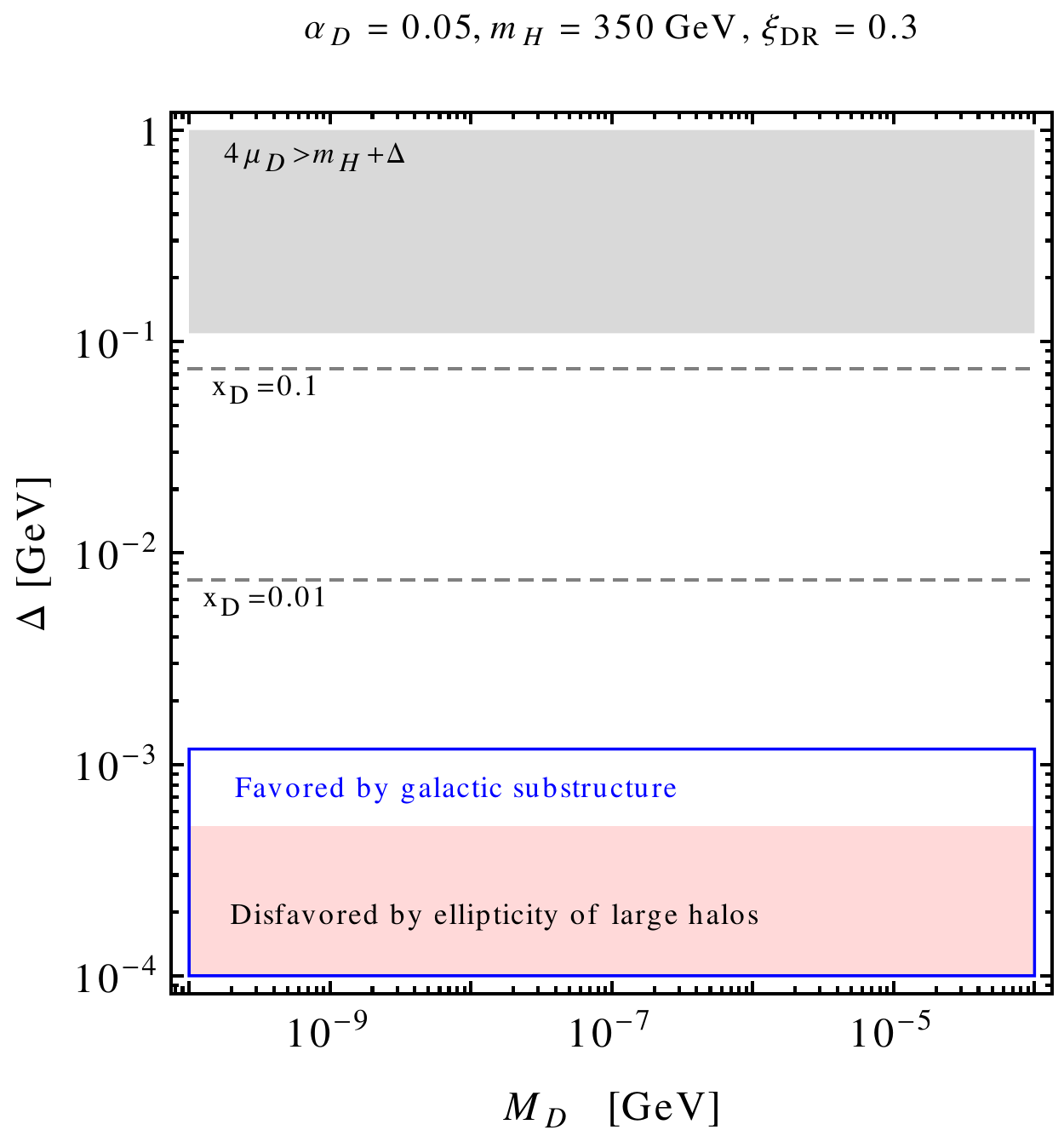}
\includegraphics[width=0.4\textwidth]{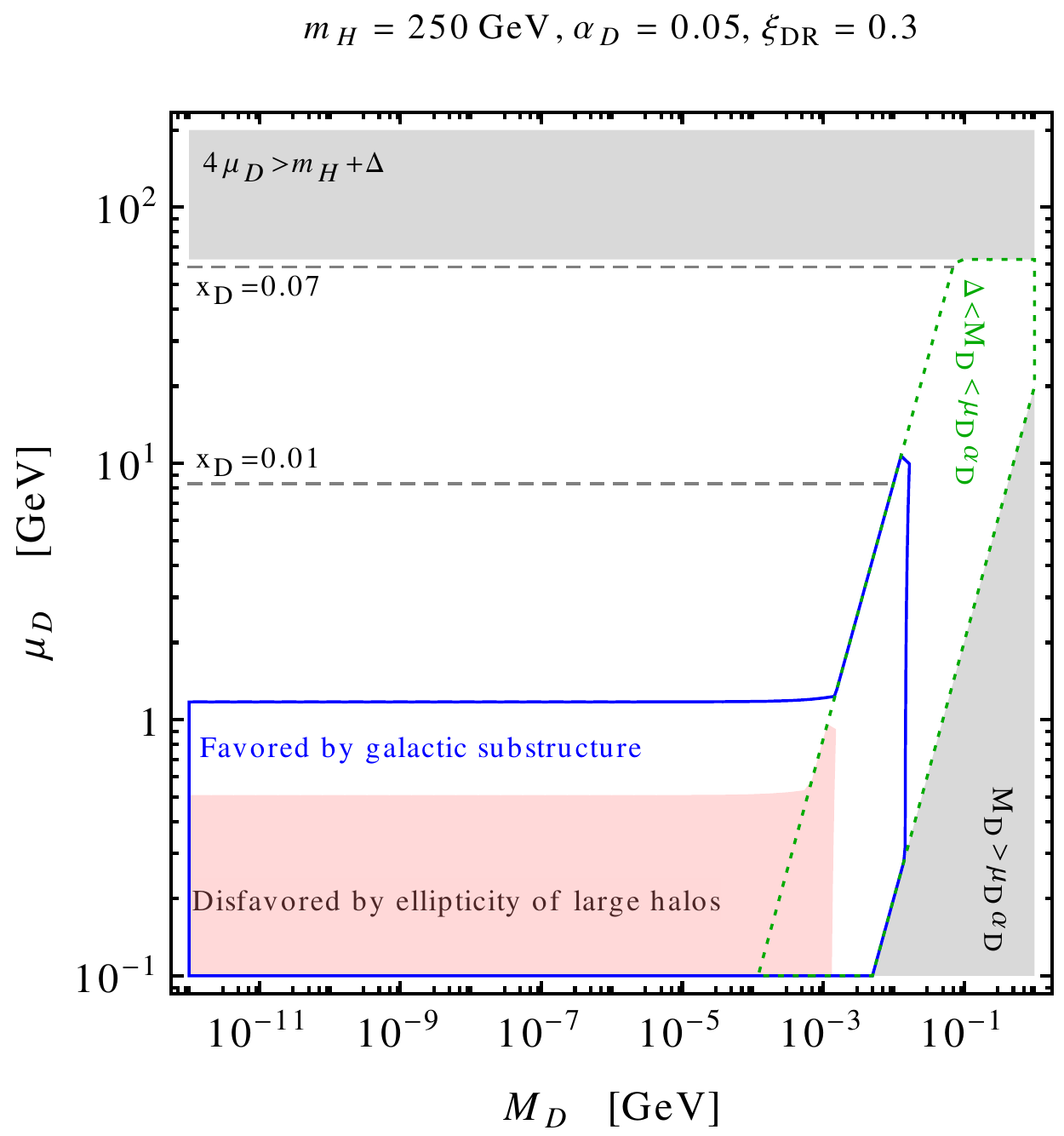}
\caption{\footnotesize  
Same as in Fig.~\ref{fig:CL_continuous MD}, using the 
approach of Ref.~\cite{CyrRacine:2012fz} for collisions involving atoms.}
\label{fig:SI_continuous MD}
 \end{figure}
%%%%%%%%%%%%%%%%%%%%%%
%%%%%%%%%%%%%%%%%%%%%%%%%%%%%%%%%%%%%%%%%%%%%%%%%%%%%%%%%%%%%%%%%%%%%%%%%%%%%%%%%%%%%%%%%

%%%%%%%%%%%%%%%%%%%%%%%%%%%%%%%%%%%%%%%%%%%%%%%%%%%%%%%%%%%%%%%%%%%%%%%%%%%%%%%%%%%%%%%%%
%%%%%%%%%%%%%%%%%%%%%%%
\begin{figure}[ht]
\centering
\includegraphics[width=0.45\textwidth]{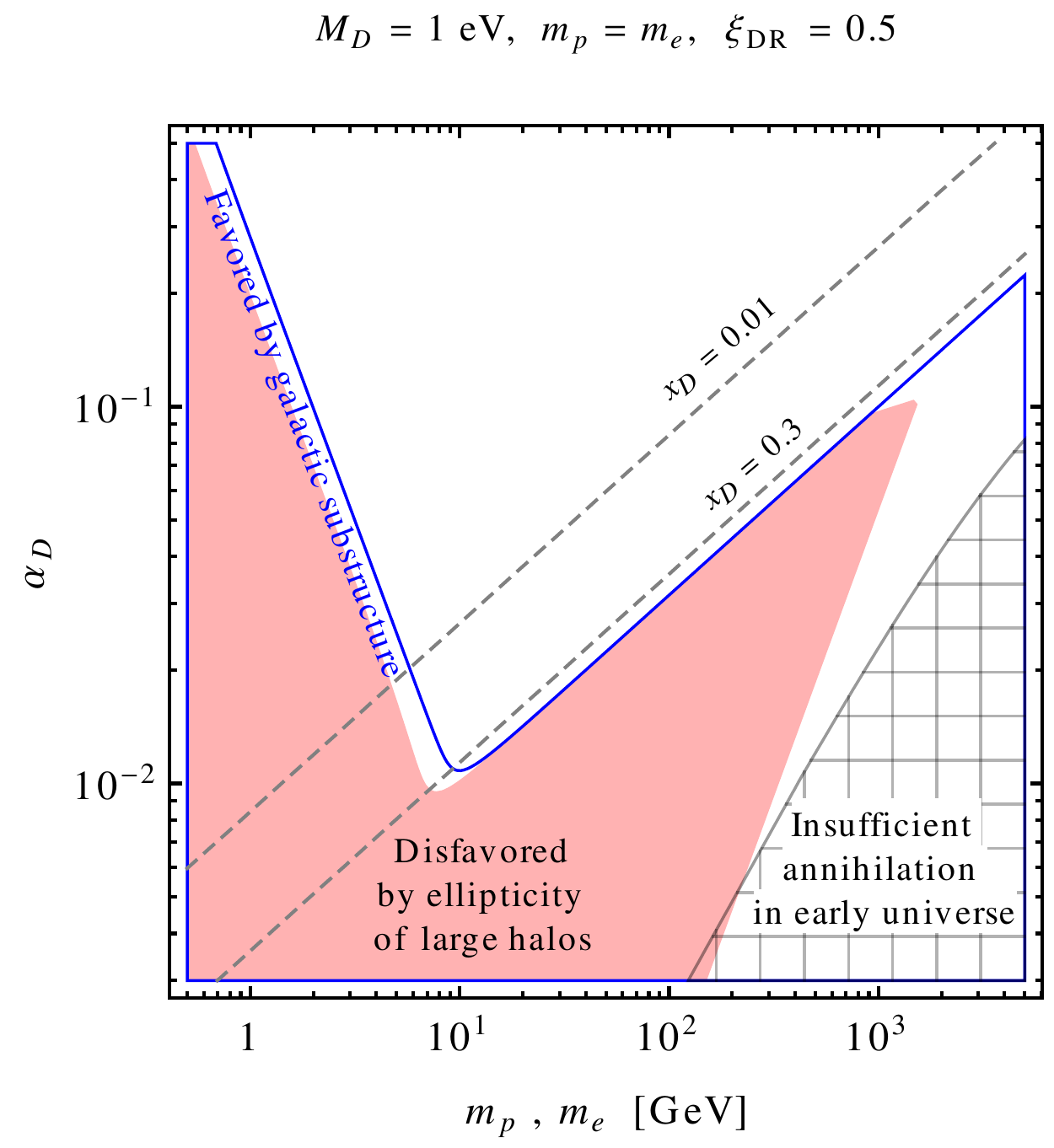}
\includegraphics[width=0.45\textwidth]{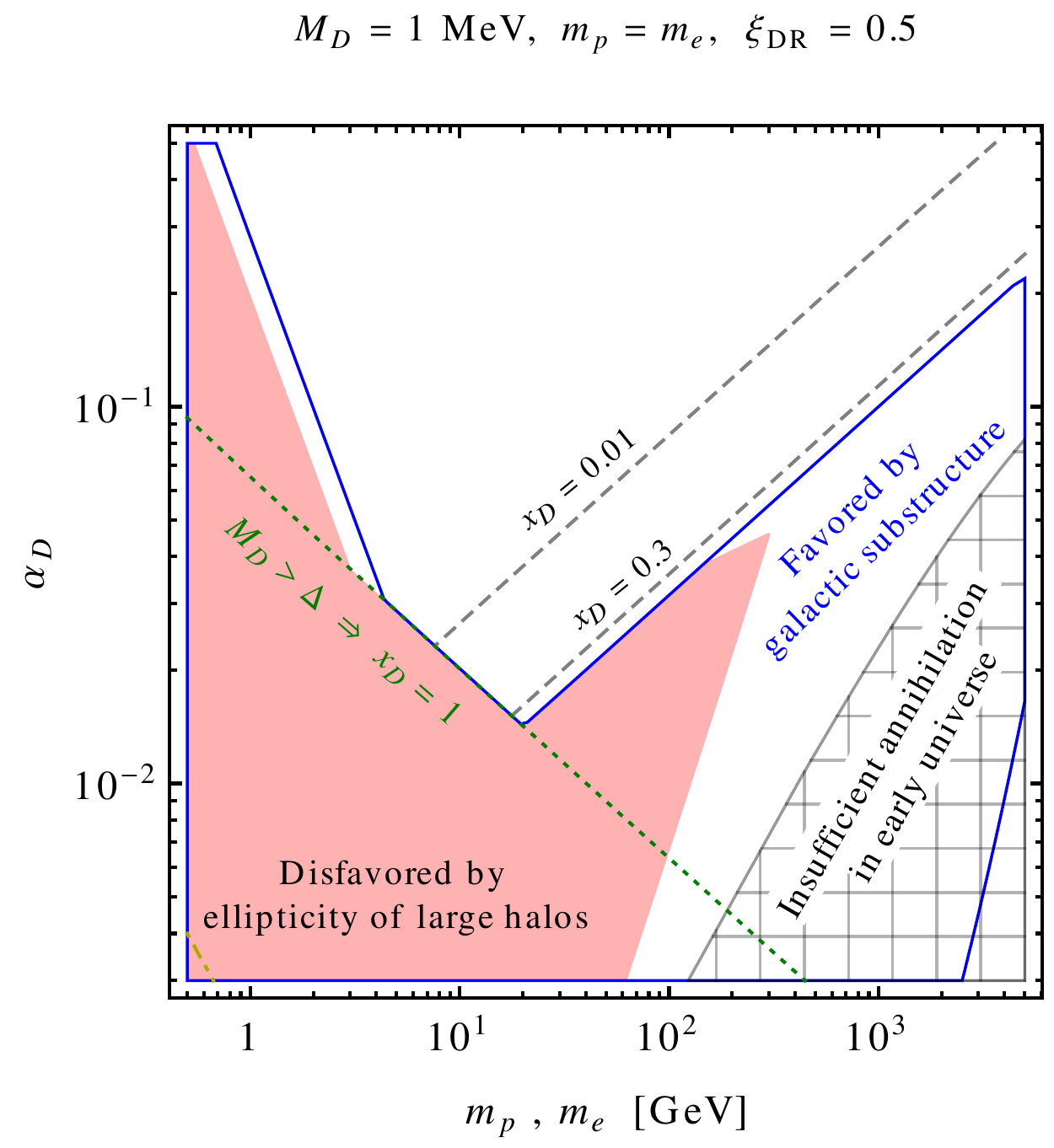}

\bigskip\bigskip
\includegraphics[width=0.45\textwidth]{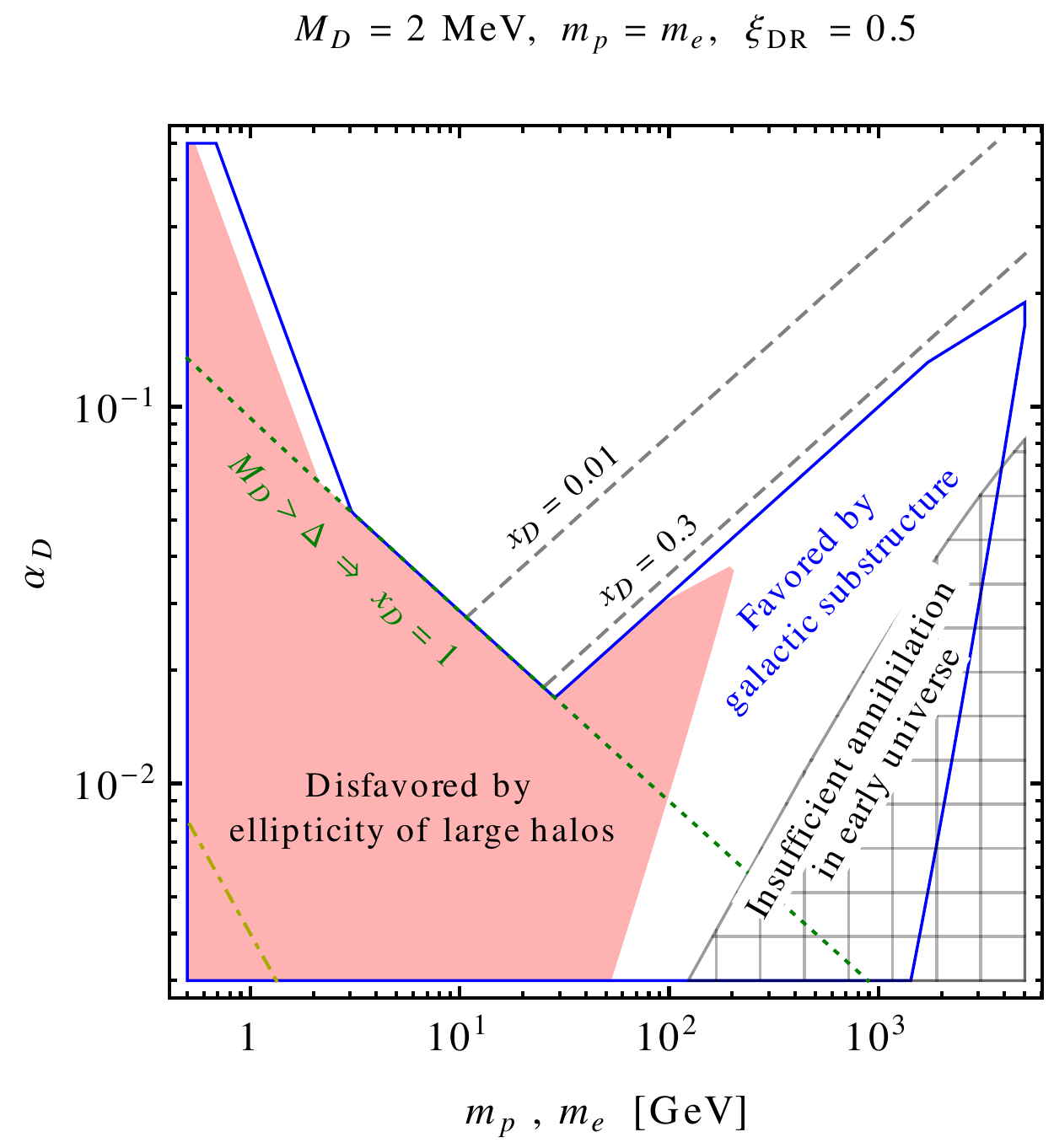}
\includegraphics[width=0.45\textwidth]{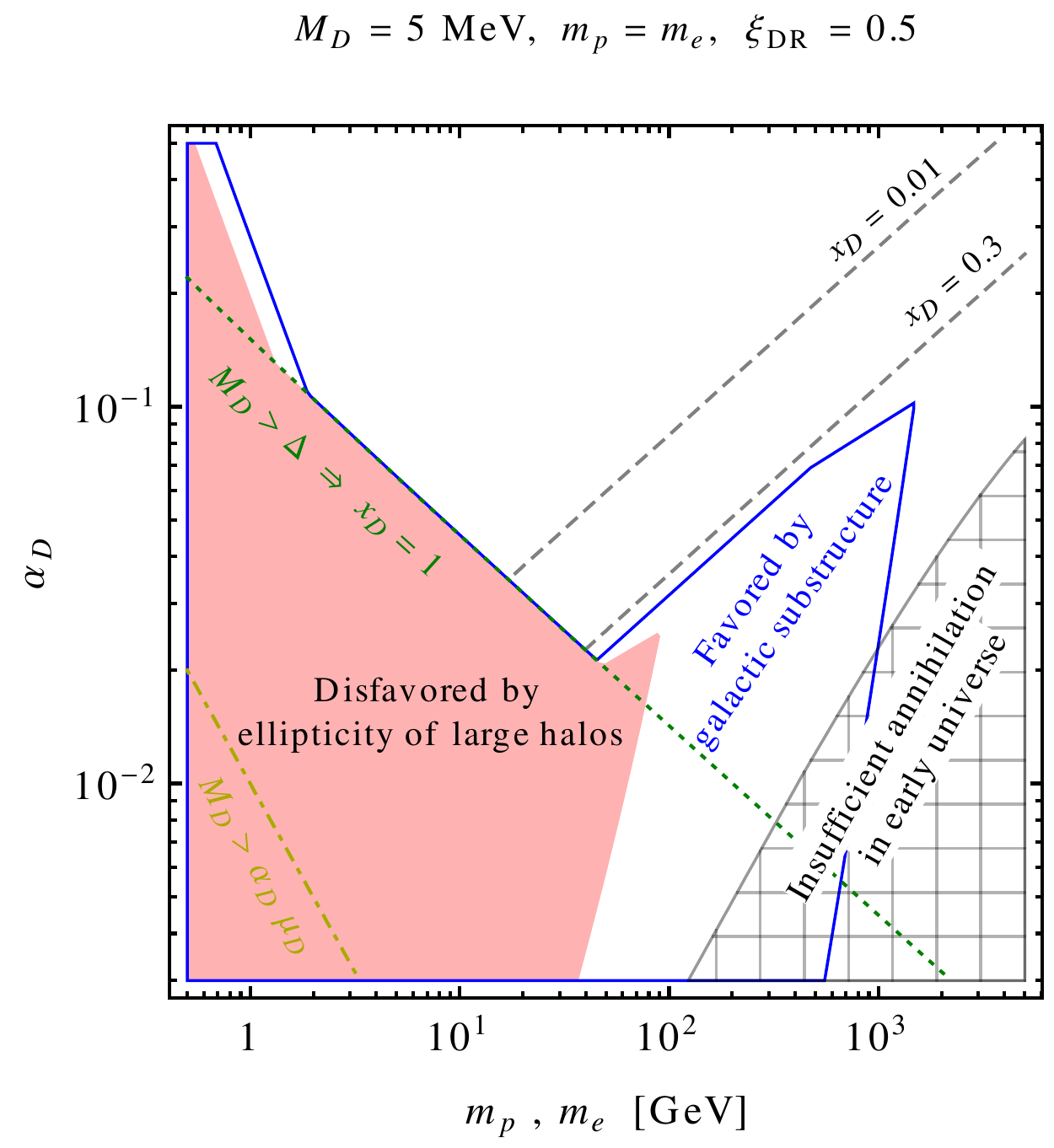}
\caption{\footnotesize  
Red-shaded regions, regions enclosed by blue solid lines and hatched 
regions  have the same meaning as in Fig.~\ref{fig:CL_alphaVmH}. Below the green dotted line, 
$M_{_D} > \Delta$, dark atoms do not form, and the ionisation fraction today is maximal. 
Below the yellow dot-dashed line, $M_{_D} > \alpha_{_D} \mu_{_D}$ and bound states do not exist.}
\label{fig:CL_me=mp small MD}
\end{figure}
%%%%%%%%%%%%%%%%%%%%%%%

%%%%%%%%%%%%%%%%%%%%%%%
\begin{figure}[ht]
\centering
\includegraphics[width=0.4\textwidth]{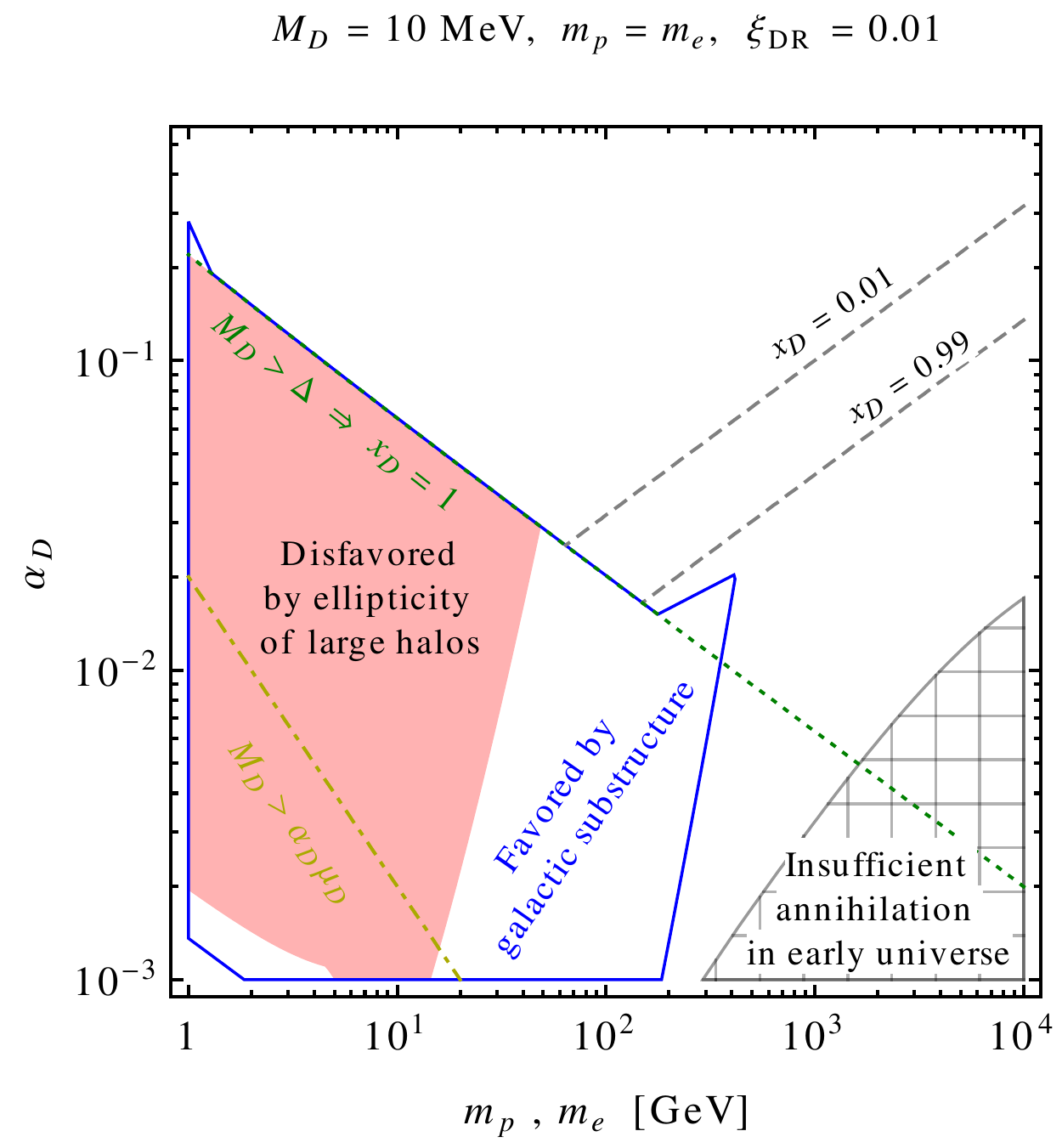}
\includegraphics[width=0.4\textwidth]{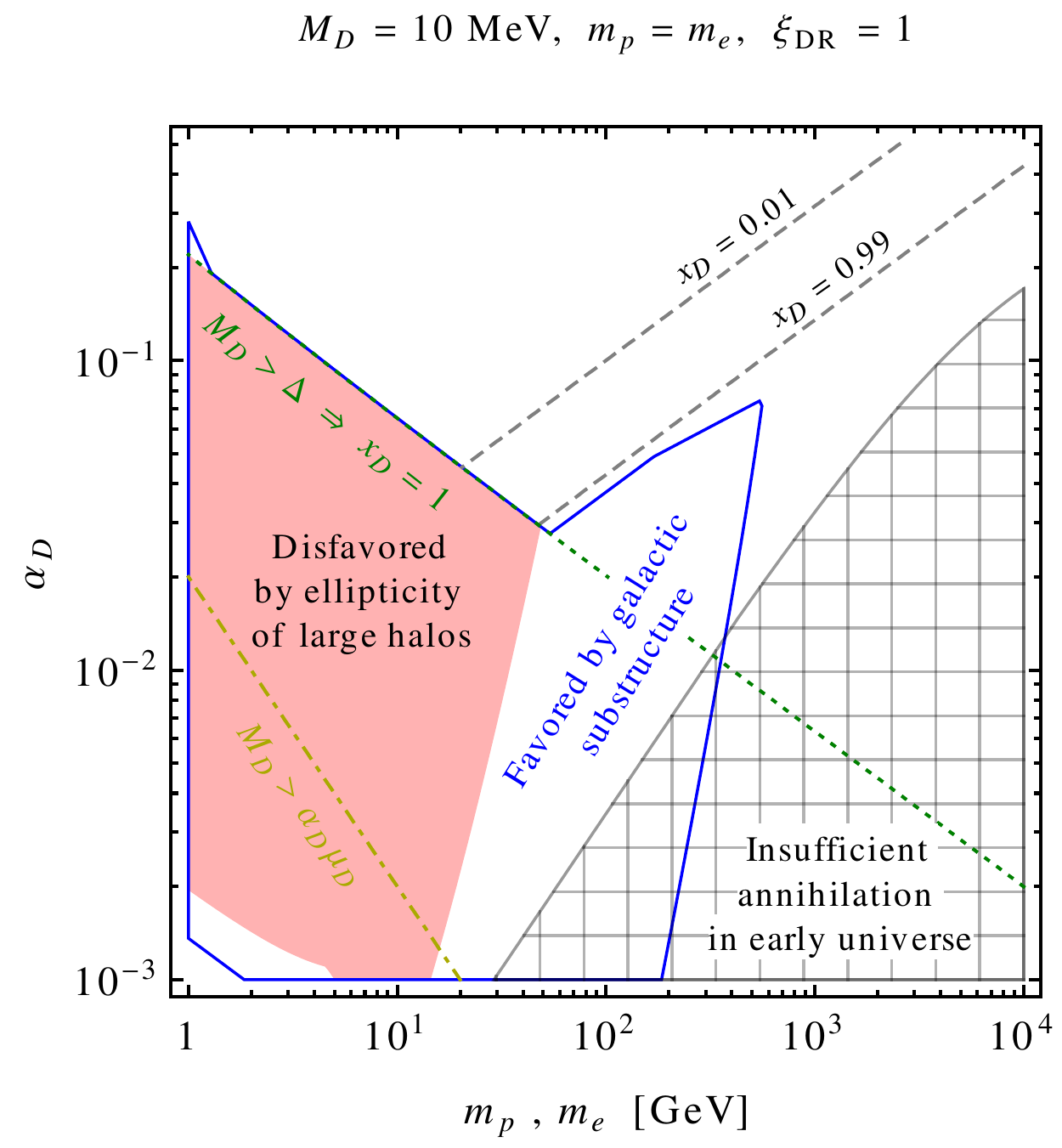}

\bigskip
\includegraphics[width=0.4\textwidth]{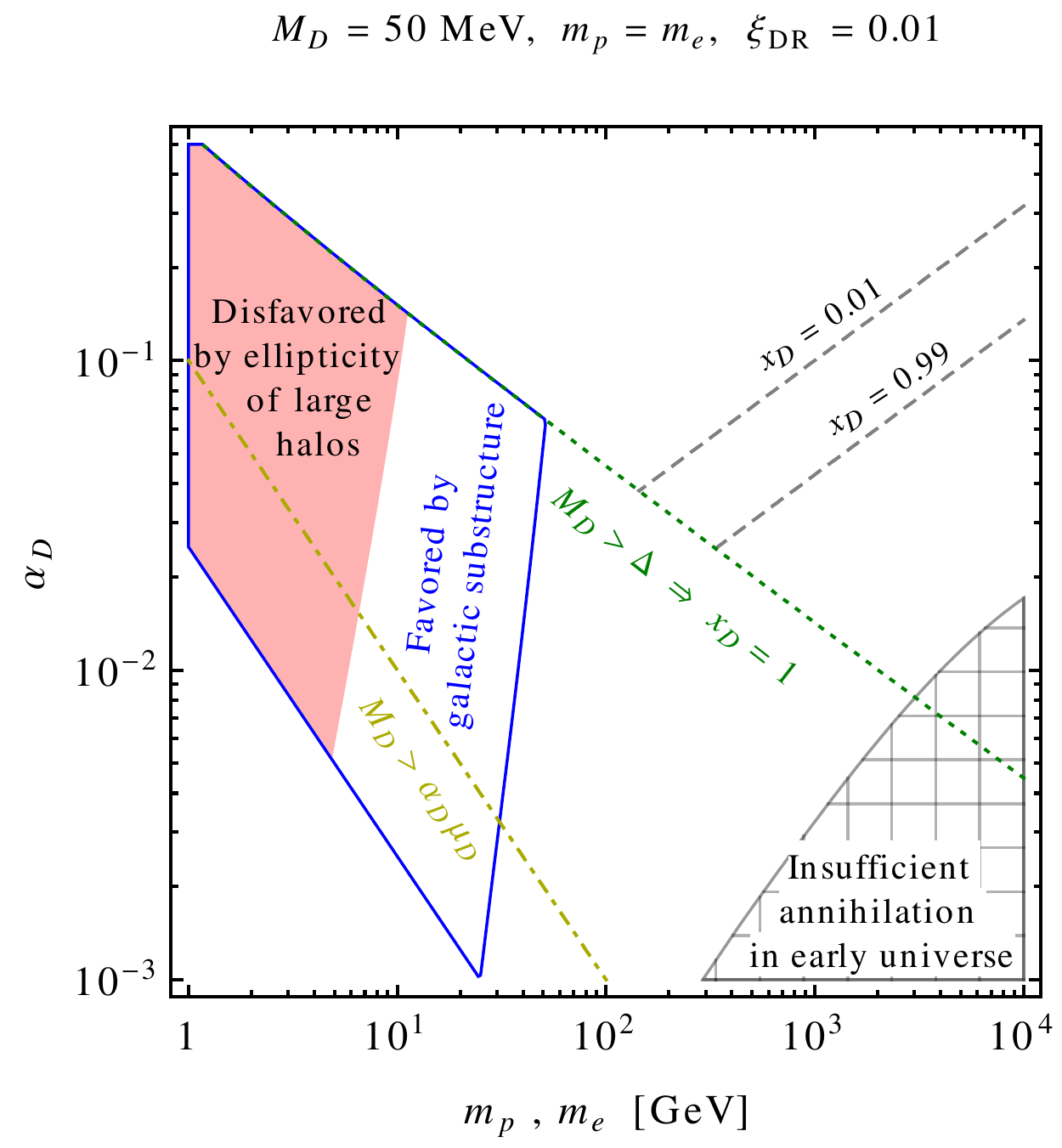}
\includegraphics[width=0.4\textwidth]{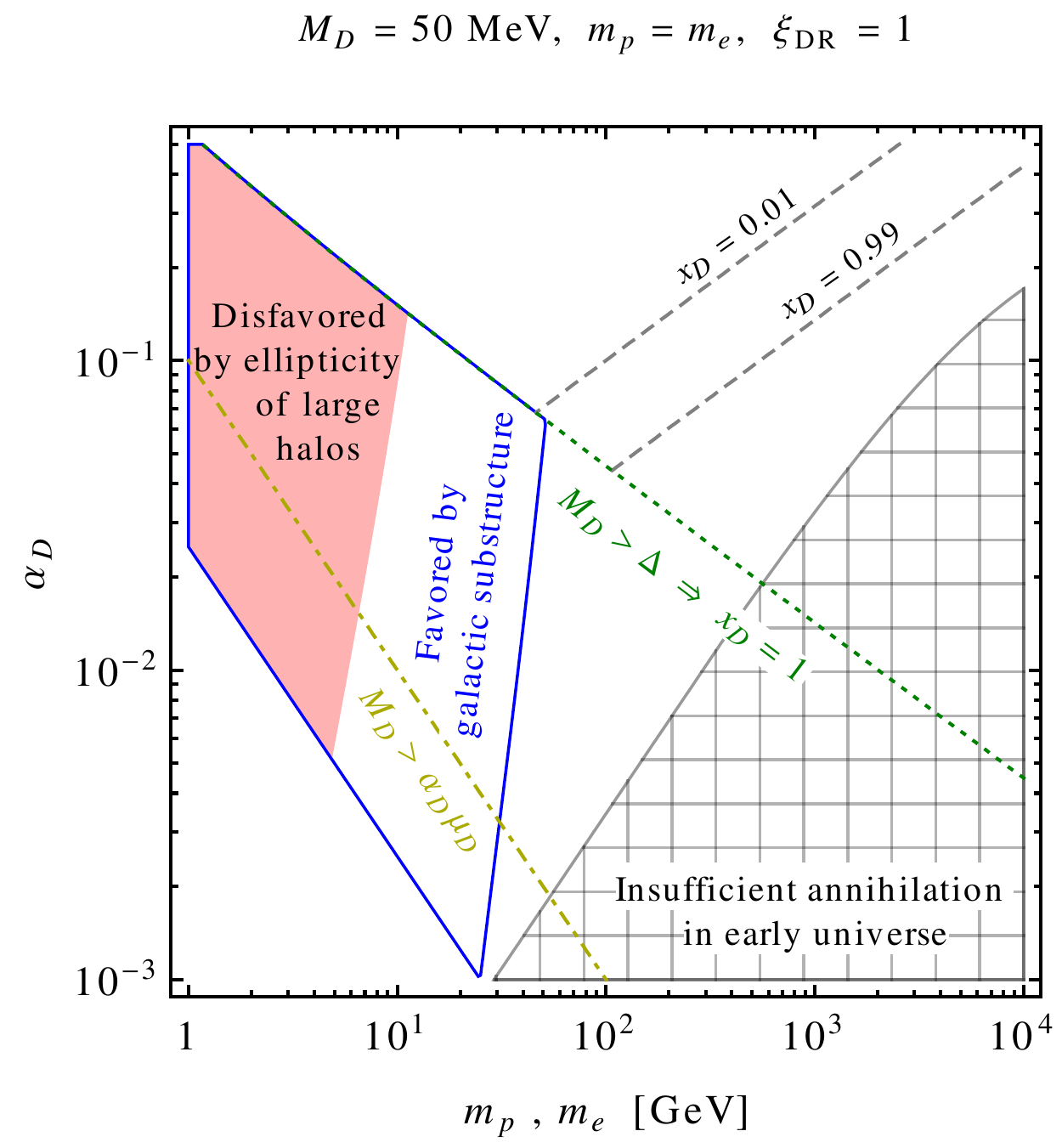}

\bigskip
\includegraphics[width=0.4\textwidth]{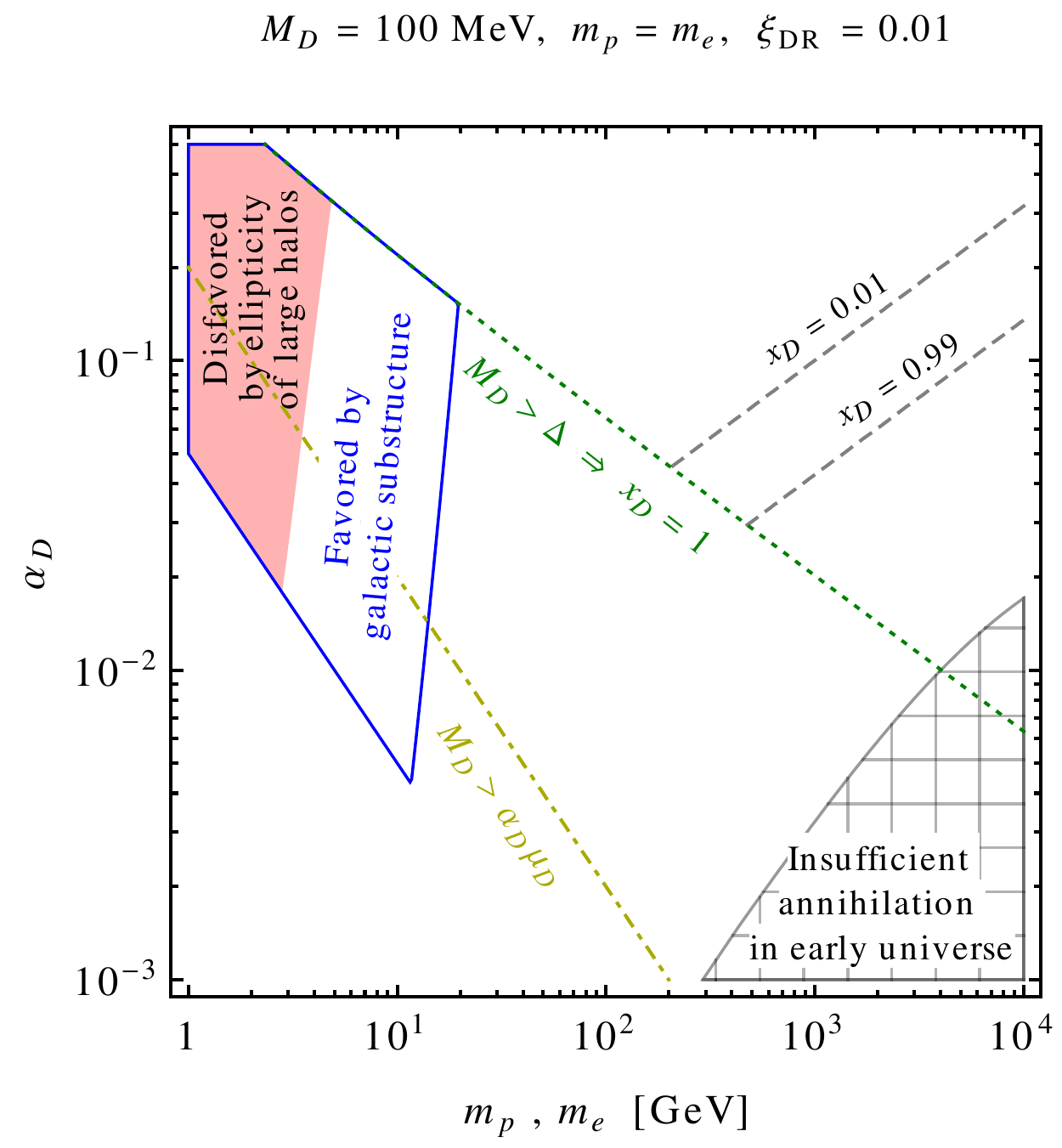}
\includegraphics[width=0.4\textwidth]{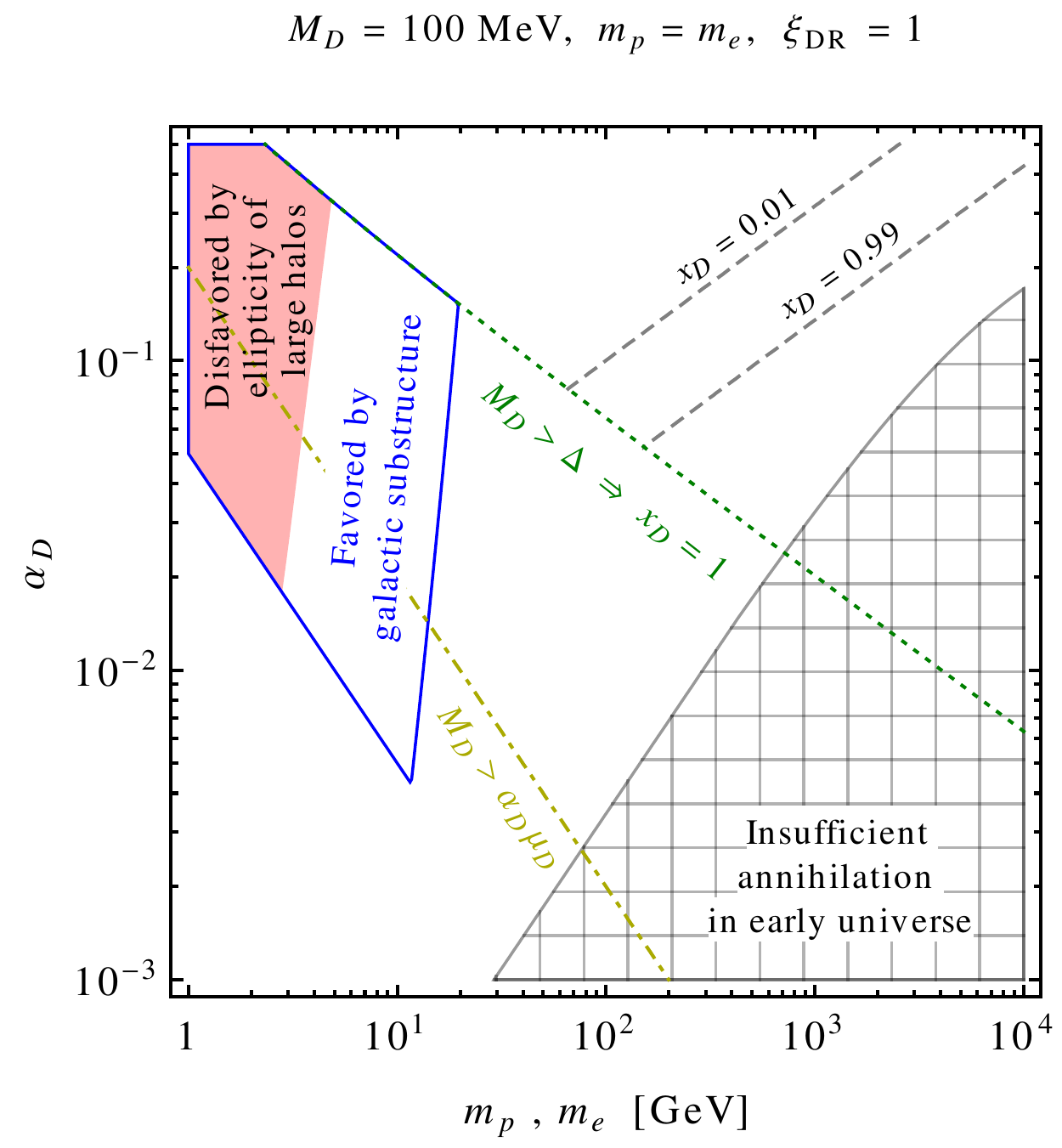}
\caption{\footnotesize  
Same as Fig.~\ref{fig:CL_me=mp small MD}, for larger values of $M_{_D}$. 
The left- and the right-column plots correspond to different values of $\xi_{_{\rm DR}}$.}
\label{fig:CL_me=mp large MD}
\end{figure}
%%%%%%%%%%%%%%%%%%%%%%

%%%%%%%%%%%%%%%%%%%%%%
\begin{figure}[ht]
\centering
\includegraphics[width=0.45\textwidth]{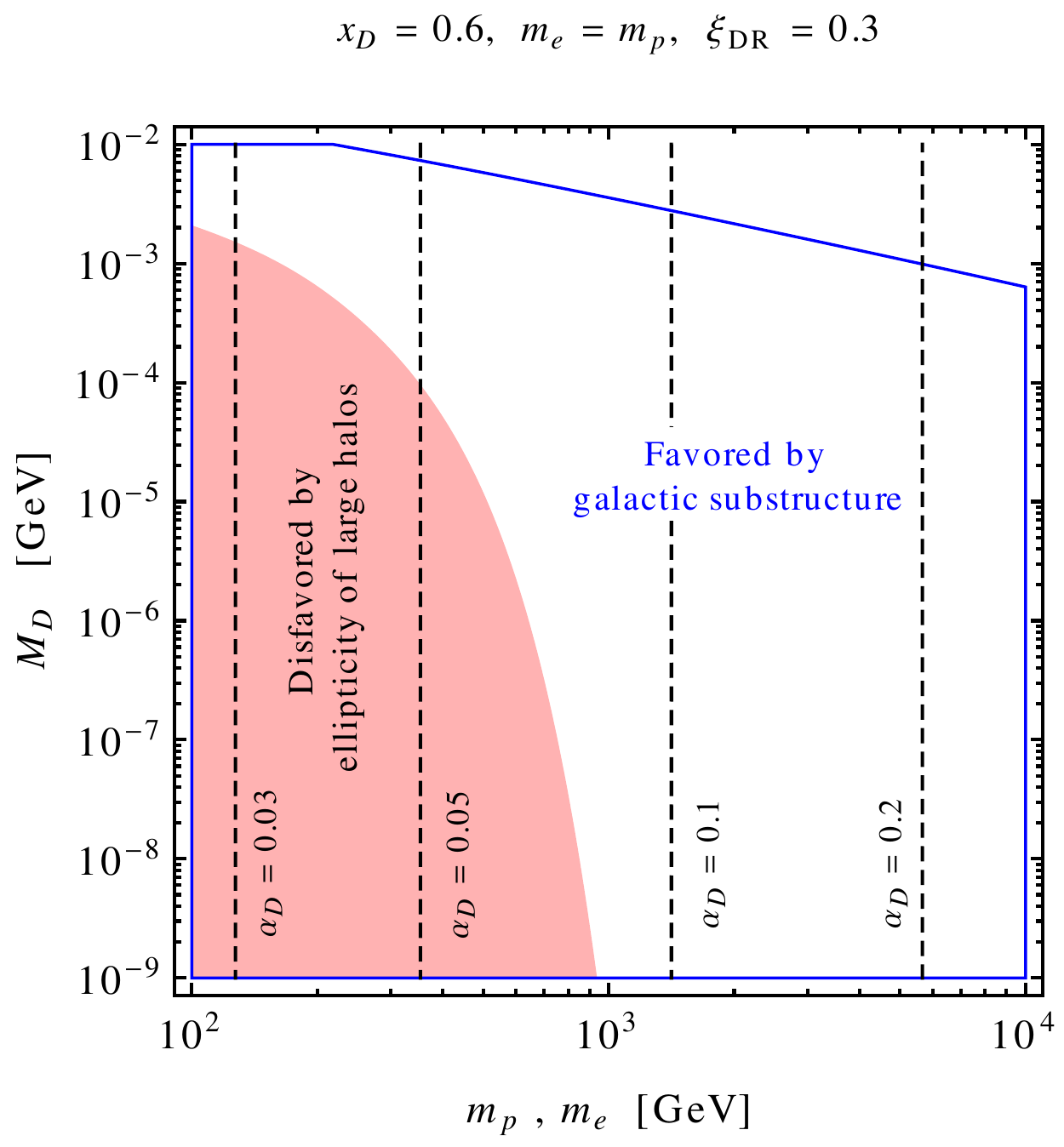}
\includegraphics[width=0.45\textwidth]{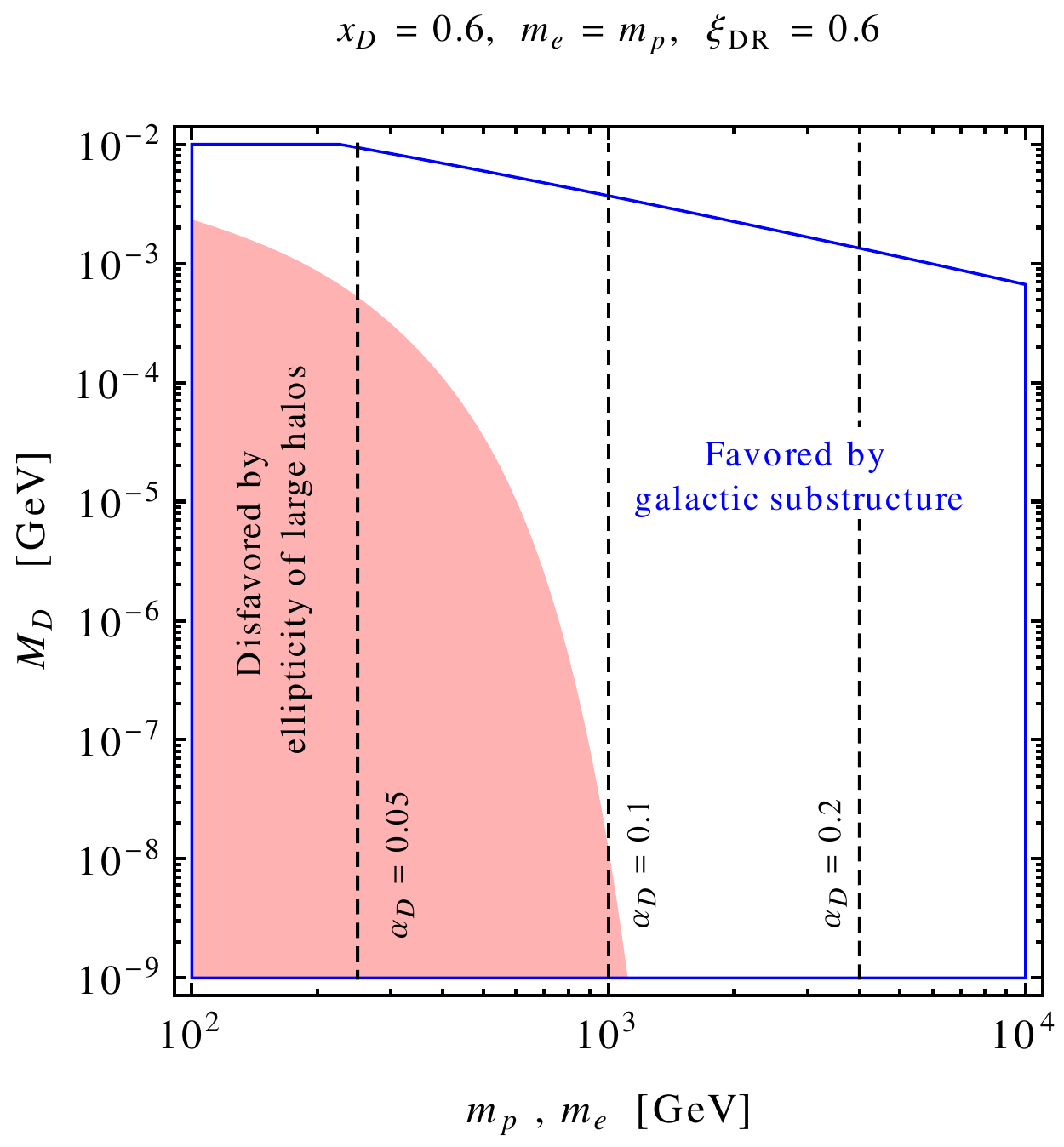}

\bigskip\bigskip
\includegraphics[width=0.45\textwidth]{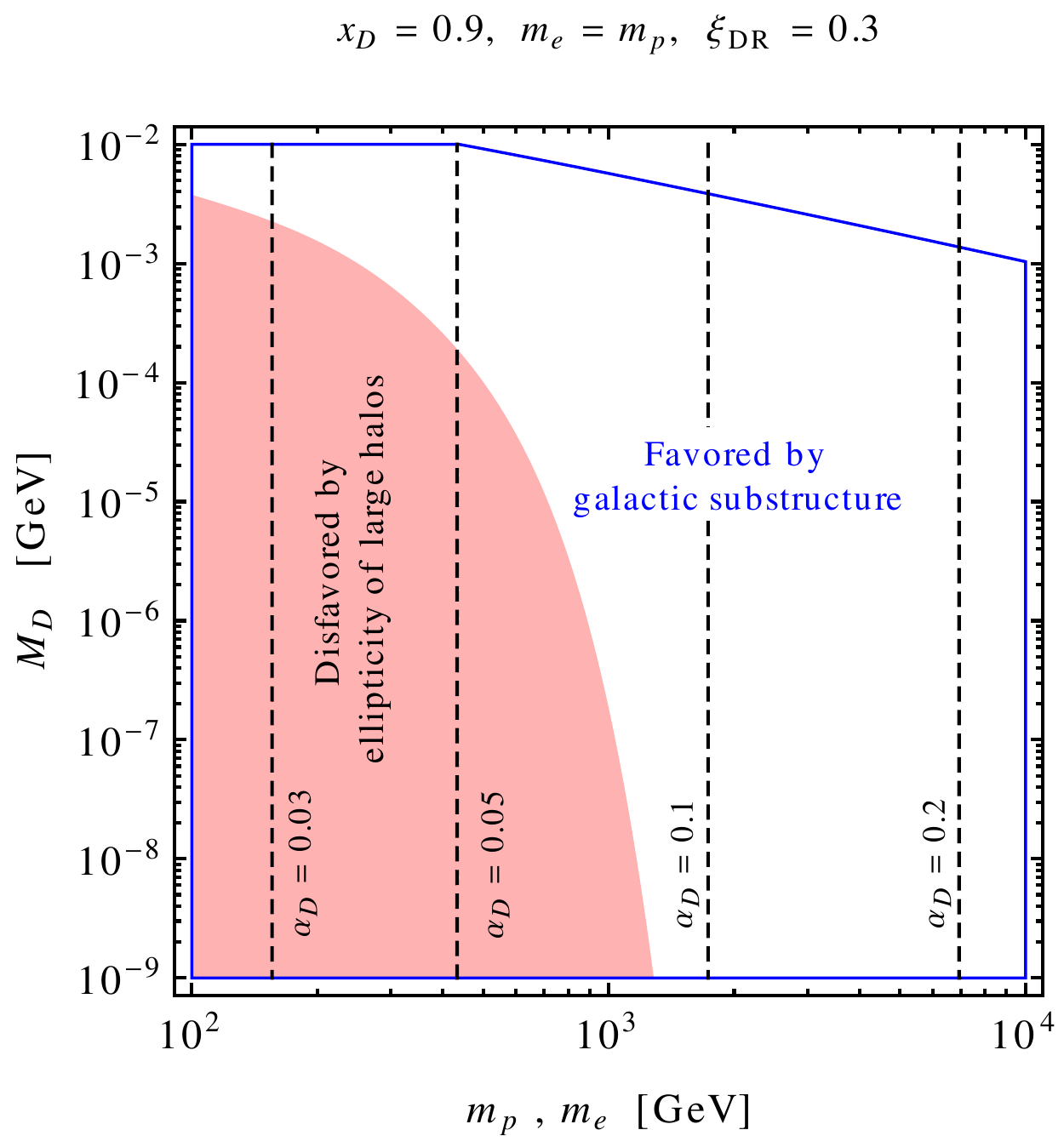}
\includegraphics[width=0.45\textwidth]{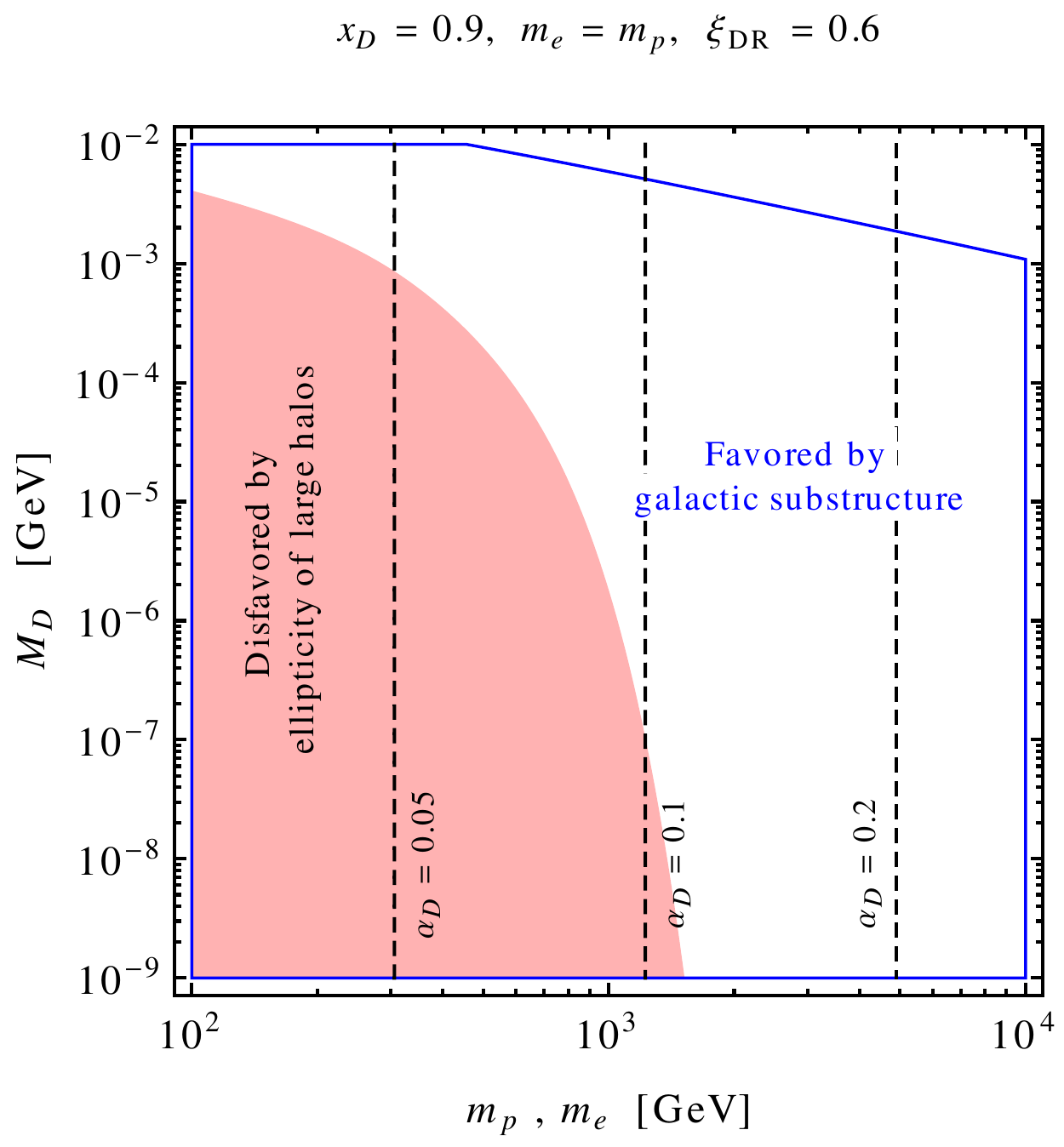}
\caption{\footnotesize  
The value of $\alpha_{_D}$ has be chosen such that the ionisation 
fraction, $x_{_D}$, takes everywhere the value mentioned in the plot label. The 
left- and the right-column plots correspond to different values of $\xi_{_{\rm DR}}$. 
Dashed vertical lines are contours of fixed $\alpha_{_D}$, mentioned on the labels.}
\label{fig:CL_fixed xD}
\end{figure}
%%%%%%%%%%%%%%%%%%%%%%

\clearpage
\bibliography{AtomicDM_arXiv_v3.bbl}

\providecommand{\href}[2]{#2}\begingroup\raggedright\begin{thebibliography}{10}

\bibitem{BoylanKolchin:2011de}
M.~Boylan-Kolchin, J.~S. Bullock, and M.~Kaplinghat, {\it {Too big to fail? The
  puzzling darkness of massive Milky Way subhaloes}},  {\em
  Mon.Not.Roy.Astron.Soc.} {\bf 415} (2011) L40,
  [\href{http://xxx.lanl.gov/abs/1103.0007}{{\tt arXiv:1103.0007}}].

\bibitem{BoylanKolchin:2011dk}
M.~Boylan-Kolchin, J.~S. Bullock, and M.~Kaplinghat, {\it {The Milky Way's
  bright satellites as an apparent failure of LCDM}},  {\em
  Mon.Not.Roy.Astron.Soc.} {\bf 422} (2012) 1203--1218,
  [\href{http://xxx.lanl.gov/abs/1111.2048}{{\tt arXiv:1111.2048}}].

\bibitem{Weinberg:2013aya}
D.~H. Weinberg, J.~S. Bullock, F.~Governato, R.~K. {de Naray}, and A.~H.~G.
  Peter, {\it {Cold dark matter: controversies on small scales}},
  \href{http://xxx.lanl.gov/abs/1306.0913}{{\tt arXiv:1306.0913}}.

\bibitem{Spergel:1999mh}
D.~N. Spergel and P.~J. Steinhardt, {\it {Observational evidence for
  selfinteracting cold dark matter}},  {\em Phys.Rev.Lett.} {\bf 84} (2000)
  3760--3763, [\href{http://xxx.lanl.gov/abs/astro-ph/9909386}{{\tt
  astro-ph/9909386}}].

\bibitem{Wandelt:2000ad}
B.~D. Wandelt, R.~Dave, G.~R. Farrar, P.~C. McGuire, D.~N. Spergel, {\em
  et.~al.}, {\it {Selfinteracting dark matter}},
  \href{http://xxx.lanl.gov/abs/astro-ph/0006344}{{\tt astro-ph/0006344}}.

\bibitem{Mohapatra:2001sx}
R.~Mohapatra, S.~Nussinov, and V.~Teplitz, {\it {Mirror matter as
  selfinteracting dark matter}},  {\em Phys.Rev.} {\bf D66} (2002) 063002,
  [\href{http://xxx.lanl.gov/abs/hep-ph/0111381}{{\tt hep-ph/0111381}}].

\bibitem{Rocha:2012jg}
M.~Rocha, A.~H. Peter, J.~S. Bullock, M.~Kaplinghat, S.~Garrison-Kimmel, {\em
  et.~al.}, {\it {Cosmological Simulations with Self-Interacting Dark Matter I:
  Constant Density Cores and Substructure}},  {\em Mon.Not.Roy.Astron.Soc.}
  {\bf 430} (2013) 81--104, [\href{http://xxx.lanl.gov/abs/1208.3025}{{\tt
  arXiv:1208.3025}}].

\bibitem{Peter:2012jh}
A.~H. Peter, M.~Rocha, J.~S. Bullock, and M.~Kaplinghat, {\it {Cosmological
  Simulations with Self-Interacting Dark Matter II: Halo Shapes vs.
  Observations}},  \href{http://xxx.lanl.gov/abs/1208.3026}{{\tt
  arXiv:1208.3026}}.

\bibitem{Vogelsberger:2012ku}
M.~Vogelsberger, J.~Zavala, and A.~Loeb, {\it {Subhaloes in Self-Interacting
  Galactic Dark Matter Haloes}},  {\em Mon.Not.Roy.Astron.Soc.} {\bf 423}
  (2012) 3740, [\href{http://xxx.lanl.gov/abs/1201.5892}{{\tt
  arXiv:1201.5892}}].

\bibitem{Vogelsberger:2012sa}
M.~Vogelsberger and J.~Zavala, {\it {Direct detection of self-interacting dark
  matter}},  {\em Mon.Not.Roy.Astron.Soc.} {\bf 430} (2013) 1722--1735,
  [\href{http://xxx.lanl.gov/abs/1211.1377}{{\tt arXiv:1211.1377}}].

\bibitem{Zavala:2012us}
J.~Zavala, M.~Vogelsberger, and M.~G. Walker, {\it {Constraining
  Self-Interacting Dark Matter with the Milky Way's dwarf spheroidals}},  {\em
  Monthly Notices of the Royal Astronomical Society: Letters} {\bf 431} (2013)
  L20--L24, [\href{http://xxx.lanl.gov/abs/1211.6426}{{\tt arXiv:1211.6426}}].

\bibitem{Faraggi:2000pv}
A.~E. Faraggi and M.~Pospelov, {\it {Selfinteracting dark matter from the
  hidden heterotic string sector}},  {\em Astropart.Phys.} {\bf 16} (2002)
  451--461, [\href{http://xxx.lanl.gov/abs/hep-ph/0008223}{{\tt
  hep-ph/0008223}}].

\bibitem{Kusenko:2001vu}
A.~Kusenko and P.~J. Steinhardt, {\it {Q ball candidates for selfinteracting
  dark matter}},  {\em Phys.Rev.Lett.} {\bf 87} (2001) 141301,
  [\href{http://xxx.lanl.gov/abs/astro-ph/0106008}{{\tt astro-ph/0106008}}].

\bibitem{Foot:2014mia}
R.~Foot, {\it {Mirror dark matter: Cosmology, galaxy structure and direct
  detection}},  \href{http://xxx.lanl.gov/abs/1401.3965}{{\tt
  arXiv:1401.3965}}.

\bibitem{Boddy:2014yra}
K.~K. Boddy, J.~L. Feng, M.~Kaplinghat, and T.~M.~P. Tait, {\it
  {Self-Interacting Dark Matter from a Non-Abelian Hidden Sector}},
  \href{http://xxx.lanl.gov/abs/1402.3629}{{\tt arXiv:1402.3629}}.

\bibitem{Hochberg:2014dra}
Y.~Hochberg, E.~Kuflik, T.~Volansky, and J.~G. Wacker, {\it {The SIMP
  Miracle}},  \href{http://xxx.lanl.gov/abs/1402.5143}{{\tt arXiv:1402.5143}}.

\bibitem{Alves:2010dd}
D.~{Spier Moreira Alves}, S.~R. Behbahani, P.~Schuster, and J.~G. Wacker, {\it
  {The Cosmology of Composite Inelastic Dark Matter}},  {\em JHEP} {\bf 1006}
  (2010) 113, [\href{http://xxx.lanl.gov/abs/1003.4729}{{\tt
  arXiv:1003.4729}}].

\bibitem{CyrRacine:2012fz}
F.-Y. Cyr-Racine and K.~Sigurdson, {\it {The Cosmology of Atomic Dark Matter}},
   {\em Phys.Rev.} {\bf D87} (2013) 103515,
  [\href{http://xxx.lanl.gov/abs/1209.5752}{{\tt arXiv:1209.5752}}].

\bibitem{Cyr-Racine:2013fsa}
F.-Y. Cyr-Racine, R.~{de Putter}, A.~Raccanelli, and K.~Sigurdson, {\it
  {Constraints on Large-Scale Dark Acoustic Oscillations from Cosmology}},
  {\em Phys.Rev.} {\bf D89} (2014) 063517,
  [\href{http://xxx.lanl.gov/abs/1310.3278}{{\tt arXiv:1310.3278}}].

\bibitem{Kaplan:2009de}
D.~E. Kaplan, G.~Z. Krnjaic, K.~R. Rehermann, and C.~M. Wells, {\it {Atomic
  Dark Matter}},  {\em JCAP} {\bf 1005} (2010) 021,
  [\href{http://xxx.lanl.gov/abs/0909.0753}{{\tt arXiv:0909.0753}}].

\bibitem{Kaplan:2011yj}
D.~E. Kaplan, G.~Z. Krnjaic, K.~R. Rehermann, and C.~M. Wells, {\it {Dark
  Atoms: Asymmetry and Direct Detection}},  {\em JCAP} {\bf 1110} (2011) 011,
  [\href{http://xxx.lanl.gov/abs/1105.2073}{{\tt arXiv:1105.2073}}].

\bibitem{Cline:2012is}
J.~M. Cline, Z.~Liu, and W.~Xue, {\it {Millicharged Atomic Dark Matter}},  {\em
  Phys.Rev.} {\bf D85} (2012) 101302,
  [\href{http://xxx.lanl.gov/abs/1201.4858}{{\tt arXiv:1201.4858}}].

\bibitem{Cline:2013pca}
J.~M. Cline, Z.~Liu, G.~Moore, and W.~Xue, {\it {Scattering properties of dark
  atoms and molecules}},  {\em Phys.Rev.} {\bf D89} (2014) 043514,
  [\href{http://xxx.lanl.gov/abs/1311.6468}{{\tt arXiv:1311.6468}}].

\bibitem{Cline:2013zca}
J.~M. Cline, Z.~Liu, G.~Moore, and W.~Xue, {\it {Composite strongly interacting
  dark matter}},  \href{http://xxx.lanl.gov/abs/1312.3325}{{\tt
  arXiv:1312.3325}}.

\bibitem{Kaplinghat:2013xca}
M.~Kaplinghat, R.~E. Keeley, T.~Linden, and H.-B. Yu, {\it {Tying Dark Matter
  to Baryons with Self-interactions}},
  \href{http://xxx.lanl.gov/abs/1311.6524}{{\tt arXiv:1311.6524}}.

\bibitem{Feng:2009mn}
J.~L. Feng, M.~Kaplinghat, H.~Tu, and H.-B. Yu, {\it {Hidden Charged Dark
  Matter}},  {\em JCAP} {\bf 0907} (2009) 004,
  [\href{http://xxx.lanl.gov/abs/0905.3039}{{\tt arXiv:0905.3039}}].

\bibitem{Feng:2009hw}
J.~L. Feng, M.~Kaplinghat, and H.-B. Yu, {\it {Halo Shape and Relic Density
  Exclusions of Sommerfeld-Enhanced Dark Matter Explanations of Cosmic Ray
  Excesses}},  {\em Phys.Rev.Lett.} {\bf 104} (2010) 151301,
  [\href{http://xxx.lanl.gov/abs/0911.0422}{{\tt arXiv:0911.0422}}].

\bibitem{Loeb:2010gj}
A.~Loeb and N.~Weiner, {\it {Cores in Dwarf Galaxies from Dark Matter with a
  Yukawa Potential}},  {\em Phys.Rev.Lett.} {\bf 106} (2011) 171302,
  [\href{http://xxx.lanl.gov/abs/1011.6374}{{\tt arXiv:1011.6374}}].

\bibitem{Davoudiasl:2012uw}
H.~Davoudiasl and R.~N. Mohapatra, {\it {On Relating the Genesis of Cosmic
  Baryons and Dark Matter}},  {\em New J.Phys.} {\bf 14} (2012) 095011,
  [\href{http://xxx.lanl.gov/abs/1203.1247}{{\tt arXiv:1203.1247}}].

\bibitem{Petraki:2013wwa}
K.~Petraki and R.~R. Volkas, {\it {Review of asymmetric dark matter}},  {\em
  Int.J.Mod.Phys.} {\bf A28} (2013) 1330028,
  [\href{http://xxx.lanl.gov/abs/1305.4939}{{\tt arXiv:1305.4939}}].

\bibitem{Zurek:2013wia}
K.~M. Zurek, {\it {Asymmetric Dark Matter: Theories, Signatures, and
  Constraints}},  {\em Phys.Rept.} {\bf 537} (2014) 91--121,
  [\href{http://xxx.lanl.gov/abs/1308.0338}{{\tt arXiv:1308.0338}}].

\bibitem{Boucenna:2013wba}
S.~Boucenna and S.~Morisi, {\it {Theories relating baryon asymmetry and dark
  matter: A mini review}},  \href{http://xxx.lanl.gov/abs/1310.1904}{{\tt
  arXiv:1310.1904}}.

\bibitem{Volkas:2013eia}
R.~R. Volkas, {\it {Brief Overview of Asymmetric Dark Matter}},  {\em Universe}
  {\bf 1} (2013), no.~3 4--16.

\bibitem{Graesser:2011wi}
M.~L. Graesser, I.~M. Shoemaker, and L.~Vecchi, {\it {Asymmetric WIMP dark
  matter}},  {\em JHEP} {\bf 1110} (2011) 110,
  [\href{http://xxx.lanl.gov/abs/1103.2771}{{\tt arXiv:1103.2771}}].

\bibitem{Bai:2010hh}
Y.~Bai, P.~J. Fox, and R.~Harnik, {\it {The Tevatron at the Frontier of Dark
  Matter Direct Detection}},  {\em JHEP} {\bf 1012} (2010) 048,
  [\href{http://xxx.lanl.gov/abs/1005.3797}{{\tt arXiv:1005.3797}}].

\bibitem{Buckley:2011kk}
M.~R. Buckley, {\it {Asymmetric Dark Matter and Effective Operators}},  {\em
  Phys.Rev.} {\bf D84} (2011) 043510,
  [\href{http://xxx.lanl.gov/abs/1104.1429}{{\tt arXiv:1104.1429}}].

\bibitem{Fox:2012ee}
P.~J. Fox, R.~Harnik, R.~Primulando, and C.-T. Yu, {\it {Taking a Razor to Dark
  Matter Parameter Space at the LHC}},  {\em Phys.Rev.} {\bf D86} (2012)
  015010, [\href{http://xxx.lanl.gov/abs/1203.1662}{{\tt arXiv:1203.1662}}].

\bibitem{MarchRussell:2012hi}
J.~March-Russell, J.~Unwin, and S.~M. West, {\it {Closing in on Asymmetric Dark
  Matter I: Model independent limits for interactions with quarks}},  {\em
  JHEP} {\bf 1208} (2012) 029, [\href{http://xxx.lanl.gov/abs/1203.4854}{{\tt
  arXiv:1203.4854}}].

\bibitem{Haisch:2012kf}
U.~Haisch, F.~Kahlhoefer, and J.~Unwin, {\it {The impact of heavy-quark loops
  on LHC dark matter searches}},  {\em JHEP} {\bf 1307} (2013) 125,
  [\href{http://xxx.lanl.gov/abs/1208.4605}{{\tt arXiv:1208.4605}}].

\bibitem{Petraki:2011mv}
K.~Petraki, M.~Trodden, and R.~R. Volkas, {\it {Visible and dark matter from a
  first-order phase transition in a baryon-symmetric universe}},  {\em JCAP}
  {\bf 1202} (2012) 044, [\href{http://xxx.lanl.gov/abs/1111.4786}{{\tt
  arXiv:1111.4786}}].

\bibitem{vonHarling:2012yn}
B.~{von Harling}, K.~Petraki, and R.~R. Volkas, {\it {Affleck-Dine dynamics and
  the dark sector of pangenesis}},  {\em JCAP} {\bf 1205} (2012) 021,
  [\href{http://xxx.lanl.gov/abs/1201.2200}{{\tt arXiv:1201.2200}}].

\bibitem{Bell:2011tn}
N.~F. Bell, K.~Petraki, I.~M. Shoemaker, and R.~R. Volkas, {\it {Pangenesis in
  a Baryon-Symmetric Universe: Dark and Visible Matter via the Affleck-Dine
  Mechanism}},  {\em Phys.Rev.} {\bf D84} (2011) 123505,
  [\href{http://xxx.lanl.gov/abs/1105.3730}{{\tt arXiv:1105.3730}}].

\bibitem{Khlopov:2007ic}
M.~Y. Khlopov and C.~Kouvaris, {\it {Strong Interactive Massive Particles from
  a Strong Coupled Theory}},  {\em Phys.Rev.} {\bf D77} (2008) 065002,
  [\href{http://xxx.lanl.gov/abs/0710.2189}{{\tt arXiv:0710.2189}}].

\bibitem{Khlopov:2008ty}
M.~Y. Khlopov and C.~Kouvaris, {\it {Composite dark matter from a model with
  composite Higgs boson}},  {\em Phys.Rev.} {\bf D78} (2008) 065040,
  [\href{http://xxx.lanl.gov/abs/0806.1191}{{\tt arXiv:0806.1191}}].

\bibitem{Belotsky:2014haa}
K.~Belotsky, M.~Khlopov, C.~Kouvaris, and M.~Laletin, {\it {Decaying Dark Atom
  constituents and cosmic positron excess}},  {\em Adv.High Energy Phys.} {\bf
  2014} (2014) 214258, [\href{http://xxx.lanl.gov/abs/1403.1212}{{\tt
  arXiv:1403.1212}}].

\bibitem{Stueckelberg:1938zz}
E.~Stueckelberg, {\it {Interaction forces in electrodynamics and in the field
  theory of nuclear forces}},  {\em Helv.Phys.Acta} {\bf 11} (1938) 299--328.

\bibitem{Kors:2005uz}
B.~Kors and P.~Nath, {\it {Aspects of the Stueckelberg extension}},  {\em JHEP}
  {\bf 0507} (2005) 069, [\href{http://xxx.lanl.gov/abs/hep-ph/0503208}{{\tt
  hep-ph/0503208}}].

\bibitem{Batell:2009yf}
B.~Batell, M.~Pospelov, and A.~Ritz, {\it {Probing a Secluded U(1) at
  B-factories}},  {\em Phys.Rev.} {\bf D79} (2009) 115008,
  [\href{http://xxx.lanl.gov/abs/0903.0363}{{\tt arXiv:0903.0363}}].

\bibitem{Linde:1978px}
A.~D. Linde, {\it {Phase Transitions in Gauge Theories and Cosmology}},  {\em
  Rept.Prog.Phys.} {\bf 42} (1979) 389.

\bibitem{Foot:1991bp}
R.~Foot, H.~Lew, and R.~Volkas, {\it {A Model with fundamental improper
  space-time symmetries}},  {\em Phys.Lett.} {\bf B272} (1991) 67--70.

\bibitem{Volkas:1988cm}
R.~Volkas, A.~Davies, and G.~C. Joshi, {\it {Naturalness of the invisible axion
  model}},  {\em Phys.Lett.} {\bf B215} (1988) 133.

\bibitem{Foot:2013hna}
R.~Foot, A.~Kobakhidze, K.~L. McDonald, and R.~R. Volkas, {\it {Poincare
  Protection for a Natural Electroweak Scale}},
  \href{http://xxx.lanl.gov/abs/1310.0223}{{\tt arXiv:1310.0223}}.

\bibitem{Ahlers:2008qc}
M.~Ahlers, J.~Jaeckel, J.~Redondo, and A.~Ringwald, {\it {Probing Hidden Sector
  Photons through the Higgs Window}},  {\em Phys.Rev.} {\bf D78} (2008) 075005,
  [\href{http://xxx.lanl.gov/abs/0807.4143}{{\tt arXiv:0807.4143}}].

\bibitem{Clarke:2013aya}
J.~D. Clarke, R.~Foot, and R.~R. Volkas, {\it {Phenomenology of a very light
  scalar (100 MeV \&lt; $m_h$ \&lt; 10 GeV) mixing with the SM Higgs}},  {\em
  JHEP} {\bf 1402} (2014) 123, [\href{http://xxx.lanl.gov/abs/1310.8042}{{\tt
  arXiv:1310.8042}}].

\bibitem{Holdom:1985ag}
B.~Holdom, {\it {Two U(1)'s and Epsilon Charge Shifts}},  {\em Phys.Lett.} {\bf
  B166} (1986) 196.

\bibitem{Foot:1991kb}
R.~Foot and X.-G. He, {\it {Comment on Z Z-prime mixing in extended gauge
  theories}},  {\em Phys.Lett.} {\bf B267} (1991) 509--512.

\bibitem{Pospelov:2008jk}
M.~Pospelov, A.~Ritz, and M.~B. Voloshin, {\it {Bosonic super-WIMPs as
  keV-scale dark matter}},  {\em Phys.Rev.} {\bf D78} (2008) 115012,
  [\href{http://xxx.lanl.gov/abs/0807.3279}{{\tt arXiv:0807.3279}}].

\bibitem{HulthenA}
L.~Hulth{\'e}n, {\it {{\"U}ber die eigenlosunger der Schr{\"o}dinger-gleichung
  des deuterons}},  {\em Ark. Mat. Astron. Fys.} {\bf 28 A} (1942), no.~5
  1--12.

\bibitem{Cassel:2009wt}
S.~Cassel, {\it {Sommerfeld factor for arbitrary partial wave processes}},
  {\em J.Phys.} {\bf G37} (2010) 105009,
  [\href{http://xxx.lanl.gov/abs/0903.5307}{{\tt arXiv:0903.5307}}].

\bibitem{Gribov:1998kb}
V.~Gribov, {\it {QCD at large and short distances (annotated version)}},  {\em
  Eur.Phys.J.} {\bf C10} (1999) 71--90,
  [\href{http://xxx.lanl.gov/abs/hep-ph/9807224}{{\tt hep-ph/9807224}}].

\bibitem{Gribov:1999ui}
V.~Gribov, {\it {The Theory of quark confinement}},  {\em Eur.Phys.J.} {\bf
  C10} (1999) 91--105, [\href{http://xxx.lanl.gov/abs/hep-ph/9902279}{{\tt
  hep-ph/9902279}}].

\bibitem{Dokshitzer:2004ie}
Y.~L. Dokshitzer and D.~E. Kharzeev, {\it {The Gribov conception of quantum
  chromodynamics}},  {\em Ann.Rev.Nucl.Part.Sci.} {\bf 54} (2004) 487--524,
  [\href{http://xxx.lanl.gov/abs/hep-ph/0404216}{{\tt hep-ph/0404216}}].

\bibitem{Carlson:1987si}
E.~D. Carlson and S.~Glashow, {\it {Nucleosynthesis versus the mirror
  universe}},  {\em Phys.Lett.} {\bf B193} (1987) 168.

\bibitem{Feng:2008mu}
J.~L. Feng, H.~Tu, and H.-B. Yu, {\it {Thermal Relics in Hidden Sectors}},
  {\em JCAP} {\bf 0810} (2008) 043,
  [\href{http://xxx.lanl.gov/abs/0808.2318}{{\tt arXiv:0808.2318}}].

\bibitem{Feng:2010zp}
J.~L. Feng, M.~Kaplinghat, and H.-B. Yu, {\it {Sommerfeld Enhancements for
  Thermal Relic Dark Matter}},  {\em Phys.Rev.} {\bf D82} (2010) 083525,
  [\href{http://xxx.lanl.gov/abs/1005.4678}{{\tt arXiv:1005.4678}}].

\bibitem{Kolb:1990vq}
E.~W. Kolb and M.~S. Turner, {\it {The Early universe}},  {\em Front.Phys.}
  {\bf 69} (1990) 1--547.

\bibitem{Cirelli:2011ac}
M.~Cirelli, P.~Panci, G.~Servant, and G.~Zaharijas, {\it {Consequences of
  DM/antiDM Oscillations for Asymmetric WIMP Dark Matter}},  {\em JCAP} {\bf
  1203} (2012) 015, [\href{http://xxx.lanl.gov/abs/1110.3809}{{\tt
  arXiv:1110.3809}}].

\bibitem{Mangano:2011ar}
G.~Mangano and P.~D. Serpico, {\it {A robust upper limit on $N_{\rm eff}$ from
  BBN, circa 2011}},  {\em Phys.Lett.} {\bf B701} (2011) 296--299,
  [\href{http://xxx.lanl.gov/abs/1103.1261}{{\tt arXiv:1103.1261}}].

\bibitem{Jaeckel:2012yz}
J.~Jaeckel, M.~Jankowiak, and M.~Spannowsky, {\it {LHC probes the hidden
  sector}},  {\em Phys.Dark Univ.} {\bf 2} (2013) 111--117,
  [\href{http://xxx.lanl.gov/abs/1212.3620}{{\tt arXiv:1212.3620}}].

\bibitem{Lees:2014xha}
{\bf BaBar Collaboration} Collaboration, J.~Lees {\em et.~al.}, {\it {Search
  for a dark photon in e+e- collisions at BABAR}},
  \href{http://xxx.lanl.gov/abs/1406.2980}{{\tt arXiv:1406.2980}}.

\bibitem{Goodsell:2009xc}
M.~Goodsell, J.~Jaeckel, J.~Redondo, and A.~Ringwald, {\it {Naturally Light
  Hidden Photons in LARGE Volume String Compactifications}},  {\em JHEP} {\bf
  0911} (2009) 027, [\href{http://xxx.lanl.gov/abs/0909.0515}{{\tt
  arXiv:0909.0515}}].

\bibitem{An:2013yua}
H.~An, M.~Pospelov, and J.~Pradler, {\it {Dark Matter Detectors as Dark Photon
  Helioscopes}},  {\em Phys.Rev.Lett.} {\bf 111} (2013) 041302,
  [\href{http://xxx.lanl.gov/abs/1304.3461}{{\tt arXiv:1304.3461}}].

\bibitem{Pospelov:2008zw}
M.~Pospelov, {\it {Secluded U(1) below the weak scale}},  {\em Phys.Rev.} {\bf
  D80} (2009) 095002, [\href{http://xxx.lanl.gov/abs/0811.1030}{{\tt
  arXiv:0811.1030}}].

\bibitem{Laha:2013gva}
R.~Laha and E.~Braaten, {\it {Direct detection of dark matter in universal
  bound states}},  {\em Phys.Rev.} {\bf D89} (2014) 103510,
  [\href{http://xxx.lanl.gov/abs/1311.6386}{{\tt arXiv:1311.6386}}].

\bibitem{Foot:2003iv}
R.~Foot, {\it {Implications of the DAMA and CRESST experiments for mirror
  matter type dark matter}},  {\em Phys.Rev.} {\bf D69} (2004) 036001,
  [\href{http://xxx.lanl.gov/abs/hep-ph/0308254}{{\tt hep-ph/0308254}}].

\bibitem{Foot:2010hu}
R.~Foot, {\it {A comprehensive analysis of the dark matter direct detection
  experiments in the mirror dark matter framework}},  {\em Phys.Rev.} {\bf D82}
  (2010) 095001, [\href{http://xxx.lanl.gov/abs/1008.0685}{{\tt
  arXiv:1008.0685}}].

\bibitem{Foot:2013msa}
R.~Foot, {\it {Direct detection experiments explained with mirror dark
  matter}},  \href{http://xxx.lanl.gov/abs/1305.4316}{{\tt arXiv:1305.4316}}.

\bibitem{Fornengo:2011sz}
N.~Fornengo, P.~Panci, and M.~Regis, {\it {Long-Range Forces in Direct Dark
  Matter Searches}},  {\em Phys.Rev.} {\bf D84} (2011) 115002,
  [\href{http://xxx.lanl.gov/abs/1108.4661}{{\tt arXiv:1108.4661}}].

\bibitem{Pospelov:2008jd}
M.~Pospelov and A.~Ritz, {\it {Astrophysical Signatures of Secluded Dark
  Matter}},  {\em Phys.Lett.} {\bf B671} (2009) 391--397,
  [\href{http://xxx.lanl.gov/abs/0810.1502}{{\tt arXiv:0810.1502}}].

\bibitem{Shepherd:2009sa}
W.~Shepherd, T.~M. Tait, and G.~Zaharijas, {\it {Bound states of weakly
  interacting dark matter}},  {\em Phys.Rev.} {\bf D79} (2009) 055022,
  [\href{http://xxx.lanl.gov/abs/0901.2125}{{\tt arXiv:0901.2125}}].

\bibitem{Pearce:2013ola}
L.~Pearce and A.~Kusenko, {\it {Indirect Detection of Self-Interacting
  Asymmetric Dark Matter}},  {\em Phys.Rev.} {\bf D87} (2013) 123531,
  [\href{http://xxx.lanl.gov/abs/1303.7294}{{\tt arXiv:1303.7294}}].

\bibitem{Foot:2012ai}
R.~Foot, {\it {Implications of mirror dark matter kinetic mixing for CMB
  anisotropies}},  {\em Phys.Lett.} {\bf B718} (2013) 745--751,
  [\href{http://xxx.lanl.gov/abs/1208.6022}{{\tt arXiv:1208.6022}}].

\bibitem{Tulin:2012wi}
S.~Tulin, H.-B. Yu, and K.~M. Zurek, {\it {Resonant Dark Forces and Small Scale
  Structure}},  {\em Phys.Rev.Lett.} {\bf 110} (2013), no.~11 111301,
  [\href{http://xxx.lanl.gov/abs/1210.0900}{{\tt arXiv:1210.0900}}].

\bibitem{Tulin:2013teo}
S.~Tulin, H.-B. Yu, and K.~M. Zurek, {\it {Beyond Collisionless Dark Matter:
  Particle Physics Dynamics for Dark Matter Halo Structure}},  {\em Phys.Rev.}
  {\bf D87} (2013) 115007, [\href{http://xxx.lanl.gov/abs/1302.3898}{{\tt
  arXiv:1302.3898}}].

\bibitem{Khrapak:2004apr}
S.~A. {Khrapak}, A.~V. {Ivlev}, G.~E. {Morfill}, S.~K. {Zhdanov}, and H.~M.
  {Thomas}, {\it {Scattering in the Attractive Yukawa Potential: Application to
  the Ion-Drag Force in Complex Plasmas}},  {\em IEEE Transactions on Plasma
  Science} {\bf 32} (Apr., 2004) 555--560.

\bibitem{Khrapak:2003jun}
S.~A. Khrapak, A.~V. Ivlev, G.~E. Morfill, and S.~K. Zhdanov, {\it {Scattering
  in the Attractive Yukawa Potential in the Limit of Strong Interaction}},
  {\em Phys. Rev. Lett.} {\bf 90} (Jun, 2003) 225002.

\bibitem{Khrapak:2004Nov}
S.~A. Khrapak, A.~V. Ivlev, and G.~E. Morfill, {\it {Momentum transfer in
  complex plasmas}},  {\em Phys. Rev. E} {\bf 70} (Nov, 2004) 056405.

\bibitem{Kahlhoefer:2013dca}
F.~Kahlhoefer, K.~Schmidt-Hoberg, M.~T. Frandsen, and S.~Sarkar, {\it
  {Colliding clusters and dark matter self-interactions}},  {\em
  Mon.Not.Roy.Astron.Soc.} {\bf 437} (2014) 2865--2881,
  [\href{http://xxx.lanl.gov/abs/1308.3419}{{\tt arXiv:1308.3419}}].

\bibitem{Buote:2002wd}
D.~A. Buote, T.~E. Jeltema, C.~R. Canizares, and G.~P. Garmire, {\it {Chandra
  evidence for a flattened, triaxial dark matter halo in the elliptical galaxy
  ngc 720}},  {\em Astrophys.J.} {\bf 577} (2002) 183--196,
  [\href{http://xxx.lanl.gov/abs/astro-ph/0205469}{{\tt astro-ph/0205469}}].

\bibitem{Battaglia:2013wqa}
G.~Battaglia, A.~Helmi, and M.~Breddels, {\it {Internal kinematics and
  dynamical models of dwarf spheroidal galaxies around the Milky Way}},  {\em
  New Astron.Rev.} {\bf 57} (2013) 52--79,
  [\href{http://xxx.lanl.gov/abs/1305.5965}{{\tt arXiv:1305.5965}}].

\end{thebibliography}\endgroup
%\bibliography{Bibliography.bib}

\end{document}